%% file: main.tex
\documentclass[journal=jnw]{CUP-JNL-DTM}

\usepackage{anyfontsize} 

\usepackage{graphicx}
\usepackage{multicol,multirow}
\usepackage{amsmath,amssymb,amsfonts}
\usepackage{mathrsfs}
\usepackage{amsthm}
\usepackage{rotating}
\usepackage{appendix}
\usepackage{ifpdf}
\usepackage[T1]{fontenc}
\usepackage{newtxtext}
\usepackage{newtxmath}
\usepackage{textcomp}
\usepackage[dvipsnames]{xcolor} 
\usepackage{mathtools} 
\usepackage{lipsum}
\usepackage[colorlinks,allcolors=blue]{hyperref}

\usepackage{xfrac} 
\usepackage{bbm} 
\usepackage{siunitx} 
\usepackage{dsfont} 

\theoremstyle{definition}

\numberwithin{equation}{section}

\jname{Journal of Nonlinear Waves}
\articletype{RESEARCH ARTICLE}
\jyear{YEAR}

\begin{document}

\begin{Frontmatter}

\title[Article Title]{Data-Driven Equation Discovery for Nonlinear Liquid Film Flows}

\author[1]{Sebastian T. Dooley}
\author[2,3]{Albert P. Bart\'{o}k}
\author[1]{James E. Sprittles}
\author[1]{Radu Cimpeanu}

\authormark{Dooley \textit{et al}.}

\address[1]{\orgdiv{Mathematics Institute}, \orgname{University of Warwick}, \orgaddress{\city{Coventry}, \postcode{CV4 7AL},  \country{United Kingdom}}.
\email{Radu.Cimpeanu@warwick.ac.uk}}

\address[2]{\orgdiv{Warwick Centre for Predictive Modelling}, \orgname{School of Engineering, University of Warwick}, \orgaddress{\city{Coventry}, \postcode{CV4 7AL},  \country{United Kingdom}}}
\address[3]{\orgname{Department of Physics, University of Warwick}, \orgaddress{\city{Coventry}, \postcode{CV4 7AL},  \country{United Kingdom}}}

\authormark{Dooley \textit{et al.}}



\abstract
{
Over the past decade data-driven equation discovery emerged as a powerful alternative to first principles-based methodologies traditionally used in mathematical modelling cycles. The approach provides a promising path towards deep, physical insight into systems that have previously evaded rigorous mathematical derivation procedures, often due to intractable complexity. The strengths of such techniques have been successfully established for many problem classes described by systems of ordinary differential equations and continue to be extended, with their reach into partial differential equation systems gaining momentum, though comparatively nascent. Herein we tackle such a frontier: elucidating the dynamics of liquid film flows, a problem space providing a rich backdrop in terms of asymptotic analytical building blocks. By leveraging expert knowledge and the ability to carefully curate data, we establish a best-case scenario for identifying the underlying governing equations. Even here, multi-collinearity, stemming from the choice of monomial basis functions in our multi-scale flow configuration, introduces complex identifiability issues. Early-time transients compound this further, as the most dynamically rich behaviour carries the largest residuals in training data. Pinpointing such vulnerabilities allows us to better define the boundaries of current discovery techniques and paves the way for the next generation of more resilient, numerically stable algorithms.
}

\end{Frontmatter}


\localtableofcontents 

\input{body_concise}

\begin{appendix}
\input{Appendix}
\end{appendix}

\begin{Backmatter}

\paragraph{Acknowledgments}
Sebastian T. Dooley is grateful for the computing resources supplied by the University of Warwick Scientific Computing Research Technology
Platform (SCRTP).

\paragraph{Funding Statement}
All co-authors gratefully acknowledge funding from the UK Engineering and Physical Sciences Research Council (EPSRC) grant EP/S022848/1 for the University of Warwick Centre for Doctoral Training in Modelling of Heterogeneous Systems (HetSys).  

\paragraph{Competing Interests}
The authors report no conflict of interest.

\paragraph{Data Availability Statement}
A compact version of the code used for this paper, along with installation instructions and documentation, can be found on GitHub via the following URL: \url{https://github.com/SebDooley/equation_discovery_liquid_film_flows}. For the purpose of open access, the authors have applied a Creative Commons Attribution (CC BY) licence to any arising Author Accepted Manuscript version.

\paragraph{Ethical Standards}
The research meets all required ethical guidelines.

\paragraph{Author Contributions} Conceptualisation: S.T.D; A.P.B; J.E.S; R.C. Investigation: S.T.D; A.P.B; J.E.S; R.C. Methodology: S.T.D; A.P.B; J.E.S; R.C. Software: S.T.D. Data curation: S.T.D. Formal analysis: S.T.D. Data visualisation: S.T.D. Writing - original draft: S.T.D; J.E.S; R.C. Writing - review \& editing: S.T.D; A.P.B; J.E.S; R.C. All authors approved the final submitted draft.


\printbibliography

\end{Backmatter}

\end{document}

%% file: body_concise.tex
\section{Introduction}
\input{intro_and_lit_review}

\section{Methodology}
\label{sec: methodology}
\input{methodology}

\section{Thin Film Equations}
\label{sec: Thin_Liquid_Films}
\input{thin_films_chapter}

\section{Surface Tension Driven Thin Film Equation}
\label{sec: nonlinear_thin_film_basic}
\input{non_gravity}

\section{Gravity and Constant Surface Tension Affected Thin Liquid Film}
\label{sec: Gravity_thin_film}
\input{gravity_chapter}

\section{Discussion and Conclusions}
\label{sec: Conclusion}
\input{conclusion}

%% file: intro_and_lit_review.tex
Liquid film flows have practical applications and experimentally realisable configurations that provide valuable data for analysis and validation \citep{reeves2025nonlinearwavedynamicschip}. Associated industrial applications include coating flows \citep{SCHMITT_battery_coating_2013}, heat transfer through falling film evaporators \citep{RIBATSKI_Jacobi_2005_Heat_Transfer}, and chemical mass transfer \citep{Hessel_Chemical_Mass_Transfer_2000}. Practical relevance is a strong motivation for modelling thin liquid film flows, which provide a canonical system that has been at the heart of modelling and the development of partial differential equation (PDE) theory for decades \citep{Craster_Matar_2009}. Modelling efforts allow researchers to probe non-observable fields, analysing flows in regimes that are difficult to access experimentally, such as at very small scales or within enclosed devices. Falling liquid films are one such thin film flow and liquid films running downhill produce beautiful natural phenomena that are scientifically interesting \citep{kalliadasis_ruyer_quil_scheid_velarde_2011}. The system can display instabilities dependent on film thickness in which, after onset, quasi-2D waves develop and propagate before resulting in 3D disordered patterns that may demonstrate chaotic behaviour \citep{RevModPhys_69_931, Thompson_Gomes_Stabilising_falling_liquid_film_flows_using_feedback_control}.

\subsection{Mathematical Modelling}
There is currently a well-established hierarchy of thin film equations  \citep{kalliadasis_ruyer_quil_scheid_velarde_2011, Tomlin_Gomes_Pavliotis_Papageorgiou_2018_OptimalCO} including the Kuramoto-Sivashinsky (KS) equation, the Benney equation (first-order), the weighted-residual equations (WRIBL) and full Navier-Stokes formulations, with each respective equation representing an increase in accuracy, but also computational cost, for solution data \citep{Cimpeanu_Gomes_Papageorgiou_2021}. The full Navier-Stokes systems are at the head of this hierarchy. Whilst well-studied, an analytical solution to such systems is famously unknown. Numerical solutions are available, but the cost of direct numerical simulation (DNS) is often prohibitively high. Meanwhile, alternative reduced-dimensional models are not universally valid long-wave models, with either very limited applicability in the case of the KS equation, or exhibiting finite-time blow-up, as for the Benney equation  \citep{scheid_Ruyer_Quil_Thiele_Kabov_Legros_Colinet_2005_Benney_validity}. Weighted residual approaches are often utilised \citep{Cimpeanu_Gomes_Papageorgiou_2021, OAHolroyd_LQR_control, Ruyer_Quil_Manneville_2000,Thompson_Tseluiko_Papageorgiou_2015} as a powerful bridge in this modelling landscape, balancing analytical tractability and numerical efficiency. However, the weighted residual approach remains fundamentally constrained by the requirement of scale separation between the film thickness and streamwise length-scales. These restrictions are overcome only in geometrically privileged configurations, such as the scenarios studies by \citet{Wray_Cimpeanu_2020,Wray_Papageorgiou_Matar_2017}. This motivates the need for development of new reduced-dimensional model derivation capabilities in this space.

Due in no small part to the vast improvement in computing hardware, researchers can routinely generate simulation data and numerically validate theoretical results \citep{Papageorgiou_2019_electric_field_review, Craster_Matar_2009, CHATZIGIANNAKIS_2021_disjoining_pressure_and_coupling_forces}. Combined with the gradual incorporation of more sophisticated modelling approaches, modern hardware has enabled a transition from reliance on simplified analytical solutions to approaches using high-fidelity numerical simulations. Existing models have progressively been supplemented with more complex physics. For example, surface-active molecules and particles may collect at fluid interfaces to yield interesting nonlinear behaviour of these complex fluid-fluid interfaces \citep{Mayer_Oswald_Papageorgiou_2026}. Whilst this combination of modelling and computation has been transformative, limitations remain. Given complex mathematical models, computational power still represents a bottleneck in many cases, particularly 3D scenarios, and model reduction is of key interest to many researchers \citep{Chang_1994, Lavalle_Vila_Blanchard_Laurent_Charru_2015}. Extending existing thin film models to more demanding flow regimes, such as from well-characterised low-Reynolds-number flows towards moderate Reynolds numbers, remains challenging. Additional physical complexity, such as the effects of surfactants \citep{swanson_strickland_shearer_daniels_2014_SURFACTANTS} or capturing film dewetting phenomena \citep{pylyp_volodin_alexey_kondyurin_2008_nano_roughness,Witelski_2020_dewetting}, compounds this further, yet lies beyond the present scope of our exploration. Collectively, these challenges indicate the need for modelling approaches that provide a route towards more general and physically representative models that can later incorporate unresolved physical effects beyond the reach of conventional modelling assumptions. Therefore, to build on previous modelling work, we can use the plethora of readily available experimental and numerical datasets they provide to facilitate such an approach.

\subsection{Equation Discovery}
The wide range of accessible datasets mark thin liquid film systems as a prime candidate for data-driven equation discovery techniques. The goal of PDE discovery is to determine a governing PDE system that explains the observations provided well and is physically interpretable. As a result, this body of work is distinct from that of surrogate modelling purely for quantitative prediction. Equation discovery can serve researchers in a variety of ways. It provides an alternative to standard first principles-based methodologies used for mathematical modelling. It can complement analytical foundations where there is good understanding for at least a subset of the solution structure. It may also be used for purely exploratory purposes. As discussed previously, there exist regimes currently outside the reach of traditional approaches involving modelling and computing discrete solutions. It is hoped that data-driven methodologies will eventually provide rapid insight into these regimes for physical interpretation and predictive capabilities.

PDE models are highly desirable as a form of governing equation to explain physical phenomena and are ubiquitous throughout scientific literature. PDEs allow the use of the full range of mathematical techniques developed over previous centuries. This includes producing analytical and numerical solutions to resolve models produced by data-driven processes. There are many well-developed mathematical tools that we can apply to PDEs, such as asymptotic analysis, which allow us to create benchmark solutions in certain parameter regimes. Some PDE systems may only model a subset of an associated parameter space well, so there is the ability to extend existing PDE models. This can be done in numerous ways, including modelling additional physics \citep{Lewin_Jones_Lockerby_Sprittles_2024}, or fitting a data-driven corrective term, such as in the case of residual networks \citep{e22020193_Chen_Dongwei, Martin_Linares_2023,CUI2024108594_hamsters}. Extending existing models provides an exciting initial use case of equation discovery. Through the careful provision of training data and physically informing algorithms, one can discover the mathematical form of errors within models. Equation discovery could also aid research by informing models that attempt to work at scales between well-studied regions \citep{Brunton_Noack_Koumoutsakos_2020}.

In principle, one could attempt to produce models over an unbounded function space and yield an arbitrarily complex model. As is commonly observed in machine learning literature \citep{Brunton_Noack_Koumoutsakos_2020}, such a model would likely fit training data well and fail to generalise for unseen, out-of-distribution conditions. An excessively complex model is also likely to have terms that result in an impractical numerical solution, due to ill-posedness or stiffness, making reliable simulation and quantitative comparison with data unfeasible \citep{kassam_trefethen_2005_stiffness_PDEs}. Thus, obtaining parsimonious models is an important goal when discovering equations and evaluating their solutions. Under this assumption, the problem of discovering PDEs from data can be formulated as one of identifying a sparse representation within a high-dimensional candidate space. This desire for sparsity results in equation discovery literature frequently referencing the problem-solving principle of Occam's razor \citep{Kaheman_Kutz_Brunton_SINDy_PI_2020} and so the sparsity assumption underlies the majority of work produced in the field \citep{Brunton_Kutz_Fasel_Zolman_2025}. This aligns with broader developments in data-driven and physics-informed modelling, where learning governing equations is posed as an optimisation problem that balances data fidelity and model complexity \citep{Kutz_2017_DL_for_FD, karniadakis_kevrekidis_lu_perdikaris_wang_yang_2021}.

Sparse regression methodologies therefore form a key branch of equation discovery literature. They were popularised with the introduction of the sparse identification of nonlinear dynamics (SINDy) \citep{Brunton_Proctor_Kutz_2016}, which was extended to constant coefficient PDEs \citep{Rudy_Brunton_Proctor_Kutz_2017} and then to parametric PDEs \citep{rudy2018datadrivenidentificationparametricpartial}. Equation discovery requires performing structural identification and determining coefficient values, so researchers must consider the loss function used and optimisation routines. SINDy popularised the use of sequentially thresholded least squares (STLSq)  to promote sparsity, with \citet{Rudy_Brunton_Proctor_Kutz_2017} recommending sequentially thresholded ridge regression (STRidge), though both works recognised alternative optimisers, such as the least absolute shrinkage and selection operator (LASSO). LASSO is an intensely-studied, convex and principled regression approach \citep{hastie_tibshirani_tibshirani_2020} that was applied by \citet{Schaeffer_2017} for equation discovery, with successes on a variety of PDEs including variants of Burgers' equation and Navier-Stokes systems. However, other works have shown mixed performance with high computational cost \citep{kaptanoglu_benchmarking_2023, Rudy_Brunton_Proctor_Kutz_2017}. A major extension to the SINDy framework was that of weak-SINDy \citep{MESSENGER2021110525}, which uses Galerkin projection and a smooth, compact test function to `move' derivatives to the test function. Doing so mitigates numerous issues relating to numerical differentiation and instead relies on numerical integration, which is a far more numerically robust process. This method is analogous to a convolution performing low-pass filtering, thus removing high-frequency dynamics/noise. More recently, there have been further developments to the family of sparse regression based methodologies including Laplace enhanced SINDy (LES-SINDy) \citep{zheng2024lessindylaplaceenhancedsparseidentification}, which uses the Laplace transform to relax assumptions of absolute integrability of candidate functions that are required when applying weak-SINDy with the Fourier transform. 
Another key methodology is that of E-SINDy \citep{Fasel_Kutz_Brunton_Brunton_2022}, which utilises bootstrapping and a frequentist framework to improve the robustness of equation discovery. Bayesian alternatives exist, such as Bayes PDE\_FIND \citep{martina_perez_simpson_baker_2021}, which uses PDE\_FIND to define prior distributions that are updated by approximate Bayesian computation. Another key approach from a Bayesian perspective is that of the BINDy framework \citep{champneys_rogers_2024}. This work uses reversible-jump Markov-chain Monte-Carlo sampling methods to overcome difficulties regarding the dimension of the parameter vector depending on the particular model it parameterises. Using this sampling method, the work allows access to the joint posterior over both models and their parameterisations, though focuses on application to ODE cases. Taken together, these developments reflect significant progress in sparse model discovery, yet operate within a shared framework of representing governing dynamics from a prescribed library of candidate functions. 

The success of sparse regression methodologies is closely tied to the completeness of the function library and whether all relevant terms are known \textit{a priori}. In many applications, this is a key issue drawing the attention of numerous researchers. Rather than fixing a function library at the start of the discovery process and searching for strong candidate systems within the span of a finite dimensional sub-space, many works involve building or altering a function library \citep{PhysRevResearch_4_023174_SGA_PDE, xu_chen_cao_tang_du_li_callaghan_zhang_2025, cranmer2020pysr}. Some methods involve fixing functional forms of library terms and optimising nonlinear parameters within the function library, such as the frequency of trigonometric terms \citep{ADAM_SINDy}. We see then that for the sparse regression methodologies we employ, the requirement of candidate library selection shifts the burden of modelling from equation derivation to library design.

Overall, this emerging field has provided success for numerous test cases that probe varying degrees of nonlinearity for ODEs and PDEs. The constructed framework is flexible and modular in the sense that a variety of extensions (many described above) have been proposed to address specific challenges and can be quickly implemented into base codes. Through careful, expert-guided sparsity tuning, the generated PDEs can explain out-of-sample data and be thought of as generalisable. In environments where a `correct' model is unclear, comparison of models is viable to attempt to uncover such a general governing PDE. Applying these equation discovery frameworks in these unknown environments remains one of the key challenges in this research space, with a significant proportion of prior development restricted to relatively well-known regions of interest.

\subsection{Data-Driven Mathematical Modelling of Liquid Film Flows}
In the above, we have highlighted how the emerging area of equation discovery has already led to significant empirical success \citep{heinrich2025rediscovering_KdV_ARXIV}. Several canonical PDE-based models have been robustly retrieved, some in the presence of significant noise \citep{Rudy_Brunton_Proctor_Kutz_2017}. We look to present the behaviour of the algorithm for more complicated reduced-order models and look to build on previous works through the study of two equations of interest to the thin liquid film community with single-equation form $h_t = \mathcal{N}(h),$ where $\mathcal{N}$ represents a highly nonlinear differential operator. Whilst there is overlap in the modelled physics, we show that there is increased complexity when recovering systems with coupled physical effects, such as gravity and surface tension. However, having discussed a wide, though non-exhaustive, list of equation discovery methods, the numerous variants and proliferation of names of new methods in the past decade are potentially symptomatic of theoretically thin grounding, and some researchers have compared machine learning practice with alchemy \citep{Hutson_2018_ALCHEMY}. Nevertheless, the development of such variants is productive in tackling particular modes of failure of prior studies. Looking towards model equations governing liquid film dynamics that lie beyond traditional benchmarks, guarantees of success are lacking and there exist limitations in the study of the practical identifiability of PDE systems \citep{salmaniw_browning_2025}. Thus, further study is motivated when looking to apply equation discovery methods to novel environments, and we deem confidence in a well understood area of modelling to be critical.

The paper is structured as follows. We begin by formulating the building blocks for an equation discovery methodology and delving into relevant numerical approaches in Section~\ref{sec: methodology}. We then refocus our attention to thin liquid films in Section~\ref{sec: Thin_Liquid_Films}, highlighting the details of existing analytical work that can inform the discovery process. Results are then discussed in Section~\ref{sec: nonlinear_thin_film_basic} for a surface tension driven flow, before Section~\ref{sec: Gravity_thin_film} where gravity is introduced to form a classical model described by competing physical mechanisms. This structure allows for the application and analysis of equation discovery for a highly nonlinear system that is extremely relevant to thin liquid film modelling. Finally, a discussion and conclusions follow as part of Section~\ref{sec: Conclusion}. 

%% file: methodology.tex
Having discussed a collection of thin liquid film models that balance complexity and fidelity, we look to construct an equation discovery framework from existing tools. This framework will allow for systematic interrogation through varying the training data provided and the targeted system for retrieval.

\subsection{Equation Discovery through Sparse Regression}
\label{sec: Eqn_Discovery_Methodology}
To perform equation discovery, we use a foundational basis derived from the combination of the STability-based Robust IDEntification of PDEs (PDE-STRIDE) framework \citep{Maddu_Cheeseman_Sbalzarini_Muller_2022_PDE_STRIDE} and the Ensemble Sparse Identification of nonlinear Dynamics (E-SINDy) \citep{Fasel_Kutz_Brunton_Brunton_2022} approach, to which we apply suitable adaptations for thin liquid film dynamics equation discovery. This includes incorporating a tolerance sampling approach \citep{Supekar_Song_Hastewell_Choi_Mietke_Dunkel_2023}, referred to as line searching, and data sub-sampling considerations \citep{ VandenHeuvel_Buenzli_Pascal_Simpson_2024}.

\subsubsection{Core Algorithm}
We begin by discussing SINDy \citep{Brunton_Proctor_Kutz_2016}. This popular method has been successfully applied to numerous classical PDE settings through the PDE\_FIND algorithm \citep{Rudy_Brunton_Proctor_Kutz_2017}. The use of SINDy assumes that physical systems, in general, have only a few relevant terms that encapsulate the dynamics of the system, which means that the governing equations are sparse in high-dimensional nonlinear function spaces.

As input, the algorithm uses a matrix of the data collected. For our scalar thin film data, this will be given by $H \in \mathbb{R}^{N_x \times N_t}$, where $N_x$ is the total number of spatial grid points and $N_t$ is the number of time steps. The entries $H_{ij}$ correspond to the interfacial height at point $x_i$ and time $t_j$, with $i=1, \dots, N_x \textrm{ and } j=1, \dots, N_t.$ Our inputted data is typically noiseless; however, in the appendix, we consider the impact of pointwise, additive Gaussian noise included as a post-process to data generation.

The methodology requires the user to specify a closed library of functions that are to be numerically approximated. This may consist of a family of polynomials to a finite, maximum degree, as well as cross products with low-order spatial derivatives. The assumed form of the partial differential equation to describe the input dataset is
\begin{equation}
    h_t(x, t) = \mathcal{N}(f_1(h), f_2(h), \dots, Q),
    \label{eq: PDE_FIND_Linear_Combo_Assume}
\end{equation}
where the operator $\mathcal{N} \in \text{span}\{ N_1, N_2, \dots, N_r \}$ for $r$ basis operators, so $$\mathcal{N} = \sum_{j=1}^r  N_j (f_1(h), f_2(h), \dots, Q)\xi_j,$$ with elements of the function library $f_i(h)$ being functions of the observed, potentially high-dimensional, state variable, and $\xi$ is a column vector containing the coefficients of each term $N_j(\dots)$ (see  \citet{Rudy_Brunton_Proctor_Kutz_2017} for additional details on this construction). The parameter $Q$ is representative of additional information that may be known about the system, such as a velocity field that is discretised on the same spatio-temporal domain. Thus, considering equation \eqref{eq: PDE_FIND_Linear_Combo_Assume} shows the importance of the choice of library, as excluded terms will not be recovered, regardless of their importance. A typical initial application of the method may involve a library of monomial-based terms such as $$\{ h^k, h^kh_{x}, h^kh_{xx}, h^kh_{xxx}, h^kh_{xxxx} \}_{k=0}^{n_{max}}$$ to provide a rich function library. For such a library, a maximum derivative order and monomial degree requires careful specification, which can incorporate \textit{a priori} known information about the physical problem to be modelled.

With an established library of basis functions, we look to form a linear system in which an assumed form of left hand side specified \textit{a priori}, such as $h_t$, may be expressed as a linear combination of the elements of the function library that can themselves be nonlinear in $h$. The nonlinearity of the basis functions enables recovery of nonlinear systems. Forming the linear system requires using the inputted data to estimate a first-order temporal derivative, $h_t(x,t)$, evaluated at each point in the spatio-temporal mesh. A simple second-order central finite difference scheme may be applied to spatial and temporal derivatives, using only local information and suitable boundary conditions.

Having estimated derivative values, a linear system may be formed by vectorising point-wise values, 
\begin{equation}
    H_t = \Theta (H, Q) \xi, \textrm{ where } H_t \in \mathbb{R}^{n\times q}, \Theta\in \mathbb{R}^{n\times p}, \xi \in \mathbb{R}^{p\times q},
    \label{eq: PDE_FIND_formed_linear_system}
\end{equation}
with $q=1$ for single-equation applications. This system can be sparsely regressed to provide coefficient values. To enforce sparsity, ridge regression and hard sequential thresholding are applied.

\subsubsection{Ridge Regression}
Ridge regression is a commonly used regularisation method in supervised machine learning \citep{Hastie01102020}. It applies a $L_2$ regularisation penalty to the least squares estimate of the regression, thus restricting the size of coefficients and in a sense rewarding sparsity in high-dimensional systems, $$\xi^{\hbox{ridge}} = \arg\min_{\xi\in\mathbb{R}^p}\{ ||\Theta\xi - H_t||_2^2+\lambda_{\hbox{ridge}}||\xi||_2^2\}.$$ The weighting of the regularisation penalty is controlled by a regularisation hyper-parameter $\lambda$, which has a value to be optimised for the environment in which it is deployed. Ridge regression makes coefficients of highly correlated variables similar to each other instead of performing selection \citep{james_witten_hastie_tibshirani_2013, hastie_tibshirani_friedman_2009}. Consequently, ridge regression alone suffers from over-fitting, as weight is assigned in all dimensions. This motivates the use of hard sequential thresholding to enforce sparsity \citep{herrity2006sparse, blumensath_davies_2008, Rudy_Brunton_Proctor_Kutz_2017}, as described in the following subsection.

\subsubsection{Sequential Thresholding}
A hard thresholding technique is used to detect coefficients with values below a specified threshold $\Delta \in \mathbb{R}_{\geq0}$ and discard the corresponding basis functions to leave a reduced support set $$\textrm{supp}(\hat{\xi}) \coloneq \{j : \hat{\xi}_j \neq 0 \}, \textrm{ where }  \hat{\xi}_j \coloneq \xi_j^{\hbox{ridge}} \cdot \mathds{1} (|\xi_j^{\hbox{ridge}} | \geq \Delta).$$ The support set is initially equivalent to the function library. This allows for regression to be performed recursively on a shrinking library of basis functions. The regression can be halted when the results stop changing or when a specified maximum number of regressions has been performed, or indeed when a further iteration would result in an empty set of remaining features and thus a null model. Previous studies have confirmed the convergence to a stable support set of this sequentially thresholded ridge regression procedure with at most $p$ steps required \citep{Zhang_Schaeffer_2019}. The result is a local minimiser of the non-convex problem. Subsequently, sequential thresholding introduces an additional hyper-parameter, in the form of the sequential thresholding tolerance, $\Delta$. As a result, the optimiser uses coefficient magnitude as a proxy for importance and naturally one may question the ability to detect features with small magnitude coefficients. To combat this, we may normalise columns of the design matrix with respect to the $L_2$ norm before sequential thresholding, which preconditions the system and aids in detecting important features with small coefficient magnitudes. A remaining valid concern may be for subtle features with substantial impacts on the long-term evolution of a system. We comment on this aspect later in subsection \ref{sec: Model_Testing_and_Selection}.

\subsubsection{Hyperparameter Optimisation}
The original PDE\_FIND algorithm employs an algorithmic approach to searching the hyper-parameter space, but results in trajectories dependent on the initial hyper-parameter values chosen \citep{huang_mabrouk_gompper_benedikt_sabass_2022}. Thus it may fail to robustly identify a globally optimal model \citep{Maddu_Cheeseman_Sbalzarini_Muller_2022_PDE_STRIDE, kaptanoglu_benchmarking_2023}.  

Instead of optimising the threshold value, we instead opt for a line search methodology. A critical tolerance is naturally defined by the tolerance value at which all features would be removed by a single application of thresholding to the ridge solution. When using a tolerance value slightly less than such a critical value, it is likely that model complexity will be low, and so a 1D line search of decreasing tolerance values starting from the critical thresholding value will produce a collection of models of typically increasing complexity. A lower bound is naturally formed by the value at which thresholding would not remove any features, although this would result in high-complexity models that we would deem to be overfitted, through the assumption of sparsity. Consequently, we terminate the line search at a potentially larger threshold value of $\epsilon \Delta_c$, where a heuristically chosen $\epsilon = 10^{-3}$ is used. A log-linear line search is chosen.

For each value of tolerance sampled, a thresholding path is taken in which features are removed in an order dependent on the path taken. Whilst convergence is assured, important terms may be removed seemingly erratically, which could result in spurious terms remaining in the final model structure. When thresholding terminates, a set of features remains that define the structure of a potential model.

\subsubsection{Bootstrapping}
Through numerical experimentation, we have encountered that the initial linear system can be very ill-conditioned due to the choice of function library and structural multi-collinearity, a property also remarked upon in previous studies \citep{Kaheman_Kutz_Brunton_SINDy_PI_2020,delgado_cano_kracht_fasel_herrmann_2025}, as the design matrix columns are constructed from the same inputted training dataset. Due to this ill-conditioning, we note that the thresholding path is sensitive to the data samples used. Therefore, we apply bootstrap sampling to reduce this sensitivity. Doing so, we generate normalised feature importance scores related to the proportion of times a feature is retained in the final generated model (when thresholding terminates). With these importance scores, we may apply an importance threshold in $[0,1]$ to retain important terms and form a final set of features that determine the candidate model. This is the case for all sampled tolerance values, so we take unique combinations of terms to generate a collection of candidate models. The cardinality of such a set of candidate models is bounded above by the number of threshold samples taken and below by $1$, due to the threshold values being sampled from $[\epsilon \Delta_c, \Delta_c]$. At this stage, one may analyse the collection of models further to remove models based on a complexity threshold or from an identifiability-based condition, for example.

\subsubsection{Recalibration}
\label{sec: methodology_recalibration}
The feature selection within the framework is complete at this stage, but the coefficients of each candidate model are yet to be fixed. We re-apply bootstrap sampling with a recalibration method, such as Levenberg-Marquardt \citep{levenberg_1944, marquardt_1963} or least squares to produce coefficient vectors to accompany the selected features and produce the set of calibrated candidate models. Once again, bootstrap sampling is used to mitigate sensitivity to the dataset samples used, due to the potential ill-conditioning of the reduced linear system, which is of particular concern to moderate to high complexity candidate models. Bootstrapping provides coefficient distributions from data, from which a median value is used to reduce sensitivity to potential anomalous results. We highlight that these distributions can be used to form a prior distribution for Bayesian equation discovery methods, and can be manually inspected for multi-modality.

One may view this bootstrapping procedure as unnecessary, because of the prior bootstrapping procedure. The feature selection bootstrap does generate weights associated with terms and, due to the importance cut-off being applied, there are a reasonable number of samples of the coefficient. However, the distribution of each coefficient is over a mixture of potentially numerous different model configurations. Such model configurations may yield zero weights in the case of exclusion and otherwise there may be competition with different subsets of the other features as model structure varies. Thus, once a fixed collection of terms is selected to generate a final model structure, the mixed distribution is no longer appropriate. To obtain coefficient distributions that are not distorted by the selection step, conditioning on the selected model and applying a separate recalibration bootstrapping procedure is required. Therefore, coefficient distributions no longer reflect covariance patterns that arise from varying feature sets. This leaves us with a set of calibrated candidate models and we may now apply a model selection criterion if a single output equation is desired. This is discussed in section \ref{sec: Model_Selection_Methodology}.

Having carefully described a roadmap for an equation discovery method suitable for our general target system up to this stage, in what follows we address associated computational details including data handling, optimisation and regression practices. 

\subsection{Preprocessing}
Without prior knowledge of any highly localised dynamics, we interpolate to standardise the inputted data on a uniform mesh with a fixed number of points. We can then apply numerical differentiation methods to uniform data and have a strong base of comparison. Interpolation methods commonly rely on the provided data points being the ground truth, so noise-corrupted data can generate poor interpolated data, especially if the magnitude of variance of the noise is larger or comparable to the magnitude of values of the data.

Our numerical experiments include interpolation as a standard pre-processing step, and we apply cubic interpolation with anti-aliasing to a uniform mesh described by $(N_x, N_t) = (512,256)$, which is still reasonably dense and allows for sufficient accuracy of the approximation of high-order derivatives with clean data. Our mesh is given by $$\left\{\frac{dx}{2}, \dots, L_X-\frac{dx}{2}\right\}\times \{0, dt, \dots, t_F\} \in [0,L_X]\times[0,t_F].$$

\subsection{Numerical Differentiation}
\label{sec: numerical_differentiation}
For strong form equation discovery, we require point-wise derivative estimates, which are found from input data $\{h(x_i,t_j) \}_{i=1, \dots, N_x}^{j=1, \dots , N_t}$. 

With a periodic spatial domain, it is natural to apply spectral differentiation. We use the fast Fourier transform to do this, enabled by the regularity of our typical solution data. Whilst training data involves nonlinear wave behaviour, it does not contain discontinuous or complicated e.g. multi-valued regimes such as those involving wave breaking that may lead to issues with spectral derivatives.

For temporal derivatives, we use central differencing methods, with equal order accuracy forward/backward methods for derivative approximations at the temporal boundaries. For clean data cases, our data is assumed to be smooth and sufficiently regular such that higher-order finite difference methods make little difference for temporal derivatives, so we use second-order accurate methods. 

\subsection{Choice of Optimiser}
It is at this stage of the process that a linear system can be formed reliably, so sparse regression is required to satisfy parsimony assumptions. 

Solving 
\begin{equation}
||\Theta\xi-b||_2^2 + \lambda_0||\xi||_0, \quad \Theta\in\mathbb{R}^{n\times p}, b\in \mathbb{R}^n, \xi \in \mathbb{R}^p
\end{equation}
is an NP-Hard problem \citep{Natarajan_1995_NP_Hard_L0}. A standard technique is to solve a convex relaxation that approximates the non-convex global solution reliably. We choose to use sequentially thresholded ridge regression (STRidge), which is an algorithm known to converge within a number of steps bounded above by the number of features of our design matrix \citep{Zhang_Schaeffer_2019}. Whilst convergence is guaranteed, it is not necessarily to the correct system structure, even when correct terms are included.

STRidge approximates best subset selection \citep{bertsimas_gurnee_2023}. The ridge component of STRidge improves the numerical stability of coefficient estimates when library terms are correlated and so multi-collinearity is handled predictably. Near-collinear terms are grouped and their effects are distributed more uniformly, although scale separation can result in varied coefficient values. Correlated library terms are a common issue in PDE discovery settings. The stability STRidge provides in these settings affects variable selection and, when applied to bootstrapped datasets, STRidge yields consistent term inclusion \citep{Fasel_Kutz_Brunton_Brunton_2022}, unlike alternatives like LASSO, which is known to select spurious terms due to correlated predictors \citep{Freijeiro_Gonzalez_Febrero_Bande_Gonzalez_Manteiga_2022}. Ridge regression also offers a simple, closed form solution. PDE discovery typically involves a tall-and-skinny design matrix, due to a small to moderate sized function library with many observations. Therefore, from a computational speed perspective, ridge regression is particularly attractive when compared to alternate convex regularisers that rely on iterative methods for solving.

A key element of these regularisation methods lies in selecting a regularisation strength, $\lambda$. We select the regularisation strength heuristically from observing the strength of performance of STRidge. The use of STRidge appeared relatively insensitive to the choice of regularisation strength compared to other hyper-parameters, such as the thresholding tolerance, in our investigation. An arguably more principled selection and common method from the machine learning community is to apply k-fold cross-validation \citep{hastie_tibshirani_friedman_2009}, however this is computationally costly. It is also common in the equation discovery community to apply a Pareto frontier analysis, \citep{Brunton_Proctor_Kutz_2016, Rudy_Brunton_Proctor_Kutz_2017} though this is commonly used for selecting just a thresholding tolerance and not jointly tuning multiple hyper-parameters. This choice has been shown to produce sparse estimates of equations that typically avoid additional spurious terms that are included by cross-validation based methods that overfit to reduce prediction error \citep{alexandre_cortiella_park_alireza_doostan_2021}. Alternatively, one may apply a Bayesian perspective to ridge regression, injecting prior knowledge of coefficient magnitudes to choose a stronger centred normal distribution prior with variance $1/\lambda$. This requires knowledge of expected coefficient magnitudes, which we do not assume.

\subsection{Improving Conditioning of the Linear System}

The formed linear system is typically initially poorly conditioned. The use of monomials within the function library worsens this. We use $L_2$ normalisation of the columns of the design matrix to reduce the condition number, which typically reduces the condition number by a large factor of $\mathcal{O}(10)$. With this normalisation, we then apply a re-scaling to enforce $||\Theta||_2=1$, so our linear system is consistent with the convergence analysis of \citet{Zhang_Schaeffer_2019}.

The full system is still commonly ill-conditioned, impeding us from obtaining results in a streamlined manner. Conditioning improves as sequential thresholding reduces the support of the regression. The initial conditioning could be improved by using a smaller function library, but this compromises the use of equation discovery for systems with unknown dynamics. In section \ref{sec: Gravity_thin_film} we discuss the use of a smaller function library spanning a rich function space. Furthermore, we also consider sampling strategies to mitigate the near-linear dependence between rows of the design matrix $\Theta$ in equation \eqref{eq: PDE_FIND_formed_linear_system} that comes from high spatio-temporal resolution required for accurate derivative approximations.

\subsection{Bootstrapping and Aggregating}
With poor conditioning likely, we are motivated to look towards bootstrap sampling to improve robustness of equation discovery. To perform bootstrap sampling, we take subsamples of rows of the design matrix $\Theta$ with replacement. This produces subsets with an expected unique proportion $1-e^{-1} \approx 63.2\% $ of the original rows and we use $200$ of these subsets to produce statistics about the predictors. The choice of $200$ bootstraps is typically deemed sufficient \citep{efron_tibshirani_1994_intro_to_bootstrap}. 
As mentioned previously, we apply bootstrap sampling to generate a set of model structures as well as to independently recalibrate each model. 

Our subsampling procedure is equivalent to generating sampling matrices $\{ S_i\}_{i=1}^{200}$, in which each $S_i$ is a row sampling matrix obtained by selecting rows of $I_n$ with replacement. With these samples, we have a linear system $S_i\Theta \xi=S_ib$ with solution $\hat{\xi} \in \mathbb{R}^p$. Previous work \citep{jones_subapsnap_2025_PRE_PRINT_ONLY}, has revealed that the residual of the sub-sampled least squares problem is related to the full least squares problem by the inequality, $$1 \leq \frac{||\Theta\hat{\xi}-b||_2}{||\Theta\xi^*-b||_2} \leq \frac{||S_i||_2}{\sigma_{\min}(S_iU)},$$ where $U$ relates to the polar decomposition $\Theta = UH$ and $\xi^*$ is the minimiser of the full least squares problem. We note that $S_iU$ with full column rank is required for a finite bound. Whilst dependent on the structure of the matrix $\Theta$, we have a large number of expected unique samples with $n\gg p$ and assume that the probability of selecting a rank-deficient subset is negligible. Thus, with a bounded residual, we have improved confidence in the quality of the estimate of model coefficients, and taking the median is now well justified.

\subsection{Model Selection}
\label{sec: Model_Selection_Methodology}
Given that the methodology is inspired by a bagging approach, we are left with multiple potential models from which to choose, thus motivating a method of model selection.

Information-theoretical approaches are becoming increasingly common in modern machine learning \citep{dong_bai_lu_fan_2022_AICc}. Typical choices include the Akaike information criterion (AIC) or the Bayesian information criterion (BIC). These approaches differ subtly, both balancing goodness-of-fit with a model complexity penalty. Typically, the BIC complexity penalty is greater per active term. Researchers indicate that these penalties are philosophically different and \citet{burnham_anderson_2010} suggest classifying such model selections based on purpose with AIC targeting efficiency, selecting the model with the best predictive accuracy, and BIC targeting consistency, selecting the true model from the candidate set as sample size grows. We consider the AIC for the general setting of equation discovery in which the true model may not be a part of the candidate set. Furthermore, both AIC and BIC assume a correctly specified likelihood function. In the following derivation, Gaussian assumptions are made, but with the nature of equation discovery the observables are themselves numerically estimated derivatives over a dense spatio-temporal mesh. Therefore, a Gaussian based likelihood is likely misspecified, although the criteria still aids heuristically.

The AIC can be derived from considering KL-divergence as a way to quantify
the difference between distributions. The metric is based on a linear combination of a model complexity parameter and quality of fit metric 
\begin{equation}
\mathrm{AIC}(\mathbf{h}; \xi) = -2ln(\mathcal{L}) + 2p,
\label{eq: AIC_simplest_form}
\end{equation}
where $\mathcal{L}$ is the likelihood function and $p$ is the number of active parameters in the model, corresponding to the number of non-zero values of the model weight vector $\xi$. A lower value of AIC indicates a better model. The residuals between predictions and data are then assumed to be Gaussian with zero-mean, thus resulting in a log-likelihood of the form
\begin{equation}
    \ln (\mathcal{L}) = -\frac{m}{2} \ln (2\pi) -\frac{m}{2} \ln(\sigma^2) - \frac{1}{2\sigma^2}\sum_{j=1}^{m} \left( \mathbf{h}(x,t) - \bar{\mathbf{h}}(x,t;\xi) \right)^2,
    \label{eq: AIC_log_likelihood}
\end{equation}
with $\sigma^2$ the estimated variance of the residuals of the $m$ measurements.
Substituting \eqref{eq: AIC_log_likelihood} into the definition of AIC gives
\begin{equation}
    \mathrm{AIC}(\mathbf{h}; \xi) = 2p + m\ln(2\pi \sigma^2)+\frac{1}{\sigma^2}\sum_{j=1}^{m}\left( \mathbf{h}(x,t) - \bar{\mathbf{h}}(x,t;\xi) \right)^2.
    \label{eq: AIC_formula}
\end{equation}
This approach assumes that the number of measurements is much larger than the number of predictors, $m \gg p$. A finite sample additive correction can be applied to improve the AIC estimate,
\begin{equation}
    \mathrm{CorrectedAIC}(\mathbf{h}; \xi) = \mathrm{AIC(\mathbf{h}; \xi)} + \frac{2(p+1)(p+2)}{m-p-2}.
    \label{eq: AICc_formula}
\end{equation}
Inspecting the correction, we see that
\begin{equation}
    \lim_{m \rightarrow \infty} \left( 2p + \frac{2(p+1)(p+2)}{m-p-2} \right) = 2p,
\end{equation}
so for a large number of measurements, the corrected result is consistent with the AIC. Consequently, we use the corrected version for model selection.

Information-theoretical scores provide a way to assess and compare the quality of models, but doing so introduces potential for misuse and misinterpretation \citep{ARNOLD_Delta_AIC_study_2010}. An example is due to the inclusion of uninformative predictors that have a benign, albeit mildly positive influence on the log-likelihood. This is exacerbated by large sample to predictor ratios, so is particularly pertinent to equation discovery, as the case of $n\gg p$ is standard. Consequently, we must be aware that it is possible to see a group of models strongly supported by these evaluation metrics that are the result of a good model being supplemented by terms with poor additional explanatory power. Hence, some higher complexity models may appear competitive by these metrics. This may be seen in results through a clustering of scores for similar complexity models. This weakness of information-theoretical approaches motivates us to also consider Pareto frontier optimisation \citep{deb2011introduction} as an alternative approach to model selection. This is a popular technique within PDE discovery literature \citep{ Rudy_Brunton_Proctor_Kutz_2017, Mangan_Kutz_Brunton_Proctor_2017} that characterises the trade-off between model complexity and quality of fit. The curve of quality of fit against model complexity may be non-injective, so we reduce this to models with the best quality of fit for a given complexity to form a frontier. From this frontier, we may still select a single model. To do so, we consider the marginal gain in error reduction, measured as the one-sided gradient from the most sparse end of the
frontier, and look for when this drops below a normalised unit threshold to provide a principled selection criterion. We discuss our application of Pareto frontier optimisation further as part of section \ref{sec: Gravity_thin_film}.

\subsection{Training Dataset Selection}
The production of physical models relies on training data that satisfy the underlying, modelled physics. We believe that data generated numerically from previously derived governing equations should provide a physically coherent foundation to work from. Varying the selection of training data can alter the discovered equations. This data-selection dependence is often either not touched upon or is not well characterised by previous empirical studies. We provide training data for feature selection and recalibration, with reserved test data for model evaluation and, when desired, selection. The sparse regression methodology scales naturally to allow for multiple inputted training datasets. When it is assumed that training datasets were generated from the same governing equation with the same parameters, then the data and corresponding basis functions may simply be vectorised and vertically stacked to form the design matrix $\Theta$, effectively appending the linear system. For inputted datasets in which the same governing equation is believed to hold but potentially different parameters are used, a block matrix setup may be used to expand the linear system \citep{rudy2018datadrivenidentificationparametricpartial}.

To guide the generation of training datasets, we must incorporate existing understanding of thin liquid film dynamics. We outline relevant modelling and analytical building blocks in this canonical setting in the following section.

%% file: thin_films_chapter.tex
We look to apply state-of-the-art equation discovery techniques to thin liquid film systems. Whilst being shown to be relevant to a host of industrial processes from coating technologies to modern cooling systems for data centres as well as the subject of continued research interest supported by modern computational resources, thin liquid films offer the key benefit of mathematical representation via reduced-order models that allow strong integration with (nonlinear) analytical solutions that lie beyond traditional benchmarks for equation discovery methods. 
Such benchmarks have typically included simplified models with a small number of polynomial terms and derivatives up to the fourth order, starting from the canonical Kuramoto-Sivashinsky equation. Its formulation, as well as more specialised counterparts such as the Benney equation and weighted residual models frequently used in the fluid dynamics community, are described in detail in subsection~\ref{subsec:thin_film_eq_discovery}. Thin liquid film models offer an excellent exploratory direction as they can be underpinned by higher-order nonlinear fluxes, multiplicative derivative terms, coupled fields and additional multi-physics mechanisms. This increases the complexity of candidate libraries of terms and provide a challenge for sparse regression methods through having more active terms that represent competing processes. This is the result of centuries of mathematical focus on such flows \citep{kalliadasis_ruyer_quil_scheid_velarde_2011} and thus provides a foundation from which modern data-driven techniques can be applied. In the next section, we highlight how we draw on this knowledge for equation discovery. 

\subsection{Analytical Setup}
\label{sec: inclined_film_setup}
\begin{figure}[htbp]
    \centering
    \includegraphics[width=0.5\linewidth]{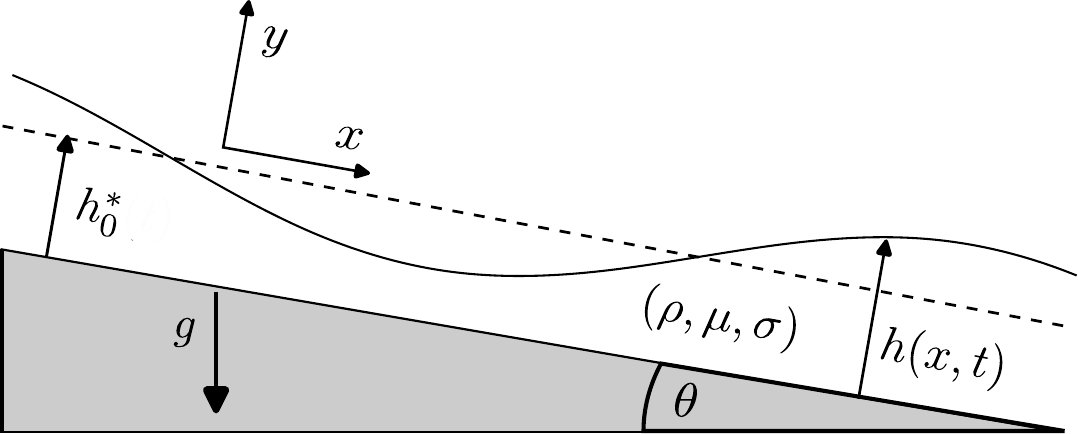}
    \caption{Schematic of a falling liquid film down an inclined plane. The inclination angle is denoted by $\theta$ and a rotated Cartesian coordinate system aligning with the flat, smooth substrate is used. The gravity vector $\mathbf{g}$, fluid properties, spatio-temporally varying interfacial height $h(x,t)$ and mean film height $h_0^*$ are also illustrated}
    \label{fig: setup_inclined_film}
\end{figure}

The setup for our problem, illustrated in figure \ref{fig: setup_inclined_film}, for the following analysis is that of an inclined thin film flow. The film height is described relative to a flat substrate with the $x$-axis along it and the $y$-axis perpendicular to it. This choice of coordinates ensures the flat substrate is located at $y=0$. Although this could in principle be tackled as a 3D problem \citep{Tomlin_Gomes_Pavliotis_Papageorgiou_2018_OptimalCO}, our framework is simplified to a 2D system to focus on the developed numerical capabilities in a tractable manner. The substrate is inclined at a fixed angle $\theta$ and the immiscible fluid is assumed to have fixed, homogeneous properties, with constant density $\rho$, dynamic viscosity $\mu$ and surface tension coefficient $\sigma$. The height of the film is described by $h(x,t)$ and is a liquid-gas boundary. The liquid is assumed to be Newtonian and the gas is modelled as hydrodynamically passive with constant pressure, $p_a$.
The only body force acting on the fluid is gravity, which acts at an angle $\theta$ to the $y$\nobreakdash-axis. Surface tension acts at the interface between the liquid and gas. Periodic boundary conditions are imposed in the streamwise direction on a periodic domain with length $L_x$.

We choose to non-dimensionalise in-line with previous works \citep{Ockendon_Ockendon_1995, Cimpeanu_Gomes_Papageorgiou_2021}, allowing for a dimensionless formulation of the governing equations to provide a universal framework applicable across different scales and setups. 
The length-scale of this problem is typically characterised by the mean film height, $h_0^*$, and its velocity scale may be characterised by a function of the mean film height $h_0^*$, gravity $g$ and the inclination angle $\theta$ \citep{RevModPhys_69_931} of the form
\begin{equation}
    x = \tilde{x}L, \quad
        y = \tilde{y}L, 
        \quad
        u = \tilde{u}U, 
        \quad
        v = \tilde{v}U, 
        \quad
        t = \tilde{t}\frac{L}{U},
        \quad
        p = \tilde{p} \frac{\mu U}{L},
        \quad
        g = \tilde{g}\frac{U^2}{L},
\end{equation}
with $L$ and $U$ chosen to satisfy
\begin{align}
        L &= h_0^* = \frac{1}{L_x}\int_0^{L_x} h(x,t)dx, \nonumber \\
        U &= U_s = \frac{\rho g h_0^{*2}\sin(\theta)}{2\mu},
\end{align}
where $U_s$ denotes Nusselt surface velocity \citep{nusselt1923warmeaustausch} and $L_x$ is the dimensional length of the domain. Additional re-scalings may then be applied to address the multi-scale nature of the problem. These scalings are often dependent upon a dimensionless long-wave parameter, $\delta \coloneq \frac{1}{L_{\delta}}$, introduced by considering solutions with characteristic wavelength $L_{\delta}$ relative to the mean fluid thickness, where $L_{\delta}$ is non-dimensional and it is assumed that $L_{\delta} \gg 1$. 
The characterisation of different flow regimes can be naturally achieved through non-dimensional groupings, which include the Reynolds, Capillary and Stokes numbers,
\begin{equation}
    \textit{Re} := \frac{\rho UL}{\mu}, 
    \quad
    \textit{Ca} := \frac{\mu U}{\sigma},
    \quad 
    \textit{St} \coloneq \frac{\rho gL^2}{\mu U}
\end{equation} 
respectively.
As well as providing excellent points of comparison between different flows, non-dimensional numbers are convenient regarding writing and manipulating equations. For equation discovery, the use of non-dimensionalisation
can minimise the number of independent parameters and simplify the form of the discovered equation. This often leads to more compact representations and can reveal simple power-law relationships, which is helpful in forming a candidate library of terms that meaningfully represents the dynamics \citep{AI_Feynman}.

To provide a concrete starting point grounded in realistic experimental conditions, typical regimes of interest in the context of this study for gravity driven water film flows are described by liquid film thicknesses of approximately $100-200\ \mu$m. In this setup, typical Reynolds number values are of $\mathcal{O}(1)$ to $\mathcal{O}(10)$, whilst common Capillary number values are $\mathcal{O}(10^{-3})$, which describe the edge of applicability of the widely used reduced-order mathematical models described previously.

\subsection{Equation Discovery and Thin Film Modelling}
\label{subsec:thin_film_eq_discovery}

Having previously discussed the history and interest in thin film modelling, we look to existing models relating to our problem setup and consider their use to augment our data-driven pipeline that was outlined in section \ref{sec: methodology}. 

A key insight from prior work is the existence of a stable steady state, the Nusselt solution, which is when a film is flat with unit dimensionless thickness \citep{nusselt1923warmeaustausch}. A known steady state impacts the equation discovery procedure. The training data used can be altered to select stable or unstable initial perturbations to evolve. Error analysis on testing datasets can vary through loss function alteration, error re-scalings, or inspection of long-time behaviour of proposed PDEs. In sections \ref{sec: nonlinear_Numerically_resolving_to_synthetic_data} and \ref{sec: nonlinear_numerical_experiments}, we consider these implications in more detail.

Section \ref{sec: Eqn_Discovery_Methodology} introduces the function library that required defining prior to discovery. The need to pre-specify this library is a criticised element of popular sparse regression methods \citep{PhysRevResearch_4_023174_SGA_PDE}. Prior thin film modelling can be used to inspire a pre-defined library and overcome this limitation. Therefore, we inspect the reduced-order models from the aforementioned mathematical model hierarchy introduced at the start of section~\ref{sec: Thin_Liquid_Films} above.

We have previously mentioned the KS equation \citep{Sivashinsky_1977, Kuramoto_1978, Sivashinsky_1980}, as a common equation discovery target and a very simplified long-wave model,
\begin{equation}
    h_t = -hh_x -h_{xx}-h_{xxxx}.
    \label{eq: KS_written_out}
\end{equation}
Through weakly nonlinear asymptotic analysis, the KS equation can be obtained from the more complex Benney equation \citep{Benney_1966_paper}. The Benney model provides a single equation model, though is commonly written as a coupled system with an explicit formulation of horizontal velocity flux,
\begin{equation}
    \begin{cases}
        h_t + q_x = 0, \\
        q(x,t) = -\frac{2}{3}h^3 - \frac{8}{15}\textit{Re}h^6h_x + \frac{2}{3}cot(\theta)h^3h_x - \frac{1}{3\hat{\textit{Ca}}}h^3h_{xxx}.
    \end{cases}
\label{eq: First_Order_Benney_Equation}
\end{equation}
It is found through linearisation about the Nusselt steady state and solving reduced systems of equations. However, the equation exhibits finite-time blow-up, motivating the need for more robust models. Using an alternative long-wave approximation, the Weighted Residuals model is a well studied reduced model for thin film flows \citep{Thompson_Gomes_Stabilising_falling_liquid_film_flows_using_feedback_control, oron_gottlieb_novbari_2009, kalliadasis_ruyer_quil_scheid_velarde_2011}
\begin{align}
    h_T + q_X &= 0, \nonumber \\
    q + \frac{2}{5}\textit{Re}\delta h^2 q_T &= \frac{2h^3}{3} - \frac{\delta h^3}{3}\left( 2h_X \cot(\theta) - \frac{h_{XXX}}{\hat{\textit{Ca}}} \right) + \textit{Re}\delta \left( \frac{18}{35} q^2h_X - \frac{34}{35}hqq_X \right),
    \label{eq: WR_derived}
\end{align}
written in non-dimensional form. This more complex formulation involves a coupling of interfacial height and flux that can no longer be combined into a single PDE. The model was derived by \citet{Ruyer_Quil_Manneville_2000}, and built upon in subsequent work \citep{Ruyer_Quil_Manneville_2002}. This coupling turns out to be vitally important for ensuring solutions adhere to physical laws and provides a substantial improvement over single-equation models \citep{Richard_Gisclon_Ruyer_Quil_Vila_2019}.

Turning our attention back to the formulation of a pre-defined function library, we inspect the terms in equations \eqref{eq: First_Order_Benney_Equation} and \eqref{eq: WR_derived}. We see models appear sparse with respect to a monomial basis and would otherwise greatly increase in complexity if an orthogonal polynomial basis were used. This indicates that the use of a monomial based function library for thin film data will likely be a strong choice. Similarly, monomials with respect to $h$ are favoured as opposed to perturbation amplitude for example, due to sparsity with respect to the basis.

Having considered prior thin film modelling and methods relating to equation discovery, we now focus on the objective of applying state-of-the-art discovery techniques to thin liquid film systems. As mentioned, models such as \eqref{eq: First_Order_Benney_Equation}-\eqref{eq: WR_derived} lie well beyond traditional benchmarks, whereas the KS equation has been recovered by a number of methodologies. We look to bridge the gap between such systems through the study of highly nonlinear, single-equation models involving high-order nonlinear fluxes.

%% file: non_gravity.tex
Our initial focus in terms of developing the aforementioned methodology in the context of a particular problem brings us to the surface tension driven thin film equation. With our thin film schematic~\ref{fig: setup_inclined_film} in mind, we apply lubrication theory in order to simplify the fully nonlinear multi-phase Navier-Stokes equations that underlie the unsimplified flow dynamics. We denote the vertical film thickness by $H$ and the horizontal streamwise length-scale by $L=L_x$, using the domain length $L_x$. It is assumed that the length-scale disparity $\epsilon$ between the horizontal streamwise length-scale and vertical film thickness directions is small, $\epsilon \coloneq \frac{H}{L} \ll 1$. We additionally assume $\epsilon^2\textit{Re} \ll 1$, where $\textit{Re}\coloneq \frac{\rho U L}{\mu}$ is the Reynolds number. These assumptions allow for a simple closed form equation that uses the low Reynolds number assumption to ignore inertial terms dependent on fluid density from the 2D Navier-Stokes equations. A derivation of equation \eqref{eq: The_Basic_Thin_Film_Eqn} and details of the non-dimensionalisation used can be found in Appendix \ref{app: Basic_Thin_Film_Derivation}.

We define the surface tension driven thin film equation as 
\begin{align}
    h_t &= -\frac{1}{3\bar{Ca}} \left( h^3h_{xxx}\right)_{x}
    \label{eq: The_Basic_Thin_Film_Eqn} \\
     &=  -\frac{1}{3\bar{Ca}}h^3h_{xxxx} - \frac{1}{\bar{Ca}}h^2h_{x}h_{xxx}, \label{eq: Chain_ruled_basic_thin_film_eqn}
\end{align}
which is written in non-dimensional form with dropped tilde decorations, and we consider the PDE with periodic spatial boundary conditions. 
This is a fourth-order nonlinear parabolic equation that represents a surface tension driven flow \citep{myers_1998_SIAM}. The non-dimensional Capillary number, $\textit{Ca} \coloneqq \frac{\mu U}{\sigma}$, represents the ratio of viscosity and surface tension effects. A re-scaled version of the Capillary number is employed, $\bar{Ca} \coloneq \epsilon^{-3}\textit{Ca}$. The model is built from a two-dimensional description of the flow, assuming invariance in the spanwise direction. We restrict our attention to films with unit mean film height, and so a corresponding stable steady state of $h \equiv 1$ is assumed to exist.

This two-term single equation model is highly nonlinear and includes what can be referred to as a derivative interaction term, $h^2h_xh_{xxx}$, yet it remains a viable candidate for testing the numerical capabilities we have developed. The complexity of this term naturally increases the size of the function library used.

Typical applications of equation discovery methods are applied to long-domain KS-based models \eqref{eq: KS_written_out}, which commonly result in chaotic behaviour for long time simulations \citep{Kalogirou_Keaveny_Papageorgiou_2015}. This is a direct contrast to the known stable steady state applicable for equation \eqref{eq: The_Basic_Thin_Film_Eqn}. Thus late time sampling of datasets may be thought to yield differing information. 

With the long-domain KS equation being well-studied, this aids in motivating the use of model \eqref{eq: The_Basic_Thin_Film_Eqn} as an alternative thin-film equation that is set in a long, periodic spatial domain, whilst minimising complexity through a single equation setup with a small number of terms. We now inspect the surface tension driven thin film equation more closely.

\subsection{Parameter Removal through Time Rescaling}
\label{sec: Parameter_removal_through_time_rescaling}
Looking to equation \eqref{eq: The_Basic_Thin_Film_Eqn}, we see that the non-dimensional parameter multiplies the differential operator and can be removed though a re-scaling $\tau \coloneq \frac{1}{3\bar{Ca}}\tilde{t}=\frac{1}{3\bar{Ca}}T^{-1}t$ to give a canonical version of the PDE,
\begin{equation}
    h_{\tau} = -\left( h^3h_{xxx}\right)_x.
    \label{eq: basic_thin_film_irreducible}
\end{equation}
Tracking the re-scaling, we see that $T_{\tau} \coloneq \frac{3\epsilon^{-4}\mu H}{\sigma}$, where $\tau = T_{\tau}t$. Previously the non-dimensional parameter $\bar{Ca}$ simply set the characteristic time scale. Additionally, a parameter sweep of equation \eqref{eq: The_Basic_Thin_Film_Eqn} over $\bar{Ca}$ for a fixed dimensionless final time, $\tilde{t}_F$, is equivalent to solving \eqref{eq: basic_thin_film_irreducible} to fixed $\tau_F = \frac{1}{3\bar{Ca}} \tilde{t}_F$.

With the most compact form of governing equation formulated, we look to generate associated datasets for equation discovery. 

\subsection{Synthetic Data Generation}
\label{sec: nonlinear_Numerically_resolving_to_synthetic_data} 
We consider the domain over which we solve equation~\eqref{eq: basic_thin_film_irreducible} with periodic boundary conditions imposed. The spatial domain is set to be unit length $[0,1]$, following from the choice of non-dimensionalisation using $L=L_x$. We now consider the temporal domain used and need to calibrate the time interval used for simulation. The long-time behaviour of the system \eqref{eq: basic_thin_film_irreducible} was studied by \citet{bertozzi_pugh_1996}, who displayed that a strictly positive initial condition with periodic boundary conditions relaxes to the constant steady state with equal mass. Thus, the dynamics are known to evolve to this stable steady state - a flat film - for all wavenumbers. With this knowledge, we balance a range of criteria for selecting the final non-dimensional time, $\tau_F$, to simulate towards as follows.

We choose to apply a slow dynamics condition, defining 
\begin{equation}
    \tau_F \coloneq \max_{\tau} \{ \tau : ||h_{\tau}||_2 < \delta \max_{\tau} \{ ||h_{\tau}||_2 \} \},
    \label{eq: Slow_Dynamics_Condition}
\end{equation}
with $\delta=0.01$. This defines $\tau_F$ to be when the average interface speed is a proportion $\delta$ of the maximum average interface speed. We may rescale proportional to $\frac{1}{\sqrt{N_x}}$ to ensure independence with respect to spatial resolution. Due to the error in $|h_{\tau}|$ appearing monotonic with respect to the temporal resolution, $d\tau$, this choice introduces some level of mesh dependence on the selected time $\tau_F$. In particular, simulations performed with a fixed number of time steps and varying temporal resolution may yield slightly different $\tau_F$ values compared with those using a fixed time resolution. Considering this and finite representation, the value $\tau_F$ is selected to within three significant figures (rounding down).

An alternative selection mechanism may be to utilise the linear dispersion relation and choose $\tau_F = \max_{\tau} \{ \tau : |h| > e^{-1} \max_x\{\left|h\right|_{\tau=0}\} \}$, thus attempting to simulate to approximately $\tau_F = k^{-4}_{min}$, with this final time approximation accuracy depending on the linearity of the system. This stopping condition is easy to implement, but generally results in over-sampling of near steady state dynamics that impacts identifiability.

With a stopping criterion selected, we have a fixed non-dimensional mesh, $[0,1]\times [0,\tau_F]$, on which to simulate the target spatio-temporal dynamics. The value of $\tau_F$ is not known \textit{a priori}, but is found by using a coarse simulation to provide a strong initial guess of $\tau_F$, before refinement to the previously mentioned precision. We choose a fixed aspect ratio $\epsilon = 1/64$ to reflect the long-wave assumption reflective of the intended thin liquid film dynamics, as well as height $H=\mathcal{O}(10^{-4})$ m for dimensional interpretation. For this $(\epsilon, H)$ pair and using water-air physical properties at room temperature, the corresponding Capillary number is defined as $\textit{Ca} = \frac{\mu L}{\sigma T_{\tau}} = 1.27\times 10^{-6}$, indicating a strong surface-tension regime.

To produce training and testing datasets, we use finite difference techniques, applying a first-order backward differentiation formula (BDF$1$) time-stepping scheme with second-order spatial central differences. The BDF$1$ scheme is implicit and L-stable, so is chosen for the numerical stability it provides. For our target equation, high-order derivatives are anticipated to introduce high levels of stiffness, making explicit methods undesirable \citep{ernst_hairer_wanner_2010}. We also note that this choice of method may introduce some numerical dissipation, though our initial conditions involve excitation of low-frequency modes. BDF$1$ provided sufficiently accurate datasets for the thin film equation studied, so higher-order methods were deemed unnecessary. Additionally, the central difference scheme used is quick to implement compared to spectral and higher-order schemes. The scheme allowed for excellent mass conservation and reduced the dispersion error from propagating waves relative to one-sided schemes \citep{olver_2013}.

Based on a combination of intuition and careful numerical experimentation, we found that the use of large perturbation training datasets was beneficial. These datasets contain $h(x,t)$ far from equilibrium, thus involving solutions with rich spectral structure that aids the underlying learning procedure in identifying the target dynamics. These types of initial conditions align intuitively with the notion of being more informative through large perturbations indicating stronger nonlinearity and multiple modes showing their interaction and dispersion. Consequently, an example of such an initial condition is
\begin{equation}
    h(x,0) = 1 + \sum_{i=1}^5 \frac{1}{4} \cos \left[  \frac{2\pi p_i}{L_X}\left(x-\frac{10i}{64}\right)\right],
    \label{eq: SuperposedModes_IC_written}
\end{equation}
where $p_i$ is the $i$-th prime and $L_X=1$ is the re-scaled domain length. This initial condition is physically valid, satisfying $h>0, \forall x\in [0, L_X]$, and satisfies the mass constraint $\int_0^{L_X} h_0 dx= 1$. 

\begin{figure}[hbtp]
    \centering
    \includegraphics[scale=0.75]{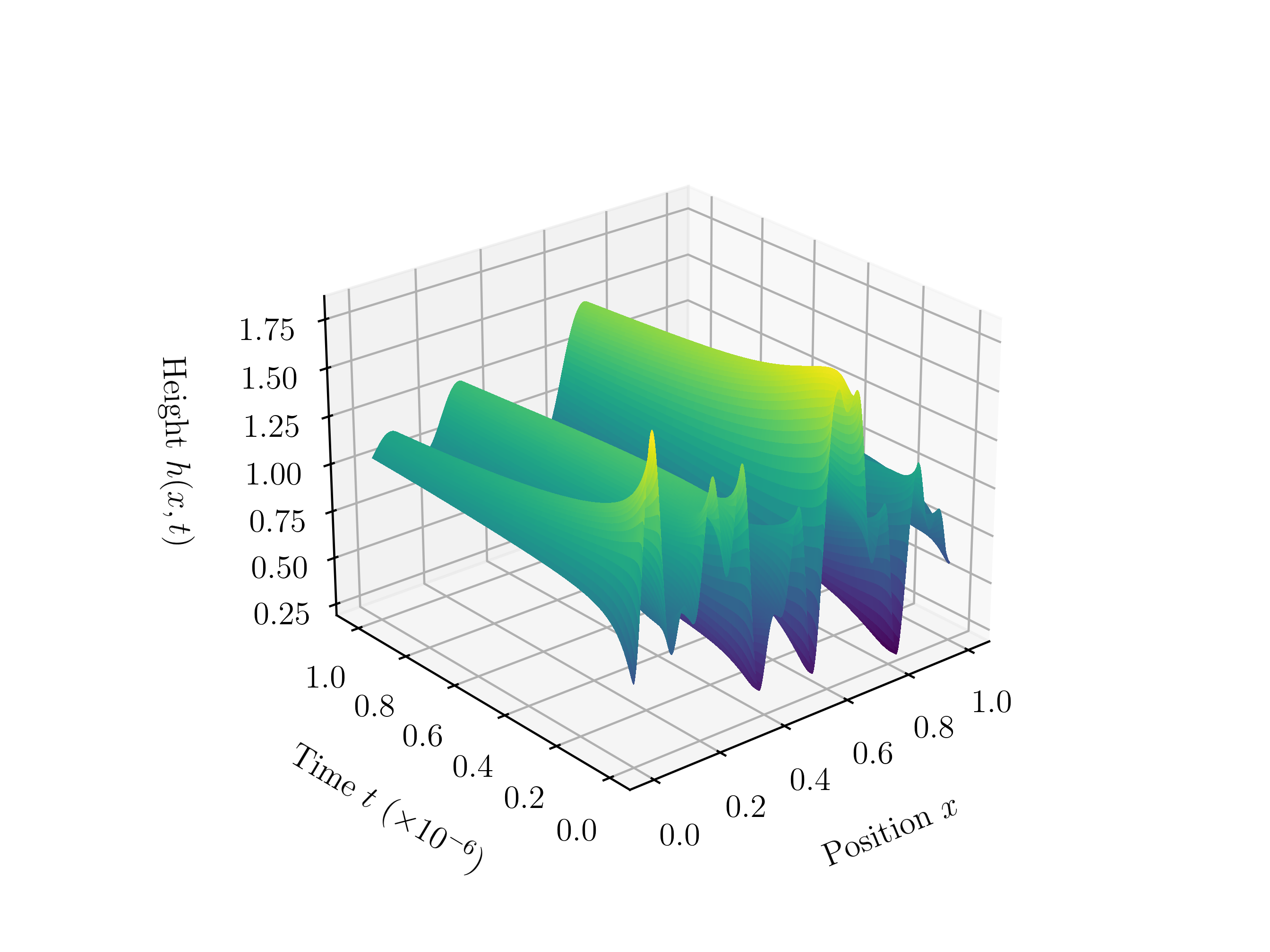} 
    \caption{The non-dimensional system \eqref{eq: basic_thin_film_irreducible} evolved from the initial condition \eqref{eq: SuperposedModes_IC_written} over spatio-temporal mesh $(\tilde{x},\tau)  \in [0,1]\times [0, t_F^* ]$. The evolution shows rapid decay of high-frequency waves and slower decay of lower frequency waves. The initial perturbation amplitude is large and all values satisfy $h(x,t)>0$}
    \label{fig: Image_of_Dataset_Data_Gen}
\end{figure}

We now solve the initial value problem (IVP) and, looking to figure \ref{fig: Image_of_Dataset_Data_Gen}, we see an example of a generated dataset, which has been evolved over the reference time span, satisfying the aforementioned stopping criteria.
We recall $T_{\tau} =\frac{3\epsilon^{-4}\mu H}{\sigma}$, which for a room temperature water-air system is $T_{\tau} \approx 0.0374 \epsilon^{-4}H$. Interpreting the final dimensional time, we see $T_{\tau}\tau_F = 6.39\times 10^{-5}s$. This is a very short evolution, showing a fast decay in amplitude of the initially excited highest frequency modes and slower decay of lower frequency mode amplitude. At the reference final time, we see that the perturbation amplitude is still large, with a maximum of $h(x,\tau)\approx 1.5$. A small number of low wavenumber modes appear to dominate the solution with energy concentrated into these modes, despite the generation of a broadband Fourier spectrum from the nonlinear evolution. Consequently, dynamics in this region have few effective degrees of freedom with limited information content.

A range of alternative periodic initial conditions can be used to specify the IVP and produce datasets. For a classical solution, one may consider $h_0 \in C_{per}^4([0,L_X])$ that are positive everywhere, $h_0>0$, and restricted to satisfy $\int_0^{L_X} h_0 dx= 1$. Thus we may consider finite Fourier sums or localised periodic bumps for example.

\subsection{Numerical Experiments}
\label{sec: nonlinear_numerical_experiments}
Having outlined the data generation process, we look to perform PDE discovery. For our numerical experiments, we use a reference final time value denoted as $\tau_F^*$, which is found by applying the slow dynamics condition \eqref{eq: Slow_Dynamics_Condition} to the irreducible PDE \eqref{eq: basic_thin_film_irreducible}, as described previously. This is equivalent to taking $1/3$ as a reference re-scaled Capillary number for the original equation \eqref{eq: The_Basic_Thin_Film_Eqn}, which produces $\mathcal{O}(1)$ coefficient values for the PDE of interest. We initially choose to look at recovering equation \eqref{eq: basic_thin_film_irreducible}, using clean data produced in the manner previously described on a dense, uniform spatio-temporal mesh.

For the following results, a single training dataset is used, shown in figure \ref{fig: Image_of_Dataset_Data_Gen} and evolved from the initial condition \eqref{eq: SuperposedModes_IC_written}. This training dataset is chosen because of its spectral structure and large initial perturbation amplitudes that result in the clear exhibition of nonlinear behaviour. 

With input training data established, we may define a function library of potential terms and the equation discovery process may be performed. In the case of a known governing equation, we may also use the equation discovery process to provide a form of validation on the input datasets by choosing a function library of only correct terms and inspecting recovered coefficients. This also provides an `easiest case' for recovery. We have found that, when fixing the model terms to the two expected terms from equation \eqref{eq: basic_thin_film_irreducible} and sample bootstrapping with ordinary least squares, there are no anomalous outlying values. We recover median values of $-0.875$ and $-2.44$ that yield the model
\begin{equation}
    h_t = -0.875 h^3h_{xxxx} - 2.44 h^2 h_x h_{xxx}.
    \label{eq: basic_nonlinear_two_term_library_result}
\end{equation}
The mean relative coefficient error is found to be $15.6\%$. This error may be explained by inspecting the residual
\begin{equation}
    r(x,t) \coloneq h_t + h^3h_{xxxx} + 3h^2h_xh_{xxx},
\end{equation}
which is large for early time data, of $\mathcal{O}(10^6)$, and quickly decays to $\mathcal{O}(10^2)$ within our studied time interval. The numerical scheme used to produce our datasets, a BDF1 scheme, typically results in numerical dissipation. This is consistent with the under-estimated coefficients seen for this case. Appendix \ref{app: residual_scaling_nonlinear} contains information on the scaling of residuals and their consistency with our data.

Within equation discovery frameworks, pipelines with suitable calibration geared towards prototypical cases may reach error levels below $1\%$ for numerous successful recoveries of known equations \citep{Rudy_Brunton_Proctor_Kutz_2017, zheng2024lessindylaplaceenhancedsparseidentification} when using clean datasets. Many studies have shown that errors increase greatly when datasets are noise-corrupted \citep{MESSENGER2021110525, Fasel_Kutz_Brunton_Brunton_2022}, with model identification commonly failing for large enough levels of noise. Similarly, using experimentally sourced datasets, larger errors of $\mathcal{O}(10\%)$ are presented \citep{schmid2024ensemble, heinrich2025rediscovering_KdV_ARXIV}. 
Therefore, we deem the obtained coefficient error to lie at a moderate level, particularly in the context of a new target nonlinear equation. We also highlight that we observed a reduction in coefficient error as the temporal resolution of the dataset was increased. The error was found via numerical experimentation to scale linearly with the temporal resolution $dt$, saturating at a level approximately $100$ times lower than the reference dataset used within the present section. Our chosen reference value in the discussion above was selected to strike a suitable balance between efficiency and sensitivity considerations, providing valuable insight into the capabilities of the suggested pipeline especially in the context in which relying on data with arbitrary frequency may not be a viable option. 

\subsubsection{Feature Selection}
\label{sec: nonlinear_feature_selection_subsection}
We perform feature selection following the discovery algorithm outlined in section \ref{sec: Eqn_Discovery_Methodology} and deviate slightly by using an extended line search interval. We line search on the interval $[1.5\epsilon \Delta_{crit}, 1.5\Delta_{crit}]$, with $\epsilon = 10^{-3}$. We call the half-line $[\Delta_{crit},\infty)$ the critical region and its intersection with the searched region is non-empty. Inspection of the intersection improves confidence that the critical tolerance is accurately determined for the full system, explaining the choice of deviation. 

Considering the bootstrap sampled case of ordinary least squares $$\arg \min_x \{ ||S^{(i)}(Ax-b)||_2\} = \hat{x} \neq x^* = \arg \min_x \{ ||Ax-b||_2 \},$$ where $S^{(i)}\in \mathbb{R}^{m_i\times n}$ sub-samples rows of the design matrix, there is typically a difference in estimators $\hat{x} \neq x^*$ between the full system and sub-sampled systems. With differences in the magnitudes of elements of the least squares solutions, it may be the case that the critical tolerance of the sub-sampled system is not equal to that of the full system, $\Delta_{crit} \neq \Delta_{crit}^{(S^{(i)})}$. This gives a misaligned sample-dependent critical region, $[\Delta_{crit}^{(S^{(i)})}, \infty)$. For the case $\Delta_{crit}^{S^{(i)}} < \Delta_{crit}$, this merely reduces effective sampling density of the line search. The extension of the sampled region is to detect the alternate case $\Delta_{crit}^{S^{(i)}} > \Delta_{crit}$, which could result in a line search that fails to sample tolerance values just below the associated critical tolerance. This being the region that would likely yield the sparsest models. This would be undesirable, due to the previously discussed typical sparsity assumptions. For the numerical experiments to follow, we do not observe the undesirable case and provide our hypothesis for the reason for this in Appendix \ref{app: crit_tol_relation_to_subsampling}.

\begin{figure}[hbtp]
    \centering
    \includegraphics[width=0.925\linewidth]{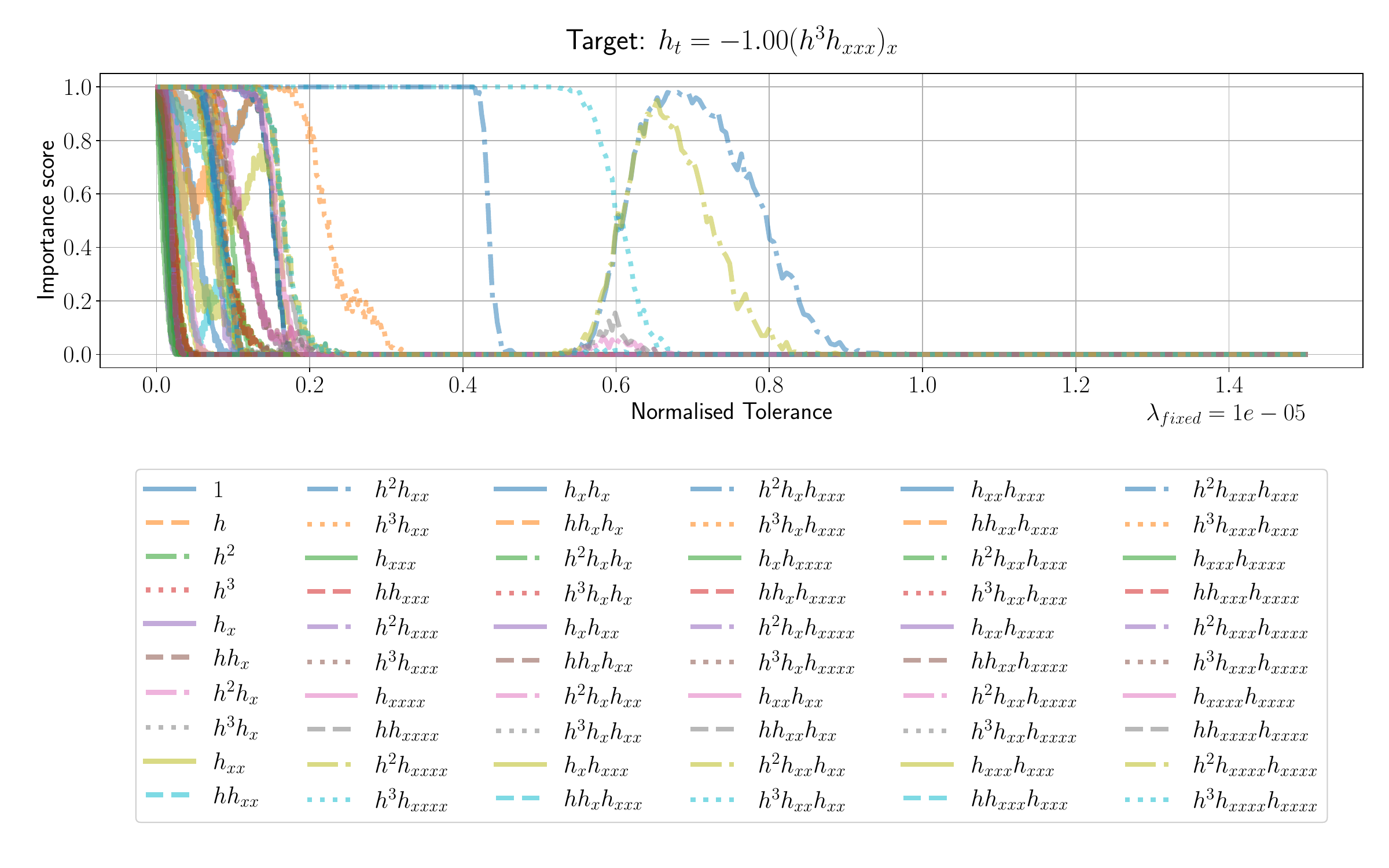}
    \caption{An example of a feature selection tolerance line search. For each term in the function library, an importance score is calculated over the range of tolerance values, relating to whether it was frequently retained from the sequentially thresholded feature selection process. The tolerance values are rescaled with respect to the critical tolerance}
    \label{fig: Discovery_Line_Graph_nonlinear_Ca_1_3}
\end{figure}

Looking to figure \ref{fig: Discovery_Line_Graph_nonlinear_Ca_1_3}, we see an overview of the line search process using the training data described in \ref{sec: nonlinear_Numerically_resolving_to_synthetic_data}. The choice of $\epsilon$ results in full-complexity models being recovered, despite the effective density of tolerance sampling being reduced due to samples greater than the critical tolerance value being used. We inspect the figure starting from high tolerance regions and moving to low tolerance regions. In the region above the critical tolerance, we see that the empty set is returned and so no candidate terms are deemed important. Below the critical tolerance value, meaningful feature selection begins. In the high tolerance region, we see the sparsest models produced, which are commonly single- or two-term models. As tolerance reduces, alternative terms may be selected. We see correlated feature selection, shown by importance score correlation. In the region $[0.55,0.65]$, we see $h^3h_{xxxx}$ gain importance, whilst $h_xh_{xxx}$ and $h^2h_xh_{xxx}$ decrease in importance at the same rate with respect to tolerance. With these features deactivated, there is another single-term model region, before a clean re-activation of the other correct term, $h^2h_{x}h_{xxx}$. \footnote{We refer to clean activations when a term's importance score against tolerance is well approximated by a sigmoid function or, equivalently, is similar to a Heaviside step function.} Thus, a region of tolerance exists in which a model with terms matching those that generated the training dataset can be selected. This is the region where researchers applying hyper-parameter optimisation approaches usually hope to find an optimised tolerance value. In lower tolerance regions, more and more terms are activated and model sparsity reduces. For a small enough tolerance value, all terms are active. 

Applying an importance cut-off heuristically set to $0.8$, we obtain a wide collection of $90$ unique linear combinations of terms from the function library. High complexity combinations form the majority of the collection and $5$ models have less than $5$ terms. Such combinations of terms require an associated weight vector to form a model $h_t = \mathcal{N}(h,Q)$. Consequently, we apply a recalibration bootstrap step, as discussed in section \ref{sec: methodology_recalibration}.

A clean dataset is used for training and there is non-zero sample variance, which we believe is reflective of numerical conditioning and coherence. With a low sample variance and unimodal coefficient distributions, we can be confident in the predictions generated. Large sample variances from recalibration can be indicative of overfitting and a multi-modal distribution affects the validity of taking the median coefficient value. 

Once recalibrated, we then have a selection of $90$ models. Although this seems to be a large set, we note that with $60$ function library terms, $2^{60}\approx 1.15\times 10^{18}$ unique combinations of terms exist, so this is a major reduction in the size of the model space to consider. Considering only sparse models with less than $5$ terms, $\sum_{k=1}^4\binom{60}{k} = 523685$ potential models exist, so the reduction to only $5$ candidates is significant. This level of reduction appears roughly in-line with other studies \citep{mangan_brunton_proctor_kutz_2016, dong_bai_lu_fan_2022_AICc}. Sizes of candidate libraries to which model selection are applied also vary significantly and appear to be strongly dependent on the underlying model complexity, as well as the size of the library. We deem the size of the candidate set we generate to be small to moderate relative to other related works when accounting for the size of function library we use \citep{Mangan_Kutz_Brunton_Proctor_2017, Maddu_Cheeseman_Sbalzarini_Muller_2022_PDE_STRIDE, wu_mcdermott_maclean_2025}.

With a significant number of candidate models remaining, we are motivated to use an automatic selection process.

\subsubsection{Model Testing and Selection}
\label{sec: Model_Testing_and_Selection}

We look to rank candidate models from the collection, which can enable the proposal of a single model if desired. A model hierarchy also enables further analysis, as one may inspect similarities in features between strong performing models. We desire a test metric that balances model complexity with model fit and so the previously mentioned finite sample corrected AIC \eqref{eq: AICc_formula} is a common choice \citep{Mangan_Kutz_Brunton_Proctor_2017}.

\begin{figure}[hbtp]
    \centering
    \includegraphics[width=0.95\linewidth]{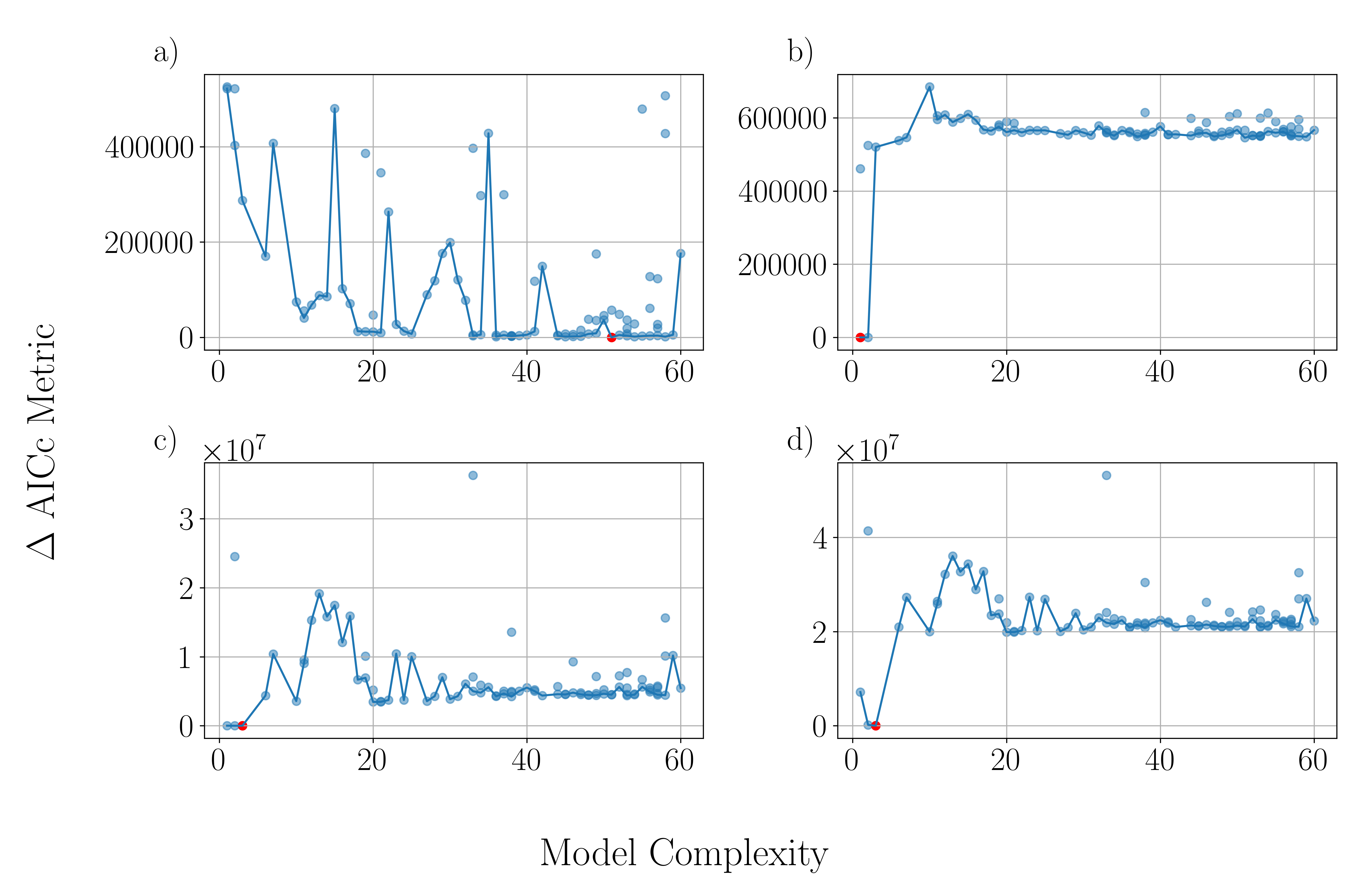}
    \caption{An array of model evaluation metrics. The overall minimum corrected AIC score (best) is highlighted in red and a line connects the minimum value for each model complexity. Additional points show scores of alternative models that were outperformed by an equally complex model. The top row uses residuals from the fitted linear system for error calculation.
    The bottom row panels involve the solution of discovered PDEs for error calculation. The left column panels use training data for evaluation, and the right column panels use separate testing data. Panel a) is produced by using training data and calculating errors using residuals of the fitted linear system. Panel b) is produced by using separate test data and calculating errors using residuals of the fitted linear system. Panel c) is produced by using training data and calculating errors with the solution of discovered PDEs against the observable, $h(x,t)$. Panel d) is produced by using separate testing data and calculating errors with the solution of discovered PDEs against the observable, $h(x,t)$}
    \label{fig: AICc_training_test_all_compared}
\end{figure}

To test models, we now utilise a reserved test dataset. The choice of test dataset is important. A poor choice includes a dataset that is too similar to the training data. Doing so would yield excellent scores for overfitted models that would not generalise and there would exist correlation between model complexity and test performance, as can be seen in the top-left panel of the upcoming figure \ref{fig: AICc_training_test_all_compared}, in which the training data is used for model evaluation. Therefore, we choose a dataset with a longer time span, $[0, 97.1\tau_F^*]$, whilst still applying the slow dynamics condition \eqref{eq: Slow_Dynamics_Condition}. The time-scale underlying this dataset may result in errors from poor quality coefficient estimates or the inclusion of overly strong/weak features to be seen, e.g. over-damping, under-damping, etc. Given the focus of interfacial flows, our chosen validation set is sinusoidal in nature with a small Gaussian bump and with mass correction applied to the initial wave profile to satisfy $\int_0^{L_X}h_0dx=1$. We note that a strong alternate validation dataset may be that of soliton data for thin film dynamics. A further consideration is that of the active modes of the test dataset. This is relevant for the dissipative system we study, which is stable for all wavenumbers. As a result, a fitted model may display instability, but after sufficient damping of the highest-frequency modes in the training data it may appear to be a good model to describe the remainder of the evolution. Therefore, such a discovered model is only conditionally valid, as its apparent stability is an artefact of the dissipative nature of the true system. This underscores the importance of testing learned models on initial conditions that span the range of dynamics they are intended to model and not just conditions seen in training data. 

With a test dataset selected, we note a choice between a solution based AIC application and a linear system based AIC application. The candidate models are generated through solving the linear system $\Theta \xi = b$, so one may use the residuals of the system resolved with recovered coefficients for the quality of fit metric. Doing so compares the linear combination of derivative estimates to the estimate of the assumed left hand side of the equation, $h_t$. As a result, errors in derivative estimation have a large effect on the resulting model rankings and is of particular concern with noisy datasets. The solution-based application requires solving the candidate PDEs for testing. For our single equation models, we note that this is typically inexpensive, but stiffness and instability to the test initial conditions can complicate numerical solutions. Instability to the initial conditions of test datasets can result in seemingly disparate errors between the derivative-based and solution-based losses. The associated loss landscapes $||\Theta_{test}\xi_{disc}-H_{t,test}||_2$ and $||h_{disc}-h_{input}||_2$ for both options can differ greatly. Working with a coarse-grained example, \citet{martina_perez_simpson_baker_2021} compares these landscapes, showing favour to the solution-based loss landscape for their application. We showcase in figure \ref{fig: AICc_training_test_all_compared} the result of using training and testing datasets, as well as the difference between the solution and derivative-based information-theoretic based metric evaluations. A striking difference can be seen between loss function scores using training or reserved testing data. Panel $(a)$ in particular shows the strong performance of high complexity models. Fitted to the training data and with numerous terms to provide flexibility, this overfitting is expected. Meanwhile, panels $(b)$ and $(d)$ show that such models may not generalise well. Contrast can also be seen between panels $(a)$ and $(c)$. The outcome of using solution data is shown to be effective, with low-complexity models being favoured in panel $(c)$. This difference is juxtaposed by the similarity of panels $(b)$ and $(d)$ in which test data is used and loss scores appear qualitatively similar.

From inspecting scores, we see strong performance from low complexity models. An example is
$$h_t = -0.699 h^2 h_{x} h_{xxx},$$
which includes one of the expected terms, so can meaningfully resolve some of the dynamics. We see a bias in the coefficient, as a consequence of underfitting. Another strong performing model includes a $\mathcal{O}(10^{-10})$ additive correction to this single-term model. A slightly higher complexity model with strong performance is

$$h_t = -0.860 h^3h_{xxxx}  -3.76 h^2h_{x}h_{xxx} + 1.23 h^3h_{x}h_{xxx}.$$

The three-term model shown contains the correct (anticipated) terms, as well as an additional term. This additional term is similar to one of the correct terms, but has a higher degree of nonlinearity. The sum of these two coefficients is approximately $-2.53$, so is close to the coefficients of the generating equation. Consequently, we see strong performance coming from this similarity, exacerbated by small perturbation regions of data. Candidate models similar to that of the correct model, but with varying degree of algebraic nonlinearity are a common occurrence. It is common as well for such collections of terms to have coefficients that sum up to a value equal in magnitude to a corresponding term in the governing equation. This is effectively redistributing the coefficient value among correlated basis functions.

\subsubsection{Recovery}
\begin{figure}[hbtp]
    \centering
    \includegraphics[width=\linewidth]{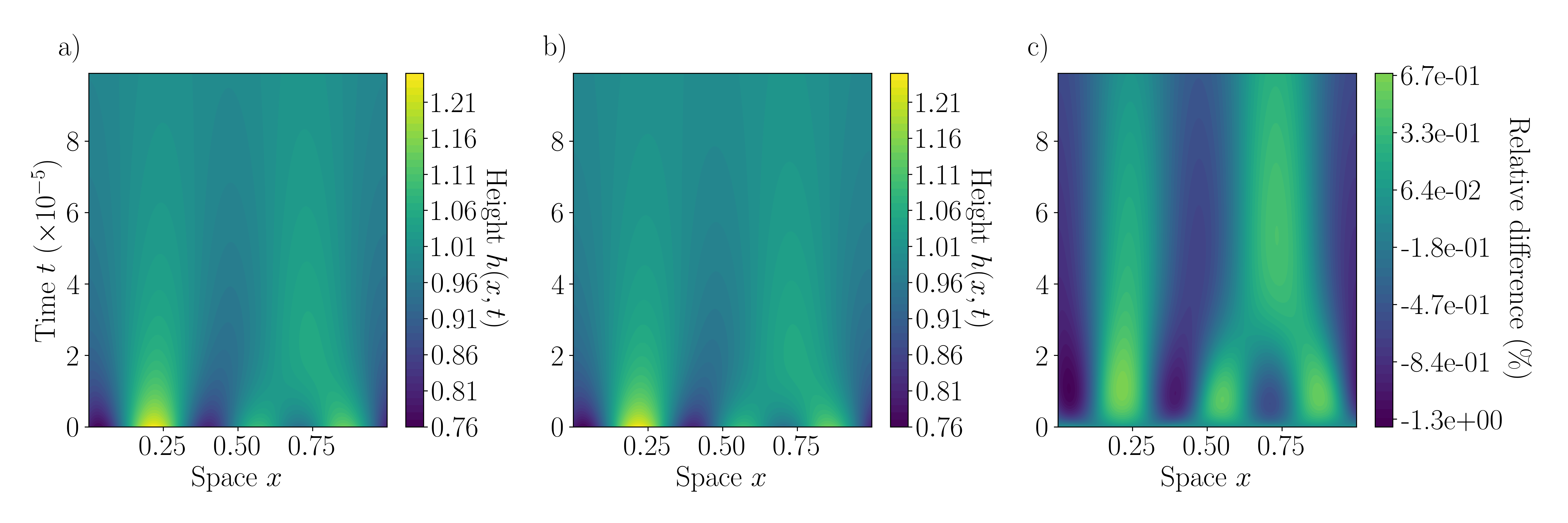}
    \caption{a) The dynamics of the discovered equation. b) The evolution of the governing equation and is the testing data used. We compare the two surfaces in panel c), displaying relative percentage error. Errors include: RMSE=$0.00413$, MAE$=0.00340$ and $||e||_{\infty} = 0.0116$}
    \label{fig: Simulated_Discovered_Equation_nonlinear_correct}
\end{figure}

We inspect the list of proposed models to see a strong performing low-complexity model $$h_t = -0.875 h^3h_{xxxx} -2.44 h^2h_{x}h_{xxx},$$ which includes all model terms used to generate the dataset with no spurious terms. When solved and compared to the reserved test dataset, this yields an $L_2$ error (normalised with respect to the null model error) of $0.0345$. We recall having previously recovered a mean relative coefficient error of $15.6\%$ in subsection \ref{sec: nonlinear_numerical_experiments}. Figure \ref{fig: Simulated_Discovered_Equation_nonlinear_correct} shows the evolution of this discovered model against reserved test data. We note this is equivalent to model \eqref{eq: basic_nonlinear_two_term_library_result}, which is expected from the methodology. 

With the thin film equation recovered using clean data, we can briefly consider the robustness of this result to artificial noise, which can be found in Appendix \ref{app: Noisy_Basic_Thin_Film_Eqn}.

\subsubsection{Time Proportion Study}
We now consider varying the training dataset and the sensitivity of equation discovery to the time-span of training data for the constant coefficient PDE \eqref{eq: basic_thin_film_irreducible}. Recall from section \ref{sec: Parameter_removal_through_time_rescaling} that we find a reference final time value $\tau_F^*$ by applying the slow dynamics condition to the irreducible PDE. The time scaling, relating to this equation, results in dimensional time values $t = T_{\tau}\tau  = \frac{3\epsilon^{-4}\mu H}{\sigma}\tau$. 

\begin{figure}[hbtp]
    \centering
    \includegraphics[width=0.99\linewidth]{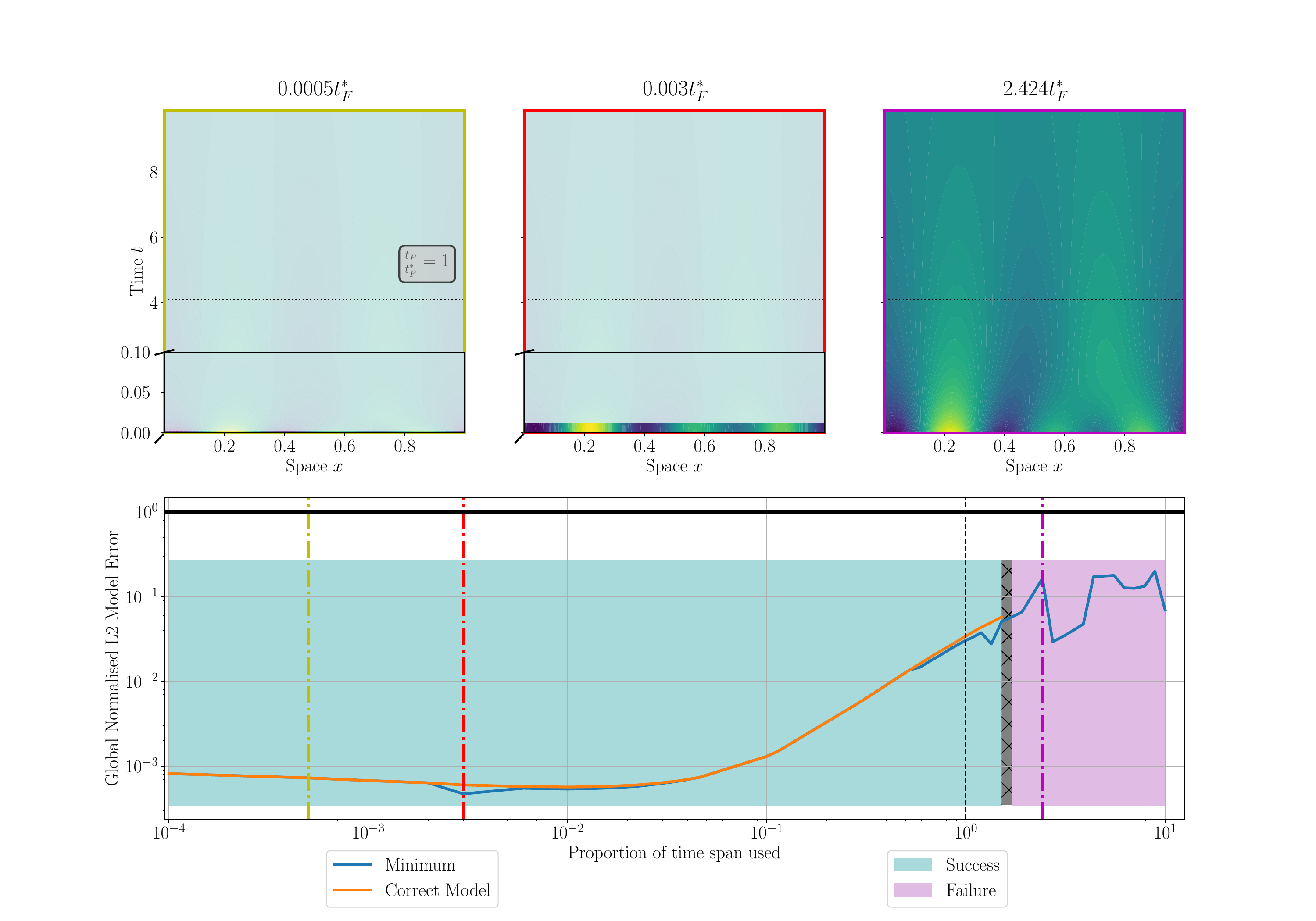}
    \caption{Varying temporal resolution in the training data, as the number of temporal samples remains constant}
    \label{fig: Time_Proportion_nonlinear_FINAL_IMAGE_variable_dt}
\end{figure}

Looking to the training datasets, we sweep final times, $\tau_F \in [0.1\tau_F^*, 10\tau_F^*]$ in a linearly spaced manner. We more coarsely sweep additional values $\tau_F \in [0.0001\tau_F^*, 0.1\tau_F^*]$. This is equivalent to taking a different proportion of the reference time span $[0, \tau_F^*]$ and we write proportions $\alpha >1$ to indicate time spans of $[0,\alpha \tau_F^*]$. For these values, we look to whether a successful recovery of the thin film equation places it among the ensemble of suggested models and inspect associated errors. During data generation, we initially ensure that the number of samples is consistent across datasets, $(N_x,N_t)=(1024,2048)$, thus varying temporal resolution appropriately with a fixed spatial resolution. The results of this are summarised in figure \ref{fig: Time_Proportion_nonlinear_FINAL_IMAGE_variable_dt}. This varying temporal resolution means that small time proportion datasets, such as those evolved to $0.1\tau_F^*$, will have finer temporal resolution and improve the accuracy at which the initial transient is resolved. With worsening temporal resolution for long time proportion datasets, such as $\tau_F^*$, adaptive time step schemes are implemented to obtain datasets, though these only affect internal data generation steps. Refinement is determined by a residual condition with fixed tolerance and a maximum refinement factor of $1/2^4$ specifies a computational budget for data generation. 

We begin by noting a large region of success for ever-decreasing proportions of the time span used. This indicates that sampling from the initial transient region with highly nonlinear dynamics is fruitful for equation discovery. This region includes rapidly decaying high-frequency modes that are well sampled with respect to time. We note that for the time spans studied, we did not find a transition to a non-successful regime, which would be eventually expected due to almost static dynamics over a very small time-span, though one may run into issues regarding machine representation of floating point numbers before this. For increasing proportions of time spans used, we see a transition to a non-successful regime. We theorise that this transition is driven by a combination of worsening representations of initial decay dynamics as the temporal resolution decreases and over-sampling of later in time dynamics after substantial relaxation of high-frequency modes.

The minimum model error for those with the expected terms appears to be for a time span proportion of $0.017$. This minimum model error is unique and model errors increase monotonically as values of the time span proportion differ from that associated with the minimum. The overall minimum error seen is from a model that uses alternative terms and is highlighted with the red, vertical dashed line. We highlight sampling in this region includes $\{ 0.002, 0.003, 0.006\} $, so there is only a single point of significant visible out-performance. This model is 
\begin{multline*}
h_t = -0.00339hh_{xxxx} + 0.00653h^2h_{xxxx} -1.00h^3h_{xxxx} +  28.8h^2h_{x}h_{x} \\ -28.2h^3h_{x}h_{x} + 0.0123hh_{x}h_{xxx} -3.01h^2h_{x}h_{xxx} + 0.000131h^3h_{x}h_{xxx}.
\end{multline*}

Inspecting around this region for systematic outperformance, we do note that there is mild out-performance by a higher complexity model for time proportion $0.006$ with $$h_t = -0.998h^3h_{xxxx} + 0.0169hh_{x}h_{xxx} -3.03h^2h_{x}h_{xxx} + 0.0163h^3h_{x}h_{xxx},$$ however the equation structure is quite different from the other out-performing model. 

Otherwise, the green success region indicates that the models containing just the correct features generally outperform other models on our test dataset for the majority of time span proportions sampled. We highlight that what is referred to as the correct model may have non-zero error due to coefficient fitting errors. Our definition of success regards only feature selection. Perhaps expectedly, we also note a spike in model error in the region transitioning from correct model identification and some non-monotonic behaviour of the error curve.

Having looked to inspect the small-time region and utilise small-time asymptotics to support our explanation for success in the low proportion region, we provide a sufficient condition for exact support recovery via thresholding given in Appendix \ref{app: Small_time_asymptotics}. While the condition is not tight in practice, it does encapsulate many requirements for successful equation discovery. Firstly, the coefficients must have a sufficiently large magnitude relative to one another, so that the associated terms can be recovered. Secondly, strong conditioning of the constructed linear system is desirable. Finally, a strong estimation of basis functions and accurate data that minimises the residual $\delta b$ is needed. Furthermore, one may even consider a system with similar magnitude coefficients and using $||\xi^*||_2 \approx \min_{j\in S^*}\{ |\xi_j^*| \} \sqrt{|S^*|}$ to show a stronger sufficient requirement for systems represented by larger true supports. This is consistent with the behaviour observed empirically. This type of bound is common in signal processing literature, relating to minimum signal strength and have been referred to as beta-min conditions \citep{buhlmann_de_2013}. However, the requirement for \textit{a priori} knowledge of the true coefficient vector renders such bounds of limited practical use.

\subsubsection{Fixed Temporal Resolution Study}
The previous study involved maintaining the size of the dataset used, and so involved a constant number of temporal samples which required varying temporal resolution. From the prior success in the low-data limit, we see that this likely optimistically distorts the effectiveness of the equation discovery method in this region. Therefore, we apply a fixed resolution scheme that involves varying the number of samples used.

To implement this, we use the prior reference dataset (proportion of time span equal to $1$ previously) to set a reference $dt$ value. We generate a large dataset with the maximum final time value of interest, $t_F = 10t_F^*$ and subsample this dataset. We note that pointwise derivative estimates are calculated using the full dataset before sub-sampling. Therefore, we choose to remove the extreme time value samples from the dataset to ensure a consistent central finite difference scheme is applied for all time proportions. We also highlight that the number of spatial samples $N_x$ is greater than the size of the function library used, so it is possible with one time sample to have a design matrix that is of full rank. 

\begin{figure}[hbtp]
    \centering
    \includegraphics[width=0.99\linewidth]{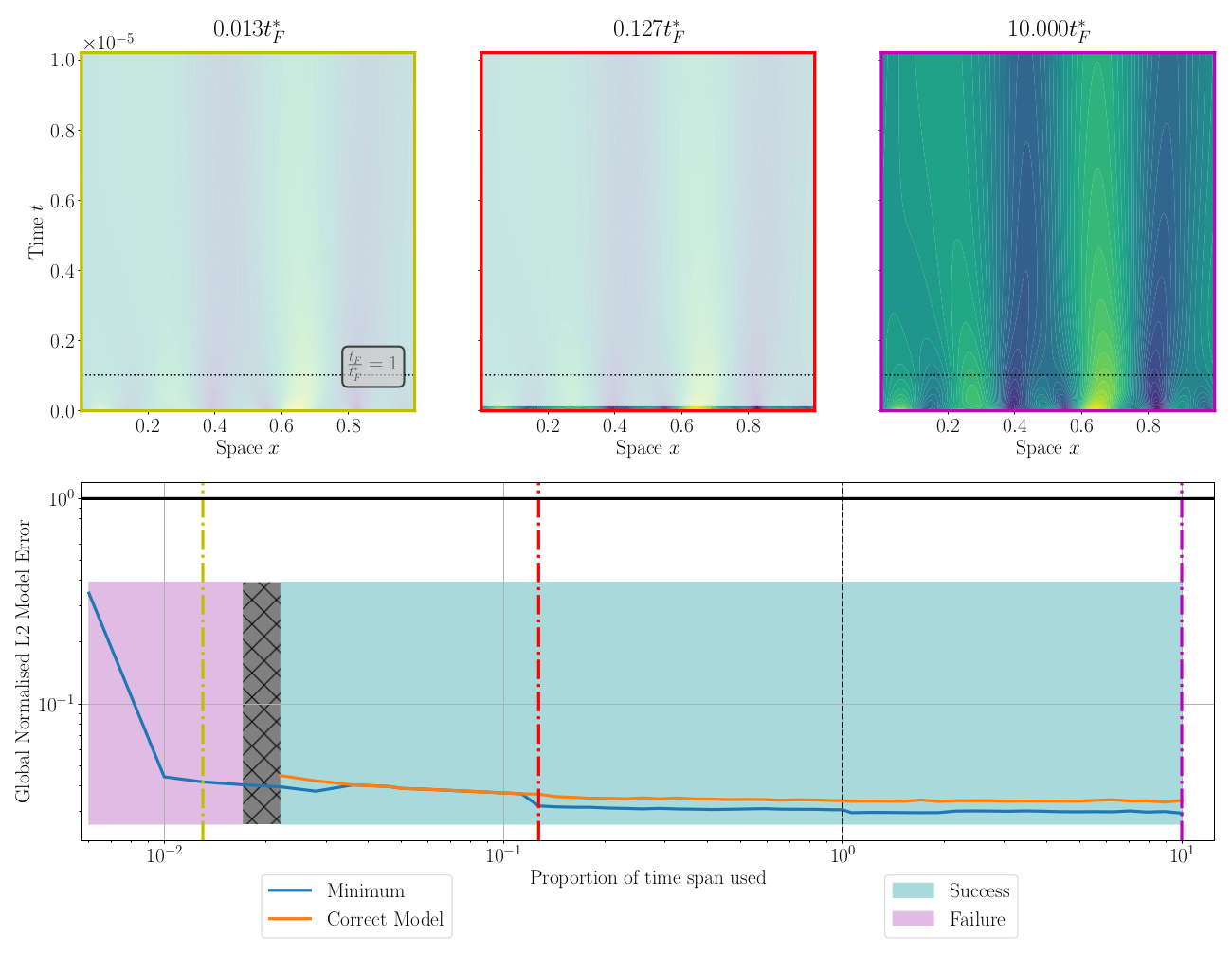}
    \caption{Fixing the temporal resolution in the training data and varying the number of temporal samples to account for this. Successful recovery continues to the maximum tested value of $10.0\tau_F^*$}
    \label{fig: FINAL_TIME_two_row_L2_fixed_resolution_training_SuperposedMode_tau_1}
\end{figure}

The results of this study can be seen in figure \ref{fig: FINAL_TIME_two_row_L2_fixed_resolution_training_SuperposedMode_tau_1}. We see a large successful region and note that there is failed recovery for low quantities of temporal data, below the proportion $0.017$. This is in direct contrast to the previous variable resolution case. Regions of outperformance to the model with correct features are visible for a range of proportions, $[0.036,0.113]^C$, though model out-performance is within distance $0.01$ with respect to the normalised error metric, with maximum outperformance of $14.3\%$. We see in the region of failure for equation discovery that the error increases slowly, remaining comparable to the successful region until a proportion of approximately $0.01$. The error increases greatly at the next sampled point. The failure region relates to single digit temporal samples being supplied, as the final time value used approaches the reference temporal resolution and limits our ability to sample more densely. 

\subsection{Key Insights}
From these numerical experiments, we see that it is highly important to have a well-resolved initial transient region. These highly nonlinear regions appear to greatly aid in structural identification of PDEs, when well-resolved.

Comparing the results of the variable and fixed temporal resolution cases, it appears that the failure for the variable resolution case was triggered due to worsening of the the temporal resolution rather than over-sampling later time dynamics. The worsening resolution had a strong effect on the early time dynamics that involved the fast decay of high-frequency modes. The failure of the fixed-resolution case appears to be from overly coarse sampling. Successes for high sample numbers indicate the transient region to be sufficiently resolved and successes from the variable resolution case over short time-spans indicate sufficiently rich dynamics when well-sampled. This is despite moderate to large residuals present within the training data for the reference temporal resolution used (see Appendix \ref{app: residual_scaling_nonlinear} for additional details). Augmenting the process with an additional training dataset used for fitting coefficient values, after structurally identifying the model with our nonlinear transient dataset could improve performance on test datasets, leading to a potentially competitive alternative to the fixed resolution case, thus aiding in model selection. A strong candidate for this additional dataset would be one evolved from an initial condition with low-frequency active modes.

We conclude at this stage that one requires a well-sampled and well-resolved transient region within training datasets to aid with the structural identification of PDEs. We further posit that this transient richness is of greater importance than sampling the relatively uninformative long-term dynamics. We are now well equipped to use this insight in order to tackle a more complex version of our problem, which incorporates the competing effect of gravity as part of the following section.

%% file: gravity_chapter.tex
Building on the previously discussed scenario where the effects of gravitational acceleration were unaccounted for, we now look towards a thin film dynamics model under gravity, and introduce
\begin{equation}
    h_t = - \left( h^3 - \Phi h^3h_{x} + h^3h_{xxx} \right)_x,
    \label{eq: thin_film_grav_case_eqn}
\end{equation}
where $\Phi \coloneqq \epsilon\cot (\theta)$. The parameter $\theta$ is the angle of inclination of the flat substrate to the horizontal and $\epsilon$ is the length-scale discrepancy, as previously. The equation is written in non-dimensional form with dropped tildes. A full derivation is included in Appendix \ref{sec: Gravity_thin_film_derivation}. 

\subsection{Model Inspection}

The choice of scales to produce the form of model seen in equation \eqref{eq: thin_film_grav_case_eqn} results in coefficients being normalised by $\sin(\theta)$, so the balance between gravitational and surface tension based hydrostatic forces is implicitly altered as the angle of inclination $\theta$ is varied. We study acute angles and see the divergence of $\Phi \rightarrow \infty$ as $\theta\rightarrow0^+$, indicating a failure of the model for a flat substrate. For the other extremal case of $\theta=\frac{\pi}{2},$ a valid model remains with $\Phi = 0$. 
With the choice of characteristic scales 
$$L = \sqrt[3]{\frac{\sigma H}{\rho g \sin (\theta) }}, \quad T = \frac{3\mu}{H^2\rho g}\sqrt[3]{\frac{\sigma H}{\rho g}}\frac{1}{\sin (\theta)^{\frac{4}{3}}},$$ 
we notice that $L \propto \sqrt[3]{\frac{H}{\sin(\theta)}}$, so $\epsilon = \frac{H}{L(\theta ; H)}$ varies with $\theta$.

\begin{figure}[hbtp]
    \centering
    \includegraphics[scale=0.5]{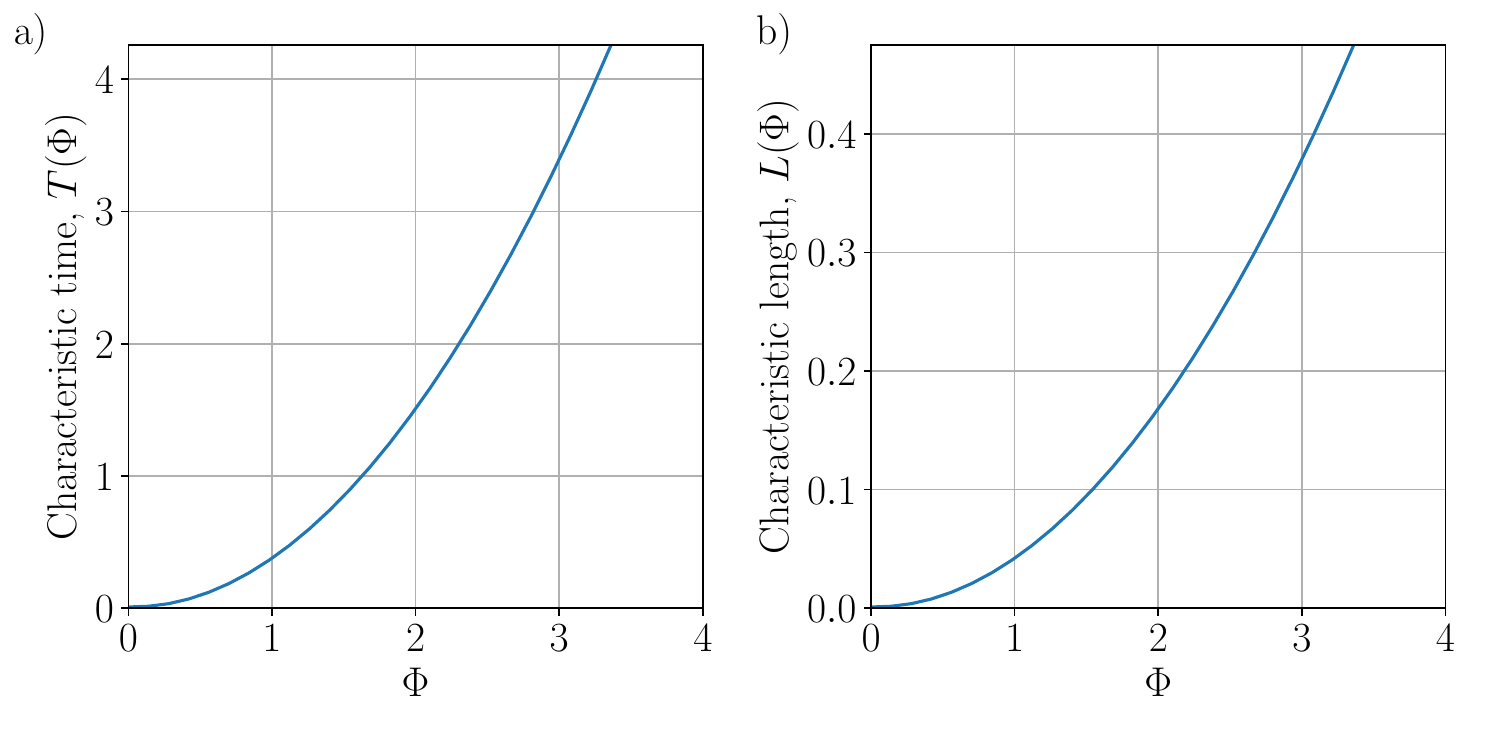} 
    \caption{a) Characteristic time, $T(\Phi)$, b) Characteristic length, $L(\Phi)$}
    \label{fig: solved_implicit_form_T_of_Phi}
\end{figure}

The PDE system depends on a single parameter $\Phi$, so for the following analysis we look to define our quantities of interest with respect to $\Phi$. To observe $L(\Phi)$ and $T(\Phi)$, we generate the quadratic  
\begin{equation}
\sin^2(\theta) + \sin^\frac{4}{3}(\theta)\Phi^2 \frac{1}{H^2}\left(\frac{\sigma H}{\rho g}\right)^{\frac{2}{3}} -1=0,
\label{eq: grav_implicit_form_T_char}
\end{equation}
to yield $\theta(\Phi)$. Taking $s \coloneqq \sin^{2/3}(\theta)$, we may solve the cubic equation to find such roots that connect $\Phi$ to $\theta$. Taking only valid values of $0<s\leq 1$, using $0<\theta \leq \frac{\pi}{2}$, we may find $L(\Phi)$ and $T(\Phi)$. The Appendix \ref{app: Existance_Uniqueness} includes a proof of the existence and uniqueness of such a root. Looking to figure \ref{fig: solved_implicit_form_T_of_Phi}, we see a graph of the relation. We note that $\epsilon(\theta(\Phi))$ is also defined from a choice of $\Phi\geq 0$.

We see that the constant solution $h= 1$ is a steady state of the system. Performing a linear stability analysis, one recovers the dispersion relation
\begin{equation}
    \omega(k) = -3ik - \Phi k^2 - k^4,
\end{equation}
with $\Re(\sigma(k)) < 0$, $\forall \theta\in \left(0, \frac{\pi}{2}\right]$ indicating a linearly stable film.

With a better understood reduced-dimensional model of the system of interest, we now look to generate synthetic datasets from the governing equation \eqref{eq: thin_film_grav_case_eqn}.

\subsection{Data Generation Considerations}
Having obtained a suitable PDE model, we look to produce a discrete numerical solution that can be used for equation discovery. A BDF1 scheme with second-order spatial central differences was used for the previously studied equation. We use this scheme again and highlight that implicit time-stepping schemes paired with second-order accurate spatial stencils have shown to provide accurate solutions of similar thin-film equations \citep{Ha_Kim_Myers_2008_thin_film_implicit_schemes_good}. The choice of parameter dependent scalings introduces a layer of complication. We produce datasets that correspond to a fixed value of $\Phi$ chosen \textit{a priori}.

With a varying $\epsilon(\theta(\Phi))$, we choose to maintain the length of the dimensional spatial domain and consequently vary the non-dimensional spatial domain with $\theta(\Phi)$. The relative sampling density is maintained between re-scalings, but the non-dimensional spatial resolution, $\tilde{dx} = \frac{L_X}{N_x}\propto \sqrt[3]{\frac{1}{\sin (\theta(\Phi) )}}$ varies, potentially affecting dataset solution quality. With spatial resolution becoming finer as $\theta(\Phi)$ decreases, the stability of the solver used for data generation is eventually impacted. This is in contrast to the previously studied equations, which allowed for static, uniform meshes to be used and produced accurate solutions for a range of parameter values. To combat this, adaptive time stepping while maintaining a similar overall computational budget is utilised. Step size adjustment is used to ensure that the solution is evaluated at pre-defined output times. After generation, these datasets are immediately interpolated to the common reference mesh size of all other training data, as is our standard pre-process to equation discovery. Solutions $h(x,t)$ remain smooth and regular with the governing PDE system lacking spatially-dependent coefficients, so we expect the generated solutions to remain valid for comparison between $\Phi$ values. Fixing the dimensional spatial domain also ensures that the geometry of the system remains consistent.

For the temporal domain of training datasets, we fix a reference non-dimensional temporal domain with respect to $\Phi$. We mark the reference non-dimensional final time time with a star $\tilde{t}_F^*$, and may then take proportions of the reference domain during experimentation. This is for consistency to allow for more direct comparison with the previous variable resolution non-gravity case. As $\theta(\Phi)$ decreases, the characteristic time-scale increases and a much greater proportion of the data is near the steady state $h=1$, which is intuitively supported by inspecting the varying characteristic time-scales and dimensional final times in table \ref{tab: Numerical_Simulation_Info_Grav_Case}. The non-dimensional final time is chosen using the slow dynamics criteria for the evolution of the training initial condition \eqref{eq: SuperposedModes_IC_written} for a small reference value $\Phi \approx 0.089$, corresponding to $\theta = \pi/3$. The fixed non-dimensional temporal domain ensures that datasets span an equivalent fraction of the characteristic time-scale. Doing so ensures we compare stages of evolution relative to the dynamics over $\Phi$.

{
\sisetup{
    scientific-notation = false,
    tight-spacing = true,
    output-exponent-marker = {},
    table-number-alignment = center,
    round-mode = figures,
    round-precision = 3,
    group-digits = false
}
\begin{table}[hbtp]
\centering
\begin{tabular}{S[table-format=1.5] S[table-format=1.5] S[table-format=1.4] S[table-format=2.1] S[table-format=1.4] S[table-format=1.5] S[table-format=1.5] S[table-format=1.7] S[table-format=1.6]}
\toprule
$\theta$ & $L_{char}$ & $\epsilon$ & $L_X$ & $L_X L_{char}$ & $T_{char}$ & $\tilde{t}_F^*$ & $t_F$ & $\Phi$ \\
\midrule
1.565 & 0.00109 & 0.161 & 10.3 & 0.0112 & 0.00976 & 0.00812 & 0.0000793 & 0.001001 \\
1.508 & 0.00109 & 0.161 & 10.3 & 0.0112 & 0.00979 & 0.00812 & 0.0000795 & 0.01012 \\
1.047 & 0.00114 & 0.153 & 9.82 & 0.0112 & 0.0118 & 0.00812 & 0.0000960 & 0.08854 \\
0.7348 & 0.00124 & 0.141 & 9.01 & 0.0112 & 0.0166 & 0.00812 & 0.000135 & 0.1558 \\
0.5156 & 0.00138 & 0.127 & 8.14 & 0.0112 & 0.0251 & 0.00812 & 0.000203 & 0.2243 \\
0.3618 & 0.00154 & 0.114 & 7.28 & 0.0112 & 0.03900 & 0.00812 & 0.000317 & 0.301 \\
0.2539 & 0.00172 & 0.102 & 6.50 & 0.0112 & 0.0616 & 0.00812 & 0.000500 & 0.3912 \\
0.1782 & 0.00194 & 0.0904 & 5.78 & 0.0112 & 0.0980 & 0.00812 & 0.000796 & 0.5019 \\
0.1250 & 0.00218 & 0.0804 & 5.14 & 0.0112 & 0.157 & 0.00812 & 0.00127 & 0.6396 \\
0.08772 & 0.00245 & 0.0715 & 4.57 & 0.0112 & 0.251 & 0.00812 & 0.00204 & 0.8125 \\
0.06155 & 0.00276 & 0.0635 & 4.07 & 0.0112 & 0.402 & 0.00812 & 0.00326 & 1.030 \\
0.04319 & 0.00310 & 0.0564 & 3.61 & 0.0112 & 0.644 & 0.00812 & 0.00523 & 1.306 \\
0.03031 & 0.00349 & 0.0502 & 3.21 & 0.0112 & 1.03 & 0.00812 & 0.00839 & 1.654 \\
0.02127 & 0.00393 & 0.0446 & 2.85 & 0.0112 & 1.66 & 0.00812 & 0.0135 & 2.096 \\
0.01492 & 0.00442 & 0.0396 & 2.54 & 0.0112 & 2.66 & 0.00812 & 0.0216 & 2.654 \\
0.01047 & 0.00497 & 0.0352 & 2.25 & 0.0112 & 4.26 & 0.00812 & 0.0346 & 3.361 \\
\bottomrule
\end{tabular}
\vspace{4pt}
\caption{Values important to the setup of the gravity affected thin film numerical simulations. Values are for training data. We see: angle of inclination to the horizontal, $\theta$, characteristic length (m), $L_{char}$, the length-scale discrepancy, $\epsilon=\frac{H}{L}$, the non-dimensional domain length $L_X = 64\epsilon$, the dimensional domain length (m), the characteristic time-scale, $T_{char}$, the reference non-dimensional final time $\tilde{t}_F^*$, dimensional final time (s) and corresponding parameter $\Phi$}
\label{tab: Numerical_Simulation_Info_Grav_Case}
\end{table}
} 

Looking to table \ref{tab: Numerical_Simulation_Info_Grav_Case}, we see how the non-dimensional spatial domain length, $L_X$, varies with the parameter $\Phi$. It is chosen so that the length will be $64$ times the length-scale discrepancy and so the dimensional length will be $64$ times the dimensional height, thus fixing the aspect ratio. The initial perturbations will then have the same frequency in dimensional space, though their frequency changes with respect to non-dimensional space.

The choices made allow for comparison of identified governing structures across regimes. The parameter-dependent scalings necessitate a choice between fixing dimensional or non-dimensional domains. 

\subsection{Alterations to Equation Discovery}
In the previous system without gravity, we recovered terms relating to a single source of dynamics. Now we target a collection of nonlinear terms arising from a dataset built on competing physical mechanisms. To do so effectively, we introduce modifications to our equation discovery process and provide reasoning for their use.

The most prominent change is that of an altered function library. Previously, we had a moderate to large $60$ term library that produced good results. Using this previous library for the now gravity-affected thin film gave poor results with advective terms commonly being missed. For the new system with gravity that we study, we introduce a `flux library' of a moderate size characterised by $15$ terms. Doing so reduces the number of competing linear dependencies when regressing and alleviates some of the destabilising effects of significant multi-collinearity. To build the flux library, we use the known structure of thin liquid film models 
\begin{align}
    h_t &= -q_x, \\
    q &= q(h),
\end{align}
in which a mass conservation equation is coupled to a horizontal velocity flux that is enslaved to interfacial height in the case of single-equation thin film models. Applying this knowledge is a way of physics informing the PDE discovery process that allows us to assume the form of the governing equation. Thus, we require only a library of terms to fit $q(h)$. With this imposed form, we see that whilst the span of the smaller flux library is strictly contained within that of the $60$ term library, this span still describes a rich set of dynamics. The library contains the terms
\begin{align}
    \bigg\{ \left( h\right)_x, \left( h^2\right)_x, \left( h^3 \right)_x, \left( h_x\right)_x, & \left( hh_x\right)_x, \left( h^2h_x\right)_x, \left(h^3h_x \right)_x, \left( h_{xx} \right)_x, \left(hh_{xx} \right)_x, \left( h^2h_{xx} \right)_x, \left( h^3h_{xx} \right)_x, \nonumber \\ & \left( h_{xxx} \right)_x, \left( hh_{xxx} \right)_x, \left( h^2h_{xxx} \right)_x, \left( h^3h_{xxx} \right)_x \bigg\}. 
\end{align} 
Inspecting a term, we see that the ratios of coefficients of terms that were previously free to vary are now fixed, such as $\left( h^3h_{xxx} \right)_x = 3h^2h_xh_{xxx} + h^3h_{xxxx}$. This results in a great improvement in the conditioning of the original system. Additionally, the smaller library size results in fewer candidate models being generated and this library size is still comparable to those used in other state-of-the-art studies \citep{heinrich2025rediscovering_KdV_ARXIV, VandenHeuvel_Buenzli_Pascal_Simpson_2024}. 

To further improve conditioning, we utilise `temporal trimming' of our dataset. We subsample the design matrix to remove rows corresponding to early/late times of the evolution. With additional convective terms we see more sustained nonlinear dynamics compared to the previous case, which may reduce the reliance on the inclusion of well-resolved initial transient dynamics. The use of temporal trimming was not required to see robust success for the previous case although, as noted in section \ref{sec: nonlinear_numerical_experiments}, it could improve coefficient estimates of the final model. For this case with gravity, the conditioning improvements provided by trimming improve structural identification. Furthermore, the temporal trimming process improves coefficient estimates and thus makes models matching the governing equation structure more competitive in model selection processes. 

The `temporal trimming' process is inspired by the data thresholding process of \citet{vandenheuvel_drovandi_simpson_2022} and their matrix pruning \citep{VandenHeuvel_Buenzli_Pascal_Simpson_2024} procedures. For this system, applying matrix pruning would mean removing extreme values of the estimated observable $h_t(x,t)$. We therefore observe significant overlap in the regions of data that would be removed by applying temporal trimming compared to matrix pruning, because of the dissipative nature of the underlying system. An additional effect is that data at the initial condition and final time are removed, which require the use of one-sided finite differences for estimation of $h_t$. This improves conditioning and performance of equation discovery.

\subsection{Numerical Experiments and Results}
Having considered the implication of varying the parameter $\Phi$ on characteristic scalings, shown in table \ref{tab: Numerical_Simulation_Info_Grav_Case}, we may proceed with a parameter sweep. We evolve the same initial condition \eqref{eq: SuperposedModes_IC_written} as before. For the system without gravity, we varied the time proportion of training data supplied. We do this for the gravity case and concurrently sweep $\Phi \in [0.001,1.654]$. Therefore, for a fixed $\Phi$ value, we supply varying training datasets. We evaluate candidate models on a testing dataset with fixed dimensional final time. The time span of the testing dataset is determined through the use of the slow dynamics criteria applied with a small reference value $\Phi \approx 0.089$, corresponding to $\theta = \pi/3$. The testing datasets simulate to $t_F = 0.1545$ s with non-dimensional final times ranging between $0.03627$ and $15.83$ for the associated characteristic times and $\Phi$ values shown in table \ref{tab: Numerical_Simulation_Info_Grav_Case}. Testing datasets are equivalent for a fixed $\Phi$.

To generate error metrics, we can compare the simulated candidate models to the inputted solution data, as previously. However, comparison between $\Phi$ values is affected by the change in simulated time intervals. Previously, we normalised errors by the null error (defined as the error of the model $h_t = 0$) with respect to the testing dataset. Applying the same process, we normalise errors with a null error dependent on $\Phi$, which ensures all errors that are less than the null error of the associated dataset are contained in $[0,1)$.

\begin{figure}[hbtp]
    \centering
    \includegraphics[width=\linewidth]{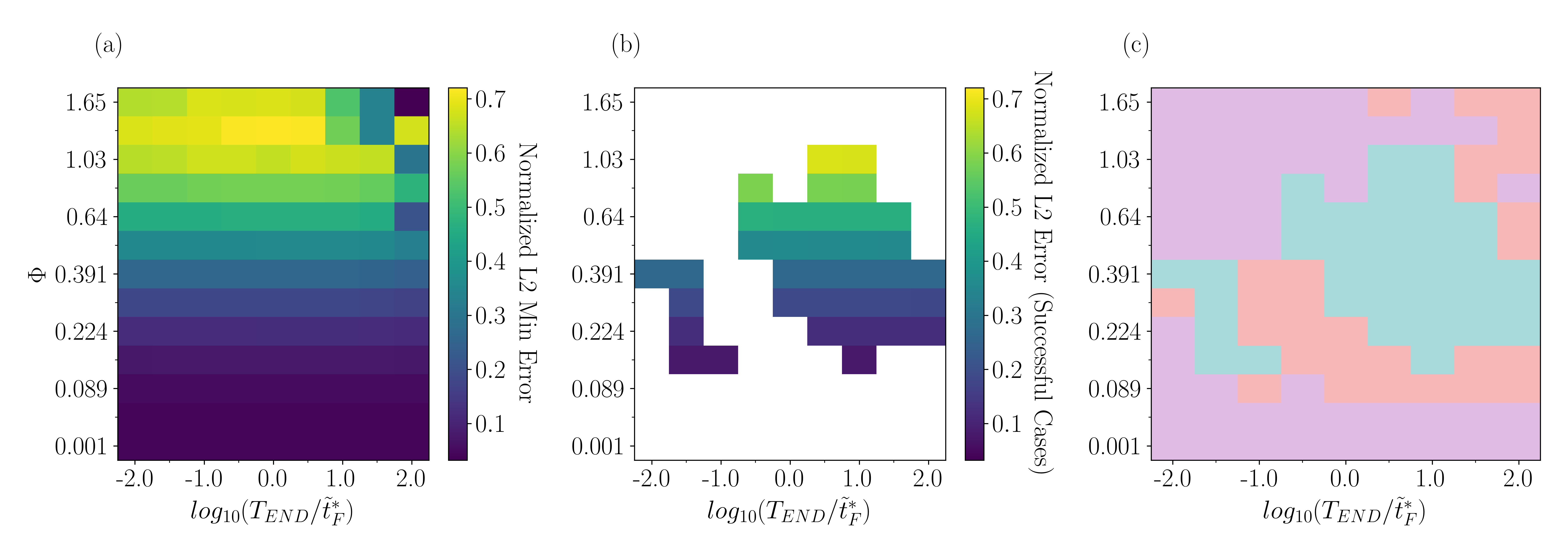}
    \caption{(a) Minimum $L_2$ error normalised with respect to the null error for corresponding $\Phi$. (b) Where valid, the normalised $L_2$ error on the testing dataset associated with the correct model. (c) The success-failure classification with near-successes also coloured}
    \label{fig: 2D_grav_plot_fixed_dim_val}
\end{figure}

Sweeping over both parameter values of $\Phi$ and final non-dimensional times of training data, we generate a set of calibrated candidate models for each pair $(\tilde{t}_F, \Phi)$ from our equation discovery process. Within each set of models, we are interested in strong performing models and models that match the governing equation used to generate datasets. A 2D array of results are shown in figure \ref{fig: 2D_grav_plot_fixed_dim_val}, presenting minimum $L_2$ errors across models for each $(\tilde{t}_F, \Phi)$ alongside $L_2$ errors of models matching the form of the known governing equation. We also colour with a success or failure classification and extend to include a near-success region, in which models with three terms for which one or more have incorrect degree of algebraic nonlinearity are classified as near-success. 
We see that the error for a fixed $\Phi$ varies minimally. This reflects the results of the case without gravity, particularly for successful regions. This may suggest a saturation of model quality is reached, requiring additional trajectories to improve errors further. The discovered success region is non-convex, however, there exist numerous $\Phi$ values for which the success region is connected by areas of near-success. The overall union of the near-success and success regions from this study is nearly convex with exceptions occurring at large $\Phi$ values and only one such exception in our displayed results.

We see errors generally increase with $\Phi$. Some recovered models align with the no gravity case. As $\Phi$ varies, such models will produce a constant error with respect to the downslope driving and an error scaling linearly with $\Phi$ for the normal hydrostatic effect of gravity. The dependence of the error on $\Phi$ may partially be explained by minimum $L_2$ error models potentially not capturing the hydrostatic effect of gravity on the film. 

For the lowest value case of $\Phi\approx 0.001$, we see models
$$h_t =  \left(-1.01 h^3h_{xxx} \right)_x$$
and
$$h_t = \left(-0.998h^3 -0.998h^3h_{xxx} \right)_x.$$
The first such model shows correspondence to the previous case without gravity. The two-term model shows the inclusion of the downslope gravitational driving, though fails to include the small normal hydrostatic effect of gravity that is present. The $L_2$ error of the two term model is lower for all training time spans studied and is typically of an order of magnitude of difference. This is expected, given that $\Phi \ll 1$ corresponds to an inclination of near $\pi/2.$ Difficulty recovering the hydrostatic term is expected for low values of $\Phi$, partially due to its weakness, but also due to the magnitude disparity in coefficients of the governing equation.

A region of erratic behaviour is also visible for high values of $\Phi$ and $\tilde{t}_F$. This correlates to a region where the design matrix formed for sparse regression has a very high condition number. The condition number increases sharply as both $\Phi$ and $\tilde{t}_F$ increase, reaching values of $\mathcal{O}(10^9)$. 

\subsubsection{Strong Performing Models}
Having performed a sweep of $(\tilde{t}_F,\Phi)$ and with a collection of models for each pair, we may use a selection method to indicate the best model for each pair. In the previous section, we used AICc analysis. We now switch to applying Pareto frontier optimisation \citep{deb2011introduction}, as mentioned in section \ref{sec: Model_Selection_Methodology}, to balance the $L_2$ error on testing data against the number of terms in the model. The switch is motivated by the additional model complexity considered, in which the additional terms are qualitatively distinct from those of the surface-tension-driven thin film equation \eqref{eq: The_Basic_Thin_Film_Eqn}. The introduction of these terms provides additional sources of multi-collinearity, as spurious advective candidates such as $h_x$ are now more correlated with terms in the governing equation \eqref{eq: thin_film_grav_case_eqn}. This is reflected in the AICc curve, which becomes notably flat for model complexities at or beyond that of the governing equation, leading to potentially unstable selections of a globally optimal model. Thus, we may interpret this behaviour as the Pareto frontier analysis providing a locally-informed criterion that increases the stability of model selection for our numerical experiments.

\begin{figure}[hbtp]
    \centering
    \includegraphics[width=0.99\linewidth]{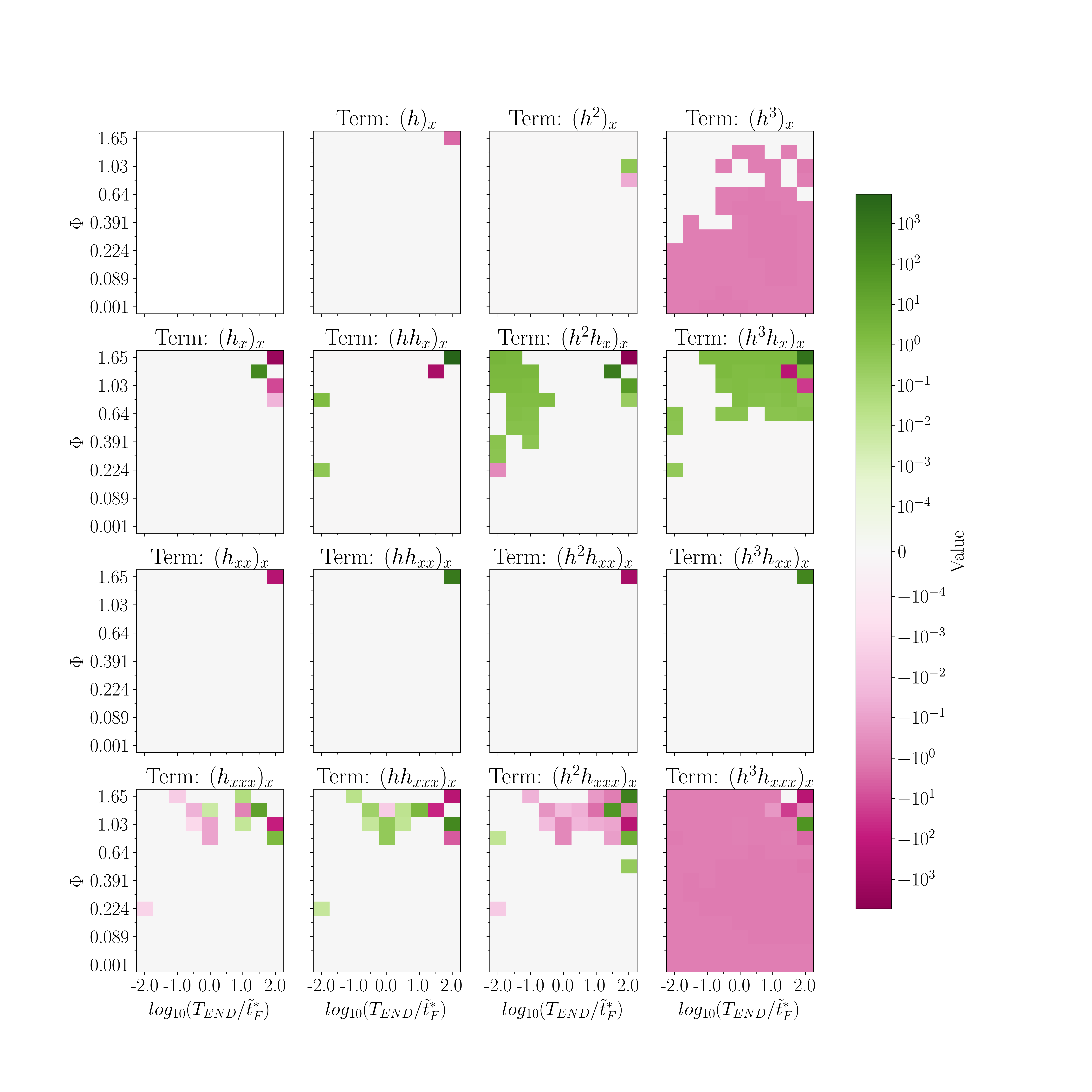}
    \caption{Heat map of low $L2$ error models selected from the Pareto frontier with coefficients proportional to colour. Grey squares indicate an unselected term with a corresponding coefficient value of $0$, as indicated by the provided colour bar. Model selection is performed using a marginal gain criterion. One-sided gradients of error with respect to model complexity are evaluated. Then, starting from the lowest complexity model, the most complex model for which the normalised marginal gain in error reduction still exceeds a unit threshold is selected
    }
    \label{fig: HeatMap_PARETO_2D_fixed_dim_val_min_L2_models}
\end{figure}

Applying this selection criterion yields figure \ref{fig: HeatMap_PARETO_2D_fixed_dim_val_min_L2_models}, which displays the model coefficient values for each term in the function library over $(\tilde{t}_F,\Phi)$ for the selected model. A diverging colour map is used to highlight cancellation of terms with differing algebraic nonlinearity, which is particularly noticeable for terms of the form $h^nh_{xxx}$. A transition between $\Phi=0.502$ and $\Phi=0.640$ appears, where strong model structures change quite sharply. This corresponds to a critical switching inclination of between $55.7^\circ$ and $62.4^\circ$ or between $0.619\pi/2$ and $0.693\pi/2$. Models change from complex, likely overfitted, suggestions with $8-15$ terms per model to more sparse suggestions. We see that the normal hydrostatic effect due to gravity ($h^3h_{x}$) is commonly not recovered for low values of $\Phi$, reflecting the use of coefficient magnitude as a proxy for importance. Training data with short time spans have seemingly produced models with greater uncertainty in the algebraic degree of $h^nh_x$. The previously mentioned conditioning correlation may relate to the performance deterioration for models corresponding to long time span training datasets.

\subsubsection{Model Recovery Frequencies}
With a vast collection of models identified over the parameter sweep, we may look at the proportion of candidate sets that include specific model structures. Table \ref{tab: grav_results} shows these proportions for the most frequently recovered models and we briefly discuss a selection of these models below.

The 15-term model is always recovered, indicating the choice of minimum tolerance value was sufficient for each case, as the least squares result without thresholding was obtained.

The single-term model corresponds to the no-gravity case and also indicates that the sampling around the critical thresholding value has occurred. We note that only a single case of $(\tilde{t}_F,\Phi) = (0.812, 1.654)$ does not recover this as the single term model and instead recovers $h_t = \alpha (h^3h_x)_x,\quad  \alpha\neq 0$. The only other one-term model recovered in the sweep was $h_t = \beta (h^2h_{xxx})_x$ for $(\tilde{t}_F,\Phi)\in \{(0.812,1.030), (0.812,1.306)\}$.

The two-term model, as previously discussed, omits the normal hydrostatic effect of gravity. The high proportion of cases with this omission may be inflated by a sample unintentionally biased towards `small' $\Phi$ values. 

The 9-term model forms an initially unexpected inclusion, though it is related to a model of the form $$h_t = \left( \alpha_0h^3 + \sum_{i=0}^3\beta_ih^ih_x + \sum_{i=0}^3\gamma_ih^ih_{xxx} \right)_x.$$ 

The next most common model is the `correct' one that aligns with the system used to generate the datasets.

Then we have a two-term model that includes the normal hydrostatic effect of gravity, but excludes the downslope gravitational driving term. And another two-term model which relates to an overfitting of the surface-tension term with competing terms of varying algebraic degree. 

\begin{table}[htbp]
    \centering
    \begin{tabular}{c c c p{10cm}}
        \toprule
        \textbf{Case} & \textbf{Complexity} & \textbf{Score} & \textbf{Terms} \\ 
        \midrule
        $(a)$ & 15 & 1.000 & \Big\{ $\left( h^3 \right)_{x}$, $\left( h^3h_{x} \right)_{x}$, $\left(h^3h_{xxx}\right)_{x}$ \Big\}, \Big( $\left(h\right)_{x}, \left( h^2 \right)_{x}$, $\left( h_{x} \right)_{x}$, $\left( hh_{x} \right)_{x}$, $\left( h^2h_{x} \right)_{x}$, $\left( h_{xxx} \right)_{x}$, $\left( hh_{xxx} \right)_{x}$, $\left(h^2h_{xxx}\right)_{x}$ \Big),  ${\left( h_{xx} \right)_{x}}$, $\left( hh_{xx} \right)_{x}$, $\left( h^2h_{xx} \right)_{x}$, $\left( h^3h_{xx} \right)_{x}$ \\
        $(b)$ & 1 & 0.992 & \Big\{ $(h^3h_{xxx})_{x}$ \Big\} \\
        $(c)$ & 2 & 0.573 & \Big\{ $\left( h^3 \right)_{x}$, $\left(h^3h_{xxx}\right)_{x}$ \Big\} \\ 
        $(d)$ & 9 & 0.453 & \Big\{ $\left( h^3 \right)_{x}$, $\left( h^3h_{x} \right)_{x}$, $\left(h^3h_{xxx}\right)_{x}$ \Big\}, \Big( $\left( h_{x} \right)_{x}$, $\left( hh_{x} \right)_{x}$, $\left( h^2h_{x} \right)_{x}$, $\left( h_{xxx} \right)_{x}$, $\left( hh_{xxx} \right)_{x}$, $\left( h^2h_{xxx} \right)_{x}$ \Big) \\
        $(e)$ & 3 & 0.308 & \Big\{ $\left( h^3 \right)_{x}$, $\left( h^3h_{x} \right)_{x}$, $\left(h^3h_{xxx}\right)_{x}$ \Big\} \\ 
        $(f)$ & 2 & 0.282 & \Big\{ $\left( h^3h_{x} \right)_{x}$, $\left(h^3h_{xxx}\right)_{x}$ \Big\} \\ 
        $(g)$ & 2 & 0.265 & \Big\{ $\left( h^3h_{xxx} \right)_{x}$ \Big\}, \Big( ${\left( h^2h_{xxx} \right)_{x}}$ \Big) \\ 
        $(h)$ & 3 & 0.205 & \Big\{ $\left( h^3 \right)_{x}$, $\left(h^3h_{xxx}\right)_{x}$ \Big\}, \Big( ${\left( h^2h_{xxx} \right)_{x}}$ \Big) \\
        $(i)$ & 10 & 0.205 & \Big\{ $\left( h^3 \right)_{x}$, $\left( h^3h_{x} \right)_{x}$, $\left(h^3h_{xxx}\right)_{x}$ \Big\}, \Big( $\left(h\right)_{x}$, $\left( h^2 \right)_{x}$, $\left( h_{x} \right)_{x}$, $\left( hh_{x} \right)_{x}$, $\left( h_{xxx} \right)_{x}$, $\left( hh_{xxx} \right)_{x}$, $\left( h^2h_{xxx} \right)_{x}$ \Big) \\
        \bottomrule
    \end{tabular}
    \vspace{4pt}
    \caption{Model Complexity, Recovery proportion of specific model structure (combination of terms, not any weights), and Associated Terms. These are models that are recovered at least 20\% of the time. Text surrounded by large curly braces indicates an expected term, matching a term in the governing equation used for data generation. Text surrounded by large parentheses indicates a term with incorrect algebraic degree. Any terms outside of curly braces or large parentheses indicate spurious terms. A visual representation of this table is provided in figure \ref{fig: model_recovery_locations}}
    \label{tab: grav_results}
\end{table}

\begin{figure}[hbtp]
    \centering
    \includegraphics[width=0.95\linewidth]{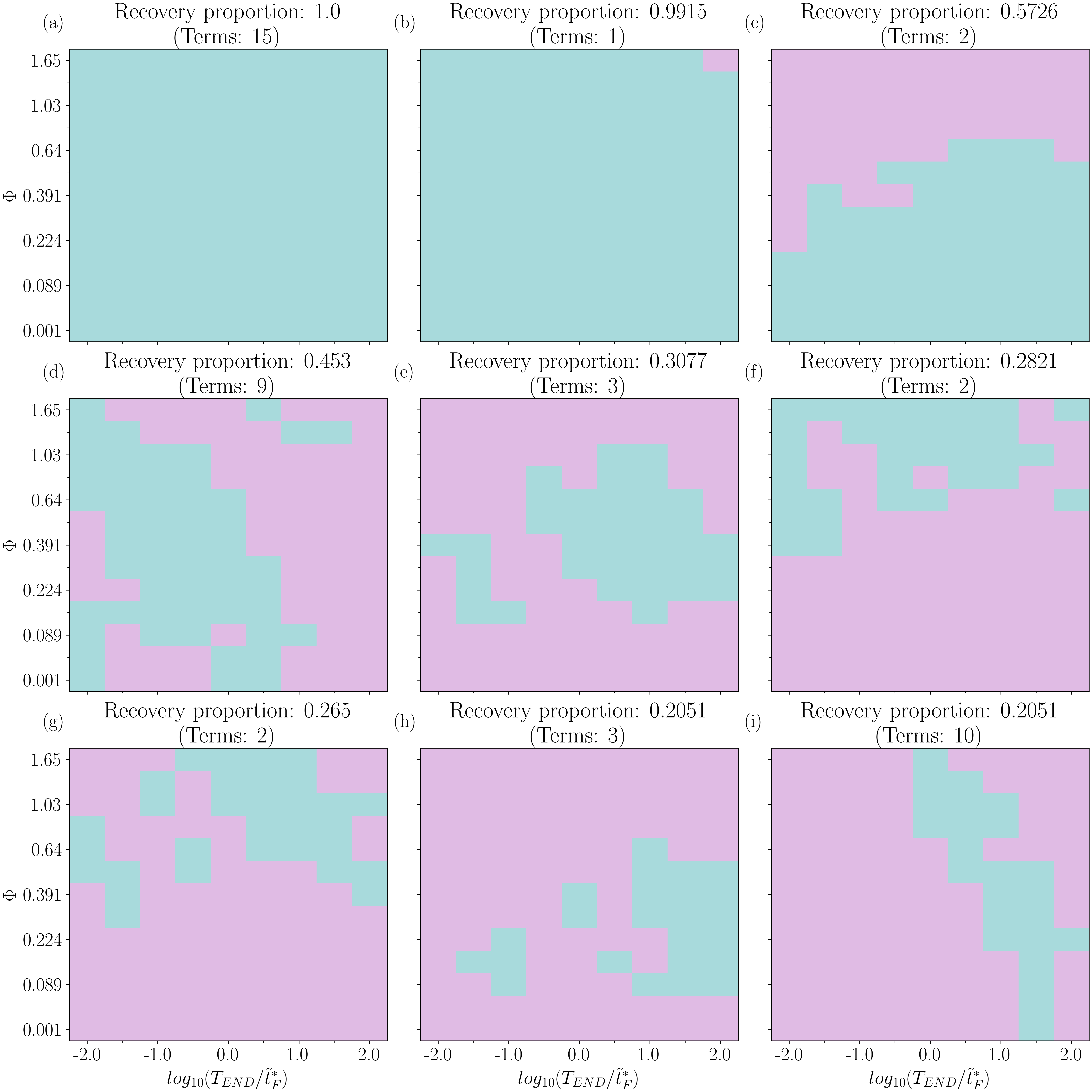}
    \caption{Panel $(a)-(i)$ correspond to the models in table \ref{tab: grav_results}. This shows a green-pink binary classification of regions in the parameter space in which the corresponding model structure was recovered. Green corresponds to a successful recovery}
    \label{fig: model_recovery_locations}
\end{figure}

We may inspect the region of recovery for each model, as shown in figure \ref{fig: model_recovery_locations}. For specific sub-panels, such as $(d), (g)$ and $(i)$, we see a banded structure of recovery. This aligns with the structured, banded variation of the condition number across the parameter space.

We see from table \ref{tab: grav_results} that $30.8\%$ of the datasets used yielded a correct model, and from figure \ref{fig: model_recovery_locations} that these recovery successes are concentrated towards intermediate $\Phi$ values with slightly longer time intervals than our reference case. Despite the use of clean data, we see a number of cases in which the correct model was not retrieved. Such cases show a variety of models with similarities to the PDE \eqref{eq: thin_film_grav_case_eqn} used for data generation. We observed robust recovery of the highly nonlinear term $(h^3h_{xxx})_x$ associated with surface tension effects. Inspecting the recovered two-term models $(c)$ and $(f)$ shows ambiguity in the selection of the secondary term with recovery locations that, on taking the union, approximately partition the explored parameter space, $(\tilde{t}_F, \Phi)$. Correct model recovery is difficult for extremal values of $\Phi$. This could be explained by the terms $(h^3)_x, (h^3h_x)_x$ requiring intermediate values of $\Phi$ to have a balanced effect on dynamics. In extremal $\Phi$ cases, one of the terms may appear to dominate and, with near-collinearity of the terms, mask the presence of the other term. 

Extending the region of successful recovery remains challenging. As highlighted, we see multi-collinearity of the terms $(h^3)_x, (h^3h_x)_x$ capturing the advective dynamics that results in competition between various representations of the governing equation. We also uncover instabilities in correct model recovery, which is reflected through the non-convex recovery region. This may relate to the initial conditioning of the design matrix $\Theta$, as there is structured, banded variation of the condition number across the parameter space. We review these challenges in the discussion that follows.

%% file: conclusion.tex
In this article we numerically investigated the applicability of equation discovery techniques to surface-tension driven flows of thin viscous liquid films, a canonical problem of wide theoretical and practical interest. We focused on sparse regression methodologies and applied them to systems both with and without the effects of gravity. The studied mathematical models were highly nonlinear and distinct from prototypical validation cases often found within equation discovery methodologies, leading to new challenges and potential solutions that were systematically explored as part of this work.

Together, the studied systems allowed us to outline the capabilities and limitations of sparse regression based equation discovery for thin liquid film models. The system without gravity shows robust identification, demonstrating that the approach is viable in this setting. The introduction of gravity reveals where the methodology is strained. This three term case, modelling two physical effects, sits below the complexity threshold of the leading first-principles models discussed in section \ref{sec: Thin_Liquid_Films}, yet we see reliable identification becomes difficult. This may point to a complexity ceiling in the sparse regression framework we applied. Such a ceiling may emerge due to the stable nature of the training data, all choices of which eventually decay to a film of constant height, limiting the informational content of the dataset.

Limitations were also identified for the gravity-enriched model. Disparities in the magnitudes of coefficients appeared to obstruct recovery of the governing equation and is consistent with the findings of \citet{fung_fasel_juniper_2025}. This issue seems to be compounded by the choice of monomials for our function library that contributed to generating initially ill-conditioned systems for regression, thus affecting the quality of identification. With monomials, we see the structure of the design matrix $\Theta(H,Q)$ is similar to that of a Vandermonde matrix, which is notoriously ill-conditioned in general \citep{Pan_2016_How_bad_are_Vandermonde}, and is well studied within the context of interpolation and quadrature applications. However, using a more numerically stable basis, such as Chebyshev polynomials, may no longer be sparse with respect to the governing equation. This is also the case for a function library constructed from monomials of the perturbation of the film $\eta(x,t) \coloneq h(x,t)-h_0$. This violation of the sparsity assumption that underpins sparse regression based equation discovery typically results in performance degradation, despite a better-conditioned design matrix being available. This once again highlights a potential complexity ceiling for the models recovered by sparse regression based equation discovery.

The use of knowledge of the system to tailor the construction of the function library alleviated some of these issues. The introduction of gravity produces a non-convex success region that points to a more fragile environment for recovery when compared to the system without gravity. Consequently, even with large volumes of clean data, a guarantee of robustness was still found to be lacking. Practical deployment of equation discovery methods require careful experimental design, such as through the choice of initial conditions, rather than relying solely on data quality and quantity.

We believe that there are numerous promising extensions to this work. More immediately, we have established a pipeline for equation discovery that is agnostic to the data source and could, in principle, be readily applied to DNS or experimental data. Our results may be particularly useful in experimental settings, where trajectory length and resolution are constrained. In the purely theoretical context, we may also look to use additional fields such as horizontal velocity flux to probe true two-equation models. Some researchers have suggested going beyond two equation models \citep{Ruyer_Quil_Manneville_2000, Richard_Gisclon_Ruyer_Quil_Vila_2019}, however we highlight two-equation models have been subjected to decades of benchmarking and tuning across parameter regimes that three-equation models have not (yet at least) undergone. Thus two-equation models provide a natural step for providing additional complexity, whilst mitigating some of the risk associated with additional stiffness and complicated numerics. With access to new datasets, we can perform exploratory work for very thin liquid films in regimes where additional micro-scale effects, potentially including stochastic contributions, become non-negligible. Conversely, in a high-fidelity simulation pipeline the developed methodology would allow us to access regimes that lie beyond the reach of traditional asymptotic approaches and their multi-scale assumption underpinnings towards films of moderate thickness. For longer-term algorithmic developments, we may be interested in taking advantage of the low-dimensional representation of our systems of interest in Fourier space, in which one could consider a system of ODEs representing the evolution of modes. Resolving the choice of dimension of this system presents a challenge and is complicated by nonlinear evolution. Another avenue of development may be the choice of sampling methodology, which could connect to randomised numerical linear algebra. Selecting specific regions of space-time to sample from or weight differently for regression promises to improve results, but sampling in a principled manner presents another challenge.

%% file: Appendix.tex
\newpage
\section{Derivation Details for a Surface Tension Driven Thin Film Equation}
\label{app: Basic_Thin_Film_Derivation}
We begin with a 2D incompressible Navier-Stokes system
\begin{equation}
    \begin{cases}
        \rho \left( u_t + [uu_x + vu_y]\right) &= -p_x + \mu (u_{xx} + u_{yy}) + \rho g\sin(\theta), \\
        
        \rho \left( v_t + [uv_x + vv_y]\right) &= -p_y +  \mu (v_{xx} + v_{yy}) - \rho g\cos(\theta), \\
        \hspace{12mm} u_x + v_y &= 0.
    \end{cases}
    \label{eq: Dim_Grouped_NS}
\end{equation}
We use the following non-dimensionalisation,
\begin{align*}
        x = \tilde{x}L, \quad
        y = \tilde{y}\epsilon L, 
        \quad
        u = \tilde{u}U, & 
        \quad
        v = \tilde{v}\epsilon U, 
        \quad
        t = \tilde{t}\frac{L}{U},
        \quad
        p = \tilde{p} \frac{\mu U}{L},
        \quad
        g = \tilde{g}\frac{U^2}{L}, \\
        \textit{Re} \coloneq \frac{\rho UL}{\mu}, 
        \quad
        &\textit{Ca} \coloneq \frac{\mu U}{\sigma},
        \textit{St} \coloneq \frac{\rho gL^2}{\mu U},
\end{align*}
where $\epsilon\coloneq \frac{H}{L} \ll 1$ and $H$ represents the vertical length-scale which is disparate from the horizontal length-scale $L$. This leads to a 2D non-dimensionalised incompressible Navier-Stokes system
\begin{equation}
    \begin{cases}
        \epsilon^2\textit{Re} \left( \tilde{u}_t + [\tilde{u}\tilde{u}_x + \tilde{v}\tilde{u}_y]\right) &= -\tilde{p}_x +  (\epsilon^2\tilde{u}_{xx} + \tilde{u}_{yy}) + \epsilon^2\textit{St}\sin(\theta), \\
        \epsilon^4\textit{Re}\left( \tilde{v}_t + [\tilde{u}\tilde{v}_x + \tilde{v}\tilde{v}_y]\right) &= -\tilde{p}_y +  (\epsilon^4\tilde{v}_{xx} + \epsilon^2\tilde{v}_{yy}) - \epsilon^3\textit{St}\cos(\theta), \\
        \hspace{12mm} \tilde{u}_x + \tilde{v}_y &= 0.
    \end{cases}
    \label{eq: ND_Grouped_NS}
\end{equation}
We then additionally assume that $\epsilon^2 \textit{Re} \ll 1$ and $\epsilon^2\textit{St} \ll 1$ to obtain the reduced system
\begin{equation}
    \begin{cases}
        -\tilde{p}_x +  \tilde{u}_{yy} 
        &= 0,
        \\
        -\tilde{p}_y &= 0, \\
        \hspace{12mm} \tilde{u}_x + \tilde{v}_y &= 0.
    \end{cases}
    \label{eq: ND_Grouped_NS_Lubrication}
\end{equation}
With this simplified system, we now consider boundary conditions. A no-slip boundary is applied at $y=0$,
\begin{equation}
    \tilde{u}=\tilde{v}=0 \textrm{ at } y = 0.
\end{equation}
A kinematic condition exists at the free surface, which after considering leading order dynamics yields
\begin{equation}
    \tilde{h}_t + \tilde{u}\tilde{h}_x = \tilde{v}.
\end{equation}
Additionally, a dynamic condition of the form
\begin{equation}
    -p_a \mathbf{n} - \mathbf{T} \cdot \mathbf{n} =  (\sigma \nabla \cdot \mathbf{n})\mathbf{n},
    \label{eq: General_dynamic_BC}
\end{equation}
is applied with $p_a$ representing atmospheric pressure, $\mathbf{n}$ the outward normal to the fluid, and the stress tensor $\mathbf{T} = -p\mathbf{I} + 2\mu \mathbf{E}$ with rate-of-strain tensor $\mathbf{E} = \textrm{sym} \left( \nabla \mathbf{u} \right)$. Utilising the length-scale disparity, one notes
\begin{align}
    \mathbf{n} = \frac{\left( -h_x,1 \right)}{\sqrt{1 + h_x^2 }} = \frac{\left( -\epsilon \tilde{h}_x,1 \right)}{\sqrt{1 + \epsilon^2\tilde{h}_x^2 }}\approx \left( -\epsilon \tilde{h}_x, 1 \right), \\
    \mathbf{t} = \frac{\left( 1, h_x \right)}{\sqrt{1+h_x^2}}\approx \left(1, \epsilon \tilde{h}_x \right), \\
    \nabla \cdot \mathbf{n} = -\frac{h_{xx}}{\left( \sqrt{1+h_x^2}\right) ^3}\approx -\frac{\epsilon}{L}\tilde{h}_{xx}.
\end{align}
Returning to condition \eqref{eq: General_dynamic_BC}, one can divide this into a tangential and normal stress balance through the use of a scalar product on \eqref{eq: General_dynamic_BC}, using $\mathbf{t}$ and $\mathbf{n}$ respectively. This evaluates to
\begin{align}
    \tilde{u}_y=0 \textrm{ at } \tilde{y}=\tilde{h}, \\
    \tilde{p} - p_a = -\frac{1}{\epsilon^{-3}\textit{Ca}}\tilde{h}_{xx}  \textrm{ at } \tilde{y}=\tilde{h},
\end{align}
for the tangential and normal stress balances respectively.

We may integrate the non-dimensional continuity equation with respect to $\tilde{y}$ from $\tilde{y}=0$ to $\tilde{y}=\tilde{h}$. Applying Leibniz's theorem then the kinematic boundary condition and no-slip condition yields,
\begin{equation}
    \left( \int_0^{\tilde{h}} \tilde{u} d\tilde{y}\right)_x + \tilde{h}_t = 0.
    \label{eq: B_Disguised_h_t_plus_q_x}
\end{equation}
Looking to the simplified, non-dimensionalised $y$-momentum equation, we see pressure is independent of $y$. Consequently, the normal stress balance implies
\begin{equation}
    \tilde{p}_x = -\frac{1}{\epsilon^{-3}\textit{Ca}}\tilde{h}_{xxx},
\end{equation}
which results in a modified $x$-momentum equation
\begin{equation}
     -\frac{1}{\epsilon^{-3}\textit{Ca}}\tilde{h}_{xxx} = \tilde{u}_{yy}.
\end{equation}
Integrating and applying boundary conditions yields a parabolic horizontal velocity profile,
\begin{equation}
    \tilde{u} = -\frac{1}{\epsilon^{-3}\textit{Ca}}\tilde{h}_{xxx}y\left( \frac{\tilde{y}}{2} - \tilde{h} \right).
\end{equation}
With this expression, we evaluate \eqref{eq: B_Disguised_h_t_plus_q_x},
\begin{equation}
    \tilde{h}_t + \frac{1}{3\epsilon^{-3}\textit{Ca}}\left( \tilde{h}^3\tilde{h}_{xxx}\right)_x = 0,
    \label{eq: Basic_thin_film_eqn_normal_Ca}
\end{equation}
to give a thin film evolution equation in the desired form. It is then convenient to define a re-scaled capillary number, $\bar{Ca} \coloneq \epsilon^{-3}\textit{Ca}$. 
Returning to a dimensional form, we obtain
\begin{equation}
    h_t = -\frac{\sigma}{3\mu} \left(h^3h_{xxx}\right)_x.
    \label{eq: dimensional_basic_thin_film_eqn}
\end{equation}


\section{Residual Scaling}
\label{app: residual_scaling_nonlinear}
We know the residuals scale as
\begin{equation}
    r(x,t) \sim \Delta t h_{tt}
\end{equation}
from the local truncation error of the BDF1 scheme used. From applying a temporal derivative to \eqref{eq: basic_thin_film_irreducible} with Schwarz's theorem we can see that
\begin{align}
    h_{tt} &\sim \partial_x(\partial_t (h^3h_{xxx})) \nonumber \\
    h_{tt} &\sim \partial_x (3h^2h_th_{xxx} + h^3h_{txxx}) \nonumber \\
    h_{tt} &\sim \partial_x (3h^2h_{xxx}(\partial_x(h^3h_{xxx})) + h^3\partial_{xxx}(\partial_x(h^3h_{xxx}))).
\end{align}
Considering $h_0(x)\coloneq h(x,0) = 1+Ae^{ikx}$ with $A<1$,
$$h_{tt}|_{t=0} \sim h_0^6 k^8,$$
so
$$r(x,0) \sim \Delta t k^8 h_0^6.$$
Using $\max_x\{h(x,0)\} \approx 1.75$ and inspecting the highest activated angular wavenumber of $k=11\times2\pi$, then $r(x,0) = \mathcal{O}(10^6)$, which is consistent with our data.

Excluding early-time data improves the coefficient estimates. Performance could also be improved by generating datasets with an alternative scheme to reduce early-time residuals.

\section{Critical Tolerance and the Relation to Row Sub-Sampling}
\label{app: crit_tol_relation_to_subsampling}

We assess why the undesirable case discussed in section \ref{sec: nonlinear_feature_selection_subsection} is not realised. To see whether the critical tolerance value of the sub-sampled system is greater, we start by considering the sign of $\max_j\{x^*_j-\hat{x}_j\}$. We may look to randomised numerical linear algebra and sub-space embeddings. Relating to regression problems, $S\in\mathbb{R}^{m\times n}$ is a sub-space embedding \citep{Woodruff_2014_sketching} of the column space of $A$ if $$| ||SAx||_2 - ||Ax||_2| < \delta, \textrm{ for some small } \delta>0, \textrm{ with high probability.}$$ In essence stating that the geometry of the column space of $A$ is approximate to $SA$. Re-calling that the row-sampling matrix $S$ is constructed by uniform random sampling with replacement, it is not guaranteed to be a sub-space embedding. However, we are aided by a number of factors. Using clean data and large over-sampling with $n\gg p$ and the expected number of uniquely sampled rows $\mathbb{E}[m] = (1-e^{-1})n \gg p$ increases the probability of a sub-space embedding, which would constrain $||x^*-\hat{x}||_2$. Furthermore, we recall the use of ridge regression and column normalisation with system re-scaling, so define $\tilde{x}^*, \hat{\tilde{x}}$ as the solutions to the altered full and sub-sampled systems respectively. The regularisation parameter $\lambda$ is kept constant, so has increased influence on the solution to the sub-sampled system, which is pre-conditioned in the same manner. Consequently explaining our observations that $\max_j\{\tilde{x}_j^*-\hat{\tilde{x}}_j\}>0$.

\section{Noisy Surface Tension Driven Thin Film Equation}
\label{app: Noisy_Basic_Thin_Film_Eqn}

To ensure robustness to errors relating to interface reconstruction and interpolation, which can be present in direct numerical simulation data, we look to the case of artificially contaminating finite difference based solution data with small additive Gaussian noise. The point-wise application of this Gaussian noise is more corrupting than the typical errors expected from DNS results \citep{Scardovelli_Zaleski_interface_reconstruction}. 

We focus on the case of $1\%$ additive Gaussian noise, defined with respect to the range of the data. We introduce the noise point-wise, $$\tilde{h}(x_i,t_j) = h(x_i,t_j) + \epsilon_{ij}, \quad \forall i,j \in [\![ N_x]\!] \times [\![ N_t]\!] 
$$ with $\epsilon_{ij}$ 
identically and independently distributed according to $$\epsilon_{ij}\sim \mathcal{N}(0, 0.01(\max_{x,t} \{ h(x,t)\}-\min_{x,t} \{ h(x,t) \} )).$$ A key consideration is whether to mass-correct, because prior system knowledge informs us of mass conservation of wave profiles over time. We can apply this to post-process the artificially corrupted data, but doing so alters the statistical properties of the noise, as $\epsilon^c_{ij{}} = \epsilon_{ij} - \frac{1}{N_x}\sum_i\epsilon_{ij}$. Consequently, the noise is no longer uncorrelated with respect to space. Our treatment means it remains uncorrelated with respect to time, so $\epsilon$ satisfies $$\mathrm{Var}(\epsilon_{ij}^c) = \sigma^2\left( 1 - \frac{1}{N_x}\right), \quad \mathrm{Cov} (\epsilon^c_{ij},\epsilon^c_{kj}) = -\frac{\sigma^2}{N_x}.$$ 

We choose to additively mass correct for all studied cases, enforcing $$\frac{1}{L_X}\sum_{i=1}^{N_x} \tilde{h}(x_i,t_j)\Delta x = 1,  \quad \forall j \in [\![ N_t]\!].$$ We study the case of adding Gaussian noise to every data-point, adding Gaussian noise to a sub-sample of data-points, and a boundary processed case, where point-wise noise is pre-multiplied by a function, $$ \frac{1}{4} \left(1 + \tanh \left( \alpha \left( x-\frac{L_X}{10}\right)\right)\right)\left( 1 - \tanh \left(\alpha \left(x-\frac{9L_X}{10}\right)\right)\right) - f_0,$$ with a steepness parameter $\alpha \in \mathbb{R}$, before being added to clean data. We use $f_0\in \mathbb{R}$ to ensure that noise is zero at the endpoints of our finite length domain. We see $f_0 \rightarrow 0$ as $\alpha L_X \rightarrow \infty$.

With noisy input data, we discuss alterations to our numerical differentiation schemes from \ref{sec: numerical_differentiation}.

\subsection{Smoothing}
\label{app: Smoothers}
Cases with noise require more careful consideration and real-world data is rarely provided in a clean format, so we are motivated to study methods of de-noising data. To de-noise data, we apply a two-stage process of de-noising then discovery, using standard techniques, instead of attempting to optimise and de-noise concurrently through solving a dual-problem. This pre-smoothing paradigm is readily applicable to systems with direct observations and predictors. For equation discovery, the observable and the linear system as a whole requires construction from data, so pre-smoothing is motivated by improving the stability of numerical differentiation. The application of low-pass spectral filters and Gaussian convolution are considered for pre-processing before applying equation discovery techniques.

We begin by considering a Gaussian convolution of data, which is a fast de-noising technique that can remove high-frequency noise and gives strong performance for cases with small, uncorrelated, non-skewed noise with zero mean \citep{d1989gaussian}. We take the 2D convolution over a finite domain $$(h*G)(x,t) = \int_0^{L_X} \int_0^{t_F} h(\tilde{x},\tilde{t})G(x-\tilde{x},t-\tilde{t})d\tilde{t}d\tilde{x},$$ and in the discrete setting this becomes a `sliding' matrix multiplication $$(h*G)_{ij}=\sum_{n=0}^{N_x-1} \sum_{m=0}^{N_t-1} h_{nm}G_{i-n,j-m}, \quad \mathrm{ where } \quad 0\leq i<N_x+N_G-1 \quad \textrm{and} \quad 0\leq j < N_t+M_G-1, $$ with a Gaussian kernel, $G\in\mathbb{R}^{N_G\times M_G}$, that approximates the function $$G(x,y) = \frac{1}{2\pi \sigma^2} e^{(-\frac{x^2+y^2}{2\sigma^2})}.$$ A commonly used Gaussian kernel is that of $G\in \mathbb{R}^{3\times 3}$ with 
\begin{equation}
    G = \frac{1}{16} \begin{pmatrix}
1 & 2 & 1\\
2 & 4 & 2\\
1 & 2& 1
\end{pmatrix}.
\label{eq: typical_Gaussian_convolution_kernel}
\end{equation}

Looking to a 1D variant, the discrete setting yields
$$(h*G)_{ij}=\sum_{m=0}^{N_t-1} h_{im}G_{j-m}, \quad \mathrm{ where } \quad 0\leq j < N_t+M_G-1, $$
with a Gaussian kernel $G \in \mathbb{R}^{M_G}$. This is analogous when applied to the spatial dimension. We use kernel $$G = \frac{1}{4} \left( 1 \quad 2\quad 1\right).$$

One consideration with the application of Gaussian blurring is that of edge handling. The data is the solution of a non-trivial IVP and the temporal boundaries are non-periodic, whereas spatial boundaries are periodic. When applying the convolution at the edges of the domain, the multiplication attempts to access values from outside of the domain. In our work, we use the periodicity of spatial values and don't apply the convolution at temporal boundaries. 

For applying equation discovery to our systems of interest, fourth-order spatial derivatives are required. Consequently, Gaussian blurring alone was found to be insufficient. To remove the noise, a larger kernel is required to approximate $G(x,y)$ with large $\sigma$, degrading the data quality too greatly.

We now consider a global Butterworth filter, which is a spectral method that involves applying the Fourier transform for low pass filtering within frequency space. Spectral filtering can use a pre-defined hard threshold, so high frequency oscillations above the threshold are discarded. Whilst this is effective at reducing aleatoric error, it requires careful consideration of the threshold value and indiscriminately removes noise and high-frequency dynamics. A typical cut-off is that of $(2/3) k_{max}$, meaning that the highest $1/3$ of frequencies are removed \citep{boyd_2001}, although tailoring this cut-off is highly recommended \citep{AHNERT_Abel_2007}. The Butterworth filter provides a smooth variant that is applied in Fourier space to smoothly reduce Fourier coefficients of high frequency signal components. This still requires a choice of cut-off wavenumber as well as a filter steepness parameter. The Butterworth filter can be applied as a 1D or 2D variant. The 2D variant accounts for joint spatial and temporal frequencies, so can respect the full spatio-temporal structure of the data. Alternatively, one may apply the 1D Butterworth filter sequentially along dimensions for a fast application of the scheme. We select sequential 1D application, due to the lack of periodicity of our data with respect to time. Spectral methods have implicit periodicity assumptions, so we apply the 1D Butterworth filter in the spatial dimension only. Some literature suggests padding to apply spectral filters to non-periodic data \citep{boyd_2002}.

Rather than manually choosing the filter hyper-parameters, we use an automatic selection scheme. To enable this, a range of filter hyper-parameter values is manually inputted and selection is done via a grid-search of values. For every tested hyper-parameter pair, the Butterworth filter is applied to the dataset and a mean square error is calculated using a comparison dataset. This comparison data is the original dataset that has been 2D Gaussian convolved \citep{li_Carvalho_2024} and had a hard spectral filter with cut-off ratio $0.5$ applied. The minimiser of this mean square error is the selected hyper-parameter pair. 

\subsection{Results}

With noisy inputted data, we apply the techniques discussed in section \ref{app: Smoothers}. To produce high quality point-wise spatial derivative estimates, we apply the 1D Butterworth filter along the spatial dimension of the noisy dataset for each point in time and automatically select filter parameters using a dense linear sampling of threshold values and a log-linear sampling of steepness values for a grid search. With this smoothing, derivative estimates are calculated using spectral differentiation. The choice of this filtering method respects the periodicity of the data and effectively removes high-frequency noise, because the dataset initial condition is composed of a sparse selection of activated frequencies with low associated wavenumbers. To produce temporal derivative estimates, we process the data separately. Taking the original noisy dataset, a 1D Gaussian convolution is applied and the observable is calculated using finite differences. The low order of the temporal derivative required, relative to the spatial derivatives, allows for a simpler de-noising process. Having applied smoothing to the training data and generating an associated linear system, the remaining equation discovery process remains unchanged.

\begin{figure}[hbtp]
    \centering
    \includegraphics[width=0.95\linewidth]{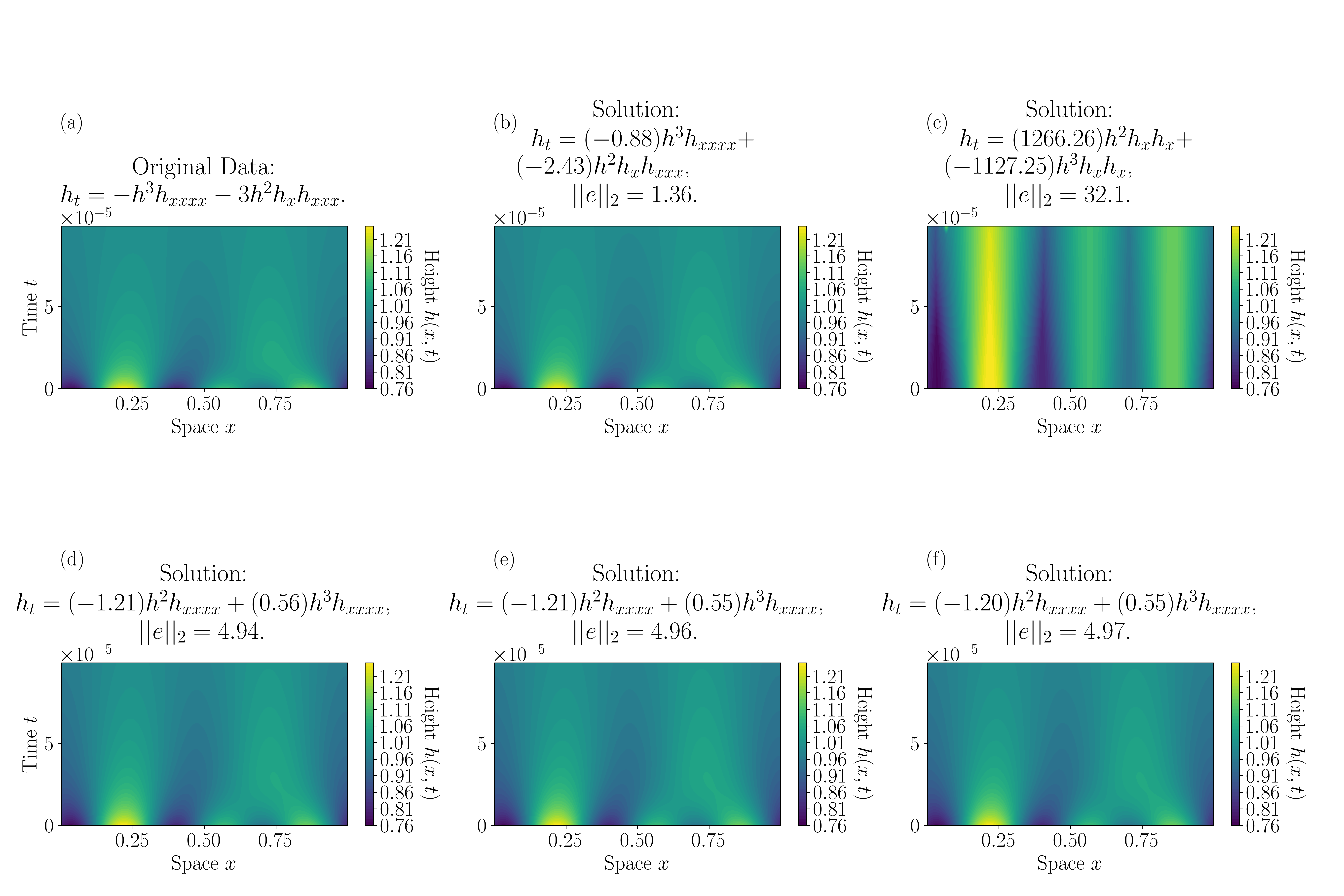}
    \caption{(a) shows the solution data for a validation initial condition solved using finite differences. 
    (b) shows the evolution of the same initial condition, using the equation discovered from clean data using our pipeline. 
    (c) shows the evolution of the same initial condition, using the equation discovered from noisy data using our pipeline without filtering/smoothing. 
    The evolution (d) relates to the equation found using data corrupted by noise at every point on the input data. 
    The evolution (e) relates to the equation found using data corrupted by noise at every point pre-multiplied by a tanh like function. So there is effectively clean data within 0.1L of the spatial boundaries, x=0, x=L and noisy data in the "bulk". 
    The evolution (f) relates to the equation found using data corrupted by noise at a uniform sub-sampling of points on the input data. Roughly every 64 points, so 16 noise affected points per wave profile.
    Coefficients rounded to 2d.p}
    \label{fig: nonlinear_a_big_two_row_comparison}
\end{figure}
All studied cases of noise corrupted data are treated in the same manner. Figure \ref{fig: nonlinear_a_big_two_row_comparison} shows an overview of our results with the best corrected AIC models shown for the noise affected cases. There is strong qualitative agreement between the original data and the discovered equation on clean datasets, as well as a low $L_2$ error to support this. The inclusion of noise with no handling, such as filtering or smoothing, yields a very poor result that is similar to a null model, $h_t = 0$. This is expected. Applying a filtering/smoothing process to noise applied point-wise improves results significantly and the recovered equations are quite distinct from the null model. We see that despite recovery of the exact form failing, equations that give lower errors than the naively processed noisy case are produced. Slightly worse performance is seen for the case of sparsely applied noise. Inspecting the sets of candidate models produced, we see single-term and two-term models proposed before a large jump in model complexity to $7$ to $12$ terms. The jump in complexity is larger for the case of sparsely applied noise and may indicate that it is important to consider applying different processes for sparse and dense noise, with knowledge of the density of noise needed \textit{a priori}. Existing techniques such as robust principal component analysis (r-PCA) have shown great success when applied to cases in which noise is sparse \citep{Candes_Li_Ma_Wright_2011, Scherl_Strom_Shang_Williams_Polagye_Brunton_2020}.

Our results reinforce that noise affected datasets present complication to strong-form equation discovery and that this is the case for thin liquid film datasets. However, we see significant improvements in model quality from the application of simple de-noising techniques to the additive Gaussian cases studied. 

\section{A Sufficient Bound for Recovery}
\label{app: Small_time_asymptotics}
We consider a PDE system $h_t = \mathcal{N}(h)$ that we may represent by $\sum_{i=1}^p \xi_i^* \phi_i(h(x,t))$ with a sparse, true coefficient vector $\xi^* \in \mathbb{R}^{p}$ and $p$ elements of a function library. We define the set $S^*\coloneq \textrm{supp}(\xi^*)$. We assume all terms describing the PDE are contained within the function library and that the formed design matrix is full-rank. We also assume the library is generated from products and derivatives of observations $h(x,t)$.

We take a multi-modal initial condition $h(x,0) = 1 + A(\cos(k_1x) + \cos(k_2x))$ for fixed $k_1,k_2 \in \mathbb{N}$ that are co-prime. For small time, we see $h(x,t) = 1 + A(\cos(k_1x) + \cos(k_2x)) + \psi(x,t)$, where $\psi(x,t)$ is a small perturbation to the initial condition. Then $\psi(x,t) = \psi_0 + t\Psi(x) + \mathcal{O}(t^2)$ with $\psi_0 \equiv 0$. So $h(x,t) = 1 + A(\cos(k_1x) + \cos(k_2x)) + t\Psi(x) + \mathcal{O}(t^2)$. Taking the temporal derivative and evaluating at $t=0$, we see $\Psi(x) = h_t|_{t=0}$ and to leading order $h_t = \Psi(x) + \mathcal{O}(t)$. 

We consider basis functions $$\phi_j(h(x,t)) = \phi_j(h(x,0)) + tD\phi_j[h_0]\mathcal{N}(h_0) + \mathcal{O}(t^2).$$

Consider $\Theta_0, b_0$ defined by $[\Theta_0]_{ij} = \phi_j(h_0(x_i))$ and $[b_0]_i = \mathcal{N}(h_0)(x_i)$. Then $\Theta_0\in \mathbb{R}^{N_x\times p}, b_0 \in \mathbb{R}^{N_x}$ are row subsampled forms of $\Theta, b$ respectively that relate to the observations of the initial condition and we further assume $N_x > p$ to allow $\Theta_0$ to also be of full-rank, given a sufficiently spectrally rich initial condition is selected.

In the small time limit, $\Theta(t_F) = \Theta_0 + \mathcal{O}(t_F)$ and $b(t_F) = b_0 + \mathcal{O}(t_F)$. We consider the leading order linear system $\Theta_0\xi = b_0$. 

We have the ridge regression solution $\hat{\xi}^{(\lambda)} = (\Theta_0^T\Theta_0 + \lambda I)^{-1}\Theta_0^Tb_0$ with regularisation constant $\lambda \in \mathbb{R}_{\geq 0}$. With small time, we assume $b_0 = \Theta_0\xi^* + \delta b$, where $\delta b \in \mathbb{R}^{N_x}$ represents the residual. So $$\hat{\xi}^{(\lambda)} = (\Theta_0^T\Theta_0 + \lambda I)^{-1}\Theta_0^T\Theta_0 \xi^* + (\Theta_0^T\Theta_0 + \lambda I)^{-1}\Theta_0^T\delta b.$$

We see
\begin{align}
(\Theta_0^T\Theta_0 + \lambda I)^{-1}\Theta_0^T\Theta_0 \xi^* &= (\Theta_0^T\Theta_0 + \lambda I)^{-1}(\Theta_0^T\Theta_0 +\lambda I - \lambda I)\xi^* \\
&= (\Theta_0^T\Theta_0 + \lambda I)^{-1}(\Theta_0^T\Theta_0 +\lambda I)\xi^* - (\Theta_0^T\Theta_0 + \lambda I)^{-1}\lambda I\xi^* \\
&= \xi^* - \lambda (\Theta_0^T\Theta_0 + \lambda I)^{-1} \xi^*,
\end{align}
so
$$\hat{\xi}^{(\lambda)} = \xi^* - \lambda (\Theta_0^T\Theta_0 + \lambda I)^{-1} \xi^* + (\Theta_0^T\Theta_0 + \lambda I)^{-1}\Theta_0^T\delta b.$$
Therefore
\begin{equation}
    ||\hat{\xi}^{(\lambda)} - \xi^*||_2 = ||- \lambda (\Theta_0^T\Theta_0 + \lambda I)^{-1} \xi^* + (\Theta_0^T\Theta_0 + \lambda I)^{-1}\Theta_0^T\delta b||_2,
\end{equation}
with triangle inequality
\begin{equation}
    ||\hat{\xi}^{(\lambda)} - \xi^*||_2 \leq || \lambda (\Theta_0^T\Theta_0 + \lambda I)^{-1} \xi^*||_2 + ||(\Theta_0^T\Theta_0 + \lambda I)^{-1}\Theta_0^T\delta b||_2,
\end{equation}
and by submultiplicativity 
\begin{equation}
    ||\hat{\xi}^{(\lambda)} - \xi^*||_2 \leq || \lambda (\Theta_0^T\Theta_0 + \lambda I)^{-1}||_2 ||\xi^*||_2 + ||(\Theta_0^T\Theta_0 + \lambda I)^{-1}\Theta_0^T||_2||\delta b||_2.
\end{equation}
With the symmetric nature of the gram matrix $\Theta_0^T\Theta_0$, we consider the largest eigenvalue of $(\Theta_0^T\Theta_0 + \lambda I)$ for the first term. For the second additive term, we consider the singular value decomposition of $\Theta_0 = U\Sigma V^T$ with orthogonal $U,V$ to give 
\begin{align}
    (\Theta_0^T\Theta_0+\lambda I)^{-1} &= (V(\Sigma^T\Sigma + \lambda I)V^T)^{-1} \\
    &= V(\Sigma^T\Sigma + \lambda I)^{-1}V^T.
\end{align}
Thus, using the orthogonality of $U$ and $V$
\begin{align}
    ||(\Theta_0^T\Theta_0+\lambda I)^{-1}\Theta_0^T||_2 &= ||V(\Sigma^T\Sigma + \lambda I)^{-1}V^T V\Sigma^TU^T||_2 \\
    &= ||V(\Sigma^T\Sigma + \lambda I)^{-1}\Sigma^TU^T||_2 \\
    &= ||(\Sigma^T\Sigma + \lambda I)^{-1}\Sigma^T||_2,
\end{align}
leaving the norm of a diagonal matrix.
\begin{align}
    ||\hat{\xi}^{(\lambda)} - \xi^*||_2 &\leq \frac{\lambda||\xi^*||_2}{\sigma_{min}(\Theta_0^T\Theta_0)+\lambda} + \max_i \left\{ \frac{\sigma_i(\Theta_0)}{\sigma_i(\Theta_0)^2 + \lambda}\right\} ||\delta b||_2 \\
    &\leq \frac{\lambda||\xi^*||_2}{\sigma_{min}(\Theta_0^T\Theta_0)+\lambda} + \frac{1}{2\sqrt{\lambda}} ||\delta b||_2. 
\end{align}
Now we know that $|\hat{\xi}^{(\lambda)}_j- \xi^*_j| = \sqrt{(\hat{\xi}^{(\lambda)}_j- \xi^*_j)^2} \leq \sqrt{\sum_j(\hat{\xi}^{(\lambda)}_j- \xi^*_j)^2}=||\hat{\xi}^{(\lambda)}- \xi^*||_2, \forall j$
and so
\begin{equation}
    |\hat{\xi}^{(\lambda)}_j- \xi^*_j| \leq \frac{\lambda||\xi^*||_2}{\sigma_{min}(\Theta_0^T\Theta_0)+\lambda} + \frac{1}{2\sqrt{\lambda}} ||\delta b||_2 ,
\end{equation}
with reverse triangle inequality
\begin{align}
    |\hat{\xi}^{(\lambda)}_j| &= |\xi^*_j+\hat{\xi}^{(\lambda)}_j-\xi^*_j|\\ &\geq \left| |\xi^*_j| - |\hat{\xi}^{(\lambda)}_j-\xi^*_j|\right| \\ &\geq |\xi^*_j| - |\hat{\xi}^{(\lambda)}_j-\xi^*_j|,
\end{align}
so
\begin{equation}
    |\hat{\xi}^{(\lambda)}_j| \geq |\xi^*_j| - \frac{\lambda||\xi^*||_2}{\sigma_{min}(\Theta_0^T\Theta_0)+\lambda} - \frac{1}{2\sqrt{\lambda}} ||\delta b||_2 .
\end{equation}

Considering also 
\begin{equation}
    |\hat{\xi}^{(\lambda)}_j|- |\xi^*_j| \leq \left||\hat{\xi}^{(\lambda)}_j|- |\xi^*_j|\right| \leq |\hat{\xi}^{(\lambda)}_j- \xi^*_j|,
\end{equation}
and taking the case of terms outside of the true support with $\xi_j^*=0$
\begin{equation}
    |\hat{\xi}^{(\lambda)}_j| \leq \frac{\lambda||\xi^*||_2}{\sigma_{min}(\Theta_0^T\Theta_0)+\lambda} + \frac{1}{2\sqrt{\lambda}} ||\delta b||_2 .
\end{equation}

Considering the lower bound on components of the ridge regression solution corresponding to the true support $S^*$ and the upper bound on the components associated with spurious terms in $(S^*)^C$, we see a gap between terms in the true support requiring
\begin{equation}
    |\xi^*_j| - \frac{\lambda||\xi^*||_2}{\sigma_{min}(\Theta_0^T\Theta_0)+\lambda} - \frac{1}{2\sqrt{\lambda}} ||\delta b||_2 > \frac{\lambda||\xi^*||_2}{\sigma_{min}(\Theta_0^T\Theta_0)+\lambda} + \frac{1}{2\sqrt{\lambda}} ||\delta b||_2 .
\end{equation}
for a strict separation. Thus, for the minimum coefficient we need
\begin{equation}
    \min_{j\in S^*}\{ |\xi_j^*| \} > \frac{2\lambda||\xi^*||_2}{\sigma_{min}(\Theta_0^T\Theta_0)+\lambda} + \frac{1}{\sqrt{\lambda}} ||\delta b||_2 \geq 0.
    \label{eq: app_Minimum_Coefficient_Condition}
\end{equation}

Consequently, we have a sufficient, though not necessary, condition. The condition is conservative and is similar to classical variable selection literature \citep{buhlmann_de_2013}, effectively specifying a sufficient signal strength. 

For our system, the condition \eqref{eq: app_Minimum_Coefficient_Condition} was not satisfied for any training datasets considered. However, we see a sufficiently spectrally rich initial condition was selected, evidenced by successful recoveries and intuitively supported through the activation of co-prime frequencies. We observed that $\Theta$ was numerically of full rank. The number of spatial samples was sufficient ($N_x=256>60=p$) and reduced aliasing effects, particularly with low-frequencies being initially excited. Furthermore, we may deem the PDE to be amenable to recovery, as the fourth-order parabolic system lacks advective terms and models a single physical effect.

\section{Derivation Details for Thin Liquid Films under Gravity}
\label{sec: Gravity_thin_film_derivation}

Once again, we begin with a 2D incompressible Navier-Stokes system
\begin{equation}
    \begin{cases}
        \rho \left( u_t + [uu_x + vu_y]\right) &= -p_x + \mu (u_{xx} + u_{yy}) + \rho g\sin(\theta), \\
        
        \rho \left( v_t + [uv_x + vv_y]\right) &= -p_y +  \mu (v_{xx} + v_{yy}) - \rho g\cos(\theta), \\
        \hspace{12mm} u_x + v_y &= 0,
    \end{cases}
    \label{eq: Dim_Grouped_NS_grav}
\end{equation}
and apply non-dimensionalisation using,
\begin{align*}
        x = \tilde{x}L, \quad
        y = \tilde{y}\epsilon L, 
        \quad
        u = \tilde{u}U,& 
        \quad
        v = \tilde{v}\epsilon U, 
        \quad
        t = \tilde{t}\frac{L}{U},
        \quad
        p = \tilde{p} \frac{\mu U}{L},
        \quad
        g = \tilde{g}\frac{U^2}{L}, \\
        \textit{Re} \coloneq \frac{\rho UL}{\mu}, 
        \quad
        &\textit{Ca} \coloneq \frac{\mu U}{\sigma},
        \textit{St} \coloneq \frac{\rho gL^2}{\mu U},
\end{align*}
where $\epsilon\coloneq \frac{H}{L} \ll 1$ and $H$ represents the vertical length-scale which is disparate from the horizontal length-scale.

This yields the 2D non-dimensional incompressible Navier-Stokes system
\begin{equation}
    \begin{cases}
        \epsilon^2\textit{Re} \left( \tilde{u}_t + [\tilde{u}\tilde{u}_x + \tilde{v}\tilde{u}_y]\right) &= -\tilde{p}_x +  (\epsilon^2\tilde{u}_{xx} + \tilde{u}_{yy}) + \epsilon^2\textit{St}\sin(\theta), \\
        \epsilon^4\textit{Re}\left( \tilde{v}_t + [\tilde{u}\tilde{v}_x + \tilde{v}\tilde{v}_y]\right) &= -\tilde{p}_y +  (\epsilon^4\tilde{v}_{xx} + \epsilon^2\tilde{v}_{yy}) - \epsilon^3\textit{St}\cos(\theta), \\
        \hspace{12mm} \tilde{u}_x + \tilde{v}_y &= 0.
    \end{cases}
    \label{eq: ND_Grouped_NS_grav}
\end{equation}

We then additionally assume that $\epsilon^2 \textit{Re} \ll 1$, $\epsilon^3 \textit{St}\cos (\theta) \sim \mathcal{O}(1)$, and $\epsilon^2 \textit{St}\sin (\theta) \sim \mathcal{O}(1)$ to obtain
\begin{equation}
    \begin{cases}
        -\tilde{p}_x +  \tilde{u}_{yy} + \epsilon^2\textit{St}\sin(\theta)
        &= 0,
        \\
        -\tilde{p}_y - \epsilon^3\textit{St} \cos(\theta) &= 0, \\
        \hspace{12mm} \tilde{u}_x + \tilde{v}_y &= 0.
    \end{cases}
    \label{eq: ND_Grouped_NS_Lubrication_more_grav}
\end{equation}

With this simplified system, we now consider boundary conditions. A no-slip boundary is applied at $y=0$,
\begin{equation}
    \tilde{u}=\tilde{v}=0 \textrm{ at } y = 0.
\end{equation}
A kinematic condition exists at the free surface, which after considering leading order dynamics gives
\begin{equation}
    \tilde{h}_t + \tilde{u}\tilde{h}_x = \tilde{v},
    \quad \textrm{at} \quad y=h(x,t).
\end{equation}
Additionally, a dynamic condition of the form
\begin{equation}
    -p_a \mathbf{n} - \mathbf{T} \cdot \mathbf{n} = \mathbf{n} \sigma \nabla \cdot \mathbf{n},
    \label{eq: General_dynamic_BC_grav}
\end{equation}
is applied with $p_a$ representing atmospheric pressure, $\mathbf{n}$ the outward normal to the fluid, and the stress tensor $\mathbf{T} = -p\mathbf{I} + 2\mu \mathbf{E}$ with rate-of-strain tensor $\mathbf{E} = \textrm{sym} \left( \nabla \mathbf{u} \right)$. Utilising the length-scale disparity, one notes
\begin{align}
    \mathbf{n} \approx \left( -\epsilon \tilde{h}_x, 1 \right), \\
    \mathbf{t} \approx \left(1, \epsilon \tilde{h}_x \right), \\
    \nabla \cdot \mathbf{n} \approx -\frac{\epsilon}{L}\tilde{h}_{xx}.
\end{align}
Returning to the dynamic condition \eqref{eq: General_dynamic_BC_grav}, one can divide this into a tangential and normal stress balance, through the use of a scalar product on \eqref{eq: General_dynamic_BC_grav} using $\mathbf{t}$ and $\mathbf{n}$ respectively. This evaluates to
\begin{align}
    \tilde{u}_y=0 \textrm{ at } \tilde{y}=\tilde{h}, \\
    \tilde{p} - p_a = -\frac{1}{\epsilon^{-3}\textit{Ca}}\tilde{h}_{xx}  \textrm{ at } \tilde{y}=\tilde{h},
\end{align}
for the tangential and normal stress balances respectively.

We may integrate the non-dimensional continuity equation with respect to $\tilde{y}$ from $\tilde{y}=0$ to $\tilde{y}=\tilde{h}$. Applying Leibniz's theorem then the kinematic boundary condition and no-slip condition yields,
\begin{equation}
    \left( \int_0^{\tilde{h}} \tilde{u} d\tilde{y}\right)_x + \tilde{h}_t = 0
    \label{eq: Disguised_h_t_plus_q_x}
\end{equation}

However, pressure is no longer independent of $y$. Inspecting the $y$-momentum equation and defining $\bar{St}=\epsilon^3\textit{St}$,
\begin{equation}
    p = -\bar{St}\cos(\theta) y + C,
\end{equation}
so we resolve the constant using the normal stress balance to evaluate the surface pressure, $p|_{y=h}$,
\begin{equation}
    p = -\bar{St}\cos(\theta) (y-h) + p_a - \frac{1}{\bar{Ca}}h_{xx}.
\end{equation}
Looking to the $x$-momentum equation is now quite different,
\begin{equation}
    - \frac{1}{\bar{Ca}}h_{xxx} + \bar{St}\cos(\theta)h_x =  \tilde{u}_{yy} + \bar{St}\epsilon^{-1}\sin(\theta).
\end{equation}
After integration we have
\begin{equation}
    \bar{St} \left( \cos(\theta)h_x - \epsilon^{-1}\sin(\theta)-\frac{1}{\bar{Ca}\bar{St}}h_{xxx} \right) \left( \frac{y^2}{2}-hy \right) = u(x,y,t).
\end{equation}
Substituting this into \eqref{eq: Disguised_h_t_plus_q_x}, yields
\begin{equation}
    h_t - \left(\frac{h^3}{3}\bar{St} \left( \cos(\theta)h_x - \epsilon^{-1}\sin(\theta)-\frac{1}{\bar{Ca}\bar{St}}h_{xxx} \right) \right)_x = 0.
\end{equation}
Re-arranging and re-scaling with $\tau \coloneq \frac{T}{\bar{Ca}}\tilde{t}$ gives
\begin{equation}
    h_{\tau} = -\frac{1}{3}\left(h^3 \left( h_{xxx} + \bar{Bo} \left( \epsilon^{-1}\sin(\theta) - h_x \cos(\theta) \right) \right) \right)_x ,
\end{equation}
where we define a Bond number $\textit{Bo} \coloneq \frac{\rho g L^2}{\sigma}$ and a re-scaled Bond number $\bar{Bo} \coloneq \epsilon^{-1}\textit{Bo}$. 
We can then dimensionalise this equation to form
\begin{equation}
    h_{t} = -\frac{1}{3\mu}\left(h^3 \left( \sigma h_{xxx} + \rho g\left( \sin(\theta) - h_x \cos(\theta) \right) \right) \right)_x .
    \label{eq: Re_dimd_thin_film_with_grav}
\end{equation}

Alternatively, with equation \eqref{eq: Re_dimd_thin_film_with_grav}, we can inspect the flux,
\begin{equation}
    q(x,t) = \frac{1}{3\mu}\left(h^3 \left( \sigma h_{xxx} + \rho g\left( \sin(\theta) - h_x \cos(\theta) \right) \right) \right),
\end{equation}
and notice that the Nusselt scaling is apparent with $Q_0 = \frac{H^3}{3\mu} \rho g \sin (\theta)$. Thus, with $q = Q_0 \tilde{q}$, one sees
\begin{equation}
    \frac{q}{Q_0} = \tilde{h}^3 - \frac{H}{L} \cot (\theta) \tilde{h}^3\tilde{h}_{x} + \frac{\sigma}{\rho g \sin (\theta) }\frac{h}{L^3} \tilde{h}^3 \tilde{h}_{xxx}.
\end{equation}
Thus, defining $L = \sqrt[3]{\frac{\sigma H}{\rho g \sin (\theta) }}$, yields
\begin{equation}
    \frac{q}{Q_0} = \tilde{h}^3 - \epsilon \cot (\theta) \tilde{h}^3\tilde{h}_{x} + \tilde{h}^3\tilde{h}_{xxx}.
\end{equation}
Then, analysing $h_t + q_x = 0$, we see
\begin{equation}
    \frac{H}{T} \tilde{h}_t = - \frac{Q_0}{L} \tilde{q}_x,
\end{equation}
which we resolve through the choice of time-scale, $T = \frac{HL}{Q_0} = \frac{3\mu L}{H^2\rho g \sin (\theta)}$.

Consequently, the final non-dimensional model is
\begin{equation}
    h_t = - \left( h^3 - \epsilon \cot (\theta) h^3h_{x} + h^3h_{xxx} \right)_x,
\end{equation}
which we may write with $\Phi \coloneq \epsilon\cot (\theta)$, as 
\begin{equation}
    h_t = - \left( h^3 - \Phi h^3h_{x} + h^3h_{xxx} \right)_x.
\end{equation}

\section{Existence and Uniqueness of a Constrained Cubic Root}
\label{app: Existance_Uniqueness}
From equation \eqref{eq: grav_implicit_form_T_char}, we analyse the cubic $f(s) = s^3 + \alpha s^2 - 1$, where $\alpha \coloneq \Phi^2\frac{1}{H^2}\left( \frac{\sigma H}{\rho g} \right)^{\frac{2}{3}}$. We inspect the regime $0 < \theta < \frac{\pi}{2}$ and see that $\Phi = \epsilon \cot(\theta)>0$, so $\alpha > 0$. Taking a derivative, $f'(s) = 3s^2 + 2\alpha s$, we see $f'(s)>0, \quad \forall s>0$. For $\theta \in \left( 0, \frac{\pi}{2} \right)$, $s=\sin^{\frac{2}{3}}(\theta) \in (0,1)$. Inspecting $f(0) = -1$ and $f(1) = \alpha > 0$, we apply the intermediate value theorem to note the existence of a root for $s \in (0,1)$. Thus, with strict monotonicity we see that the root $s\in (0,1)$ is also unique for this cubic. 

For $\theta = \frac{\pi}{2}$, $\Phi=0$, so $\alpha=0$. $f(s) = s^3-1$ with real root $s=1$.